\documentclass[aps,rmp,twocolumn,showpacs,superscriptaddress]{revtex4}
\usepackage{graphicx}
\usepackage{amsmath}

\begin{document}

\title{Antiferromagnetic order and spin dynamics in iron-based superconductors}

\author{Pengcheng Dai}
\email{pdai@rice.edu}
\affiliation{
Department of Physics and Astronomy, Rice University, Houston, Texas 77005, USA
}


\begin{abstract}
High-transition temperature (high-$T_c$)
superconductivity in the iron pnictides/chalcogenides emerges from the suppression of the static antiferromagnetic order in their parent compounds, similar to
copper oxides superconductors.
This raises a fundamental question concerning the role of magnetism in the superconductivity of these materials.
Neutron scattering, a powerful probe to study the magnetic order and spin dynamics, plays an essential role in determining the relationship between magnetism and superconductivity in high-$T_c$ superconductors.
The rapid development of modern neutron time-of-flight spectrometers allows a direct determination of
the spin dynamical properties of iron-based superconductors throughout the entire Brillouin zone.
In this review, we present an overview of the neutron scattering results on iron-based superconductors, focusing on the evolution of spin excitation spectra as a function of electron/hole-doping and
isoelectronic substitution. We compare spin dynamical properties of iron-based superconductors with
those of copper oxide and heavy fermion superconductors, and discuss the common features of spin excitations
in these three families of unconventional superconductors and their relationship with superconductivity.
\end{abstract}

\maketitle

\tableofcontents

\section{Introduction}

The interplay between magnetism and superconductivity has a long history. For example,
it is well known that superconductivity
in conventional Bardeen-Cooper-Schrieffer (BCS) superconductors \cite{bcs} such as the element lanthanum
 can be suppressed by as little as 1\% gadolinium substitution as the magnetic impurity \cite{btm}.
Within the BCS framework, magnetic impurities can act as pairing breaking agents rapidly suppressing superconductivity \cite{balatsky}.
In the case of unconventional superconductors such as
 heavy fermions \cite{steglich,Stewart,qmsi}, copper oxides \cite{bednorz,PALee}, and iron pnictides/chalcogenides \cite{kamihara06,kamihara,mawkuen2,canfield,RLGreene,johnston,GRStewart,JSWen1},
the observation that superconductivity always appears near the static
antiferromagnetic (AF) order \cite{uemura} suggests that magnetism may be a common thread for
understanding the microscopic origin of unconventional superconductors and
high-transition temperature (high-$T_c$) superconductivity \cite{DJScalapino}.
Based on this premise, much work has focused on studying the interplay between magnetism and superconductivity,
particularly the high-$T_c$ copper oxide superconductors (cuprates) since its discovery in 1986
\cite{birgeneau,kivelson,armitage,Yamada,Tranquada2}. Although understanding the magnetism and
its relationship with superconductivity in cuprates is
still an area of active research \cite{Yamada,Tranquada2}, the discovery of AF order in the parent compounds of
 iron-based superconductors in 2008 \cite{cruz,qhuang,jzhao1} provided a new opportunity to study the interplay between magnetism and superconductivity.
There are already several review articles summarizing the general progress in the field of iron-based superconductors \cite{canfield,RLGreene,johnston,GRStewart,JSWen1,mazin2011n,hirschfeld,chubukov,fwang09,dagotto}.
The interplay between magnetism and superconductivity studied by neutron scattering has been
briefly discussed as well \cite{lumsden10,PCDai1}.

The purpose of this review article is to present a comprehensive account of
the AF order, spin excitations, and their relationship with superconductivity in iron pnictides and chalcogenides.
Since magnetism is generally believed to play an important role in the electron pairing mechanism of high-$T_c$ superconductivity \cite{DJScalapino},
it is important to summarize the progress in the field over the past several years and compare the outcome with previous work on
high-$T_c$ copper oxides and heavy fermion superconductors.
Neutrons--with their wavelengths comparable to the atomic spacing and their spins directly probing the unpaired electrons in solids--have
played a unique role in determining the magnetic properties of high-$T_c$ superconductors. Soon after the discovery of iron pnictide superconductor
LaFeAsO$_{1-x}$F$_x$ with $T_c=26$ K \cite{kamihara}, neutron and X-ray scattering experiments have discovered that its parent compound LaFeAsO
exhibits a tetragonal-to-orthorhombic structural distortion followed by a collinear AF order \cite{cruz,nomura}.
Since the presence of a collinear AF structure in LaFeAsO was predicted earlier in the band structure calculations as due to
a spin-density-wave order arising from nesting of the hole and electron Fermi surfaces \cite{jdong},
its confirmation by neutron scattering and the semi-metallic nature
of these materials \cite{kamihara} provided strong evidence for the itinerant origin of the magnetism in the iron based superconductors \cite{mazin2011n,hirschfeld}.
This is fundamentally different from the parent compounds of copper oxide superconductors,
which are Mott insulators because the Coulomb repulsive energy cost $U$ of having two electrons (or holes) on the same
site is much larger than the electron hopping energy $t$ \cite{PALee}.  For a Mott insulator, the static AF order arises from a saving in
energy of $4t^2/U$ via virtual hopping and is due to electron correlation effects.  Since the Mott insulating state of copper oxides is believed to play an essential
role in the pseudogap physics and
mechanism of high-$T_c$ superconductivity \cite{PALee}, it is interesting to ask whether iron-based superconductors are also close to a Mott insulator and determine the effect of
electron correlations and local moments to their electronic properties
and spin dynamics \cite{qsi,cfang,ckxu,khaule,Qazilbash}.

From the experimental point of view, a systematic determination of the magnetic structures
and the doping evolution of spin excitations
in different classes of iron-based superconductors and
their associated materials will form the basis to establish whether magnetism is responsible for high-$T_c$ superconductivity.
For copper oxides, superconductivity can be induced by charge carrier doping (electrons or holes) into the CuO$_2$,
resulting complicated phase diagrams with
many incipient states competing with superconductivity \cite{birgeneau,kivelson,armitage,Yamada,Tranquada2}.  The undoped copper oxides
such as La$_2$CuO$_4$
\cite{vaknin} and YBa$_2$Cu$_3$O$_{6+x}$ \cite{tranquada88} are simple antiferromagnets with the
neighboring spins oppositely aligned.  When holes are doped into the parent compounds, the static
AF order is gradually suppressed, but spin excitations (or short-range spin fluctuations) survive across the entire
superconducting dome and couple with superconductivity
via a collective magnetic excitation termed neutron spin resonance \cite{Yamada,Tranquada2,eschrig}.
However, inelastic neutron scattering experiments
designed to study the doping evolution of spin excitations were only carried out on the La$_{2-x}$Sr$_x$CuO$_4$ family of cuprates
across the entire phase diagram from the undoped parent compounds to heavily
overdoped non-superconducting samples \cite{Yamada2,Hayden1}.  There are no comprehensive measurements throughout the entire phase diagram on other cuprates due to material limitations (for example, YBa$_2$Cu$_3$O$_{6+x}$ cannot be hole overdoped to completely suppress superconductivity) or
the difficulty in growing large single crystals suitable for inelastic neutron scattering experiments.

In the case of iron-based superconductors, there are two major classes of materials, the
iron pnictides and iron chalcogenides \cite{johnston,GRStewart}.
Compared with the hole-doped La$_{2-x}$Sr$_x$CuO$_4$
copper oxide superconductors, where superconductivity can only be induced via substituting the trivalent La by the divalent element Ba or Sr, superconductivity in iron pnictide such as
BaFe$_2$As$_2$ \cite{rotter} can be induced by ionic substitution at any element site.
These include Ba by K/Na to form hole-doped Ba$_{1-x}$K$_x$Fe$_2$As$_2$ \cite{Johrendt1,Pramanik}, Fe by Co or Ni
to have electron-doped BaFe$_{2-x}T_x$As$_2$ ($T=$Co, Ni) \cite{Mandrus1,ZAXu1}, and
As by P in the isovalent (or isoelectronic) doped compounds BaFe$_2$As$_{2-x}$P$_x$ \cite{ZAXu2}.
While K or Na doping to BaFe$_2$As$_2$ induces the same numbers of holes to the FeAs layer,
 the effect of Ni-doping is expected to introduce twice the number of electrons
into the FeAs layer as that of Co-doping from naive electron counting point of view \cite{johnston,GRStewart}.
Since large sized single crystals can be grown by self-flux method in many of these cases \cite{canfield}, doped BaFe$_2$As$_2$ materials provide
a unique opportunity to study the evolution of the static AF order and spin excitations as a function of hole, electron, and isovalent
doping throughout the entire AF order to superconductivity phase diagram, and determine their
 connection with superconductivity.
These experiments, together with
neutron scattering studies of related materials \cite{Ishikodo09,Shamoto10,Wakimoto10,Yogesh,Johnston11,Simonson,Lamsal,kim11,marty11,inosov13,AETaylor2013},
 will establish the common features in the magnetic order and spin excitations in different families of iron-based superconductors.  The outcome, together with the results from
high-$T_c$ copper oxide and heavy fermion superconductors, and can form a basis to determine if magnetism is indeed responsible
for superconductivity in these materials \cite{DJScalapino}.

Compared with other techniques suitable to study magnetism in solids including Muon spin rotation ($\mu$SR) \cite{uemura},
 nuclear magnetic resonance (NMR) \cite{alloul},
and resonant inelastic X-ray scattering (RIXS) \cite{ament}, neutron scattering has several unique advantages:
(1) The neutron itself is a charge 0 fermion with a spin $S=1/2$, resulting a magnetic dipole moment
which can interact with unpaired electrons and magnetism in solids; (2) The charge neutrality of the neutron
renders it a weakly interacting probe with well-known scattering cross sections; (3)
 The available wavelength and energies of neutrons as a scattering probe are ideally
suited to study static magnetic order and spin excitations in solids. The general scattering principle involved is simply to measure the
number of neutrons scattered into a given solid angle at a known energy ($E=\hbar\omega$, where $\hbar$ is
the reduced Planck's constant and $\omega$ is the angular frequency) and momentum transfer (${\bf Q}$).  The laws of
conservation of momentum and energy are satisfied via ${\bf Q}={\bf k_i}-{\bf k_f}$ and $E=\hbar\omega=E_i-E_f$, where $k_i$, $E_i=\hbar^2k_i^2/2m$, $k_f$, and $E_f=\hbar^2k_f^2/2m$ are the
incident neutron wave vector, energy, outgoing neutron wave vector, and energy, respectively, and $m$ is the mass of a neutron.
The coherent magnetic scattering cross section from a system with a single species of magnetic atoms is then \cite{gyxu13},
$$
{d^2\sigma\over d\Omega dE}= {N\over\hbar}{k_f\over k_i}p^2e^{-2W}\sum_{\alpha,\beta}
(\delta_{\alpha,\beta}-\tilde{Q}_\alpha\tilde{Q}_\beta)S^{\alpha\beta}({\bf Q},\omega).
$$
Here $N$ is the number of unit cells, $p=({\gamma r_0\over 2})^2g^2f({\bf Q})^2$ (where
${\gamma r_0\over 2}=0.2695\times 10^{-12}$ cm, $g\approx 2$ is the electron spin $g$-factor, and $f({\bf Q})$
is the magnetic form factor), $e^{-2W}$ is the Debye-Waller factor, $\alpha,\beta$ are the Cartesian coordinates
$x$, $y$, and $z$, and $\tilde{Q}_\alpha$, $\tilde{Q}_\beta$ are the projections of the unit wave vector $\tilde{Q}$ onto the
Cartesian axes. $S^{\alpha\beta}({\bf Q},\omega)$ is the dynamic spin correlation function, and is
associated with the imaginary part of the dynamic susceptibility $\chi_{\alpha\beta}^{\prime\prime}({\bf Q},\omega)$ via the
fluctuation-dissipation theorem:
$$\chi^{\prime\prime}_{\alpha\beta}({\bf Q},\omega)=g^2\mu^2_{\rm B}{\pi\over\hbar}(1-e^{-\hbar\omega/k_BT})S^{\alpha\beta}({\bf Q},\omega).$$
For a paramagnet with isotropic spin excitations, $S^{xx}({\bf Q},\omega)=S^{yy}({\bf Q},\omega)=S^{zz}({\bf Q},\omega)$.
Since neutron scattering is only sensitive to spin (fluctuations) direction
perpendicular to the wave-vector transfer ${\bf Q}$, the $S({\bf Q},\omega)$ of an isotropic paramagnet
measured by unpolarized neutron scattering experiments (see Section III. F for neutron polarization analysis)
is related to $S^{zz}({\bf Q},\omega)$ via $S({\bf Q},\omega)=2S^{zz}({\bf Q},\omega)$.
By measuring $S({\bf Q},\omega)$
in absolute units via phonon or vanadium normalization \cite{gyxu13}, one can estimate the energy dependence of
the $\bf Q$-averaged or the local dynamic susceptibility
$\chi^{\prime\prime}(\omega)=\int_{\rm BZ}\chi^{\prime\prime}({\bf Q},\omega)d{\bf Q}/\int_{\rm BZ}d{\bf Q}$
within a Brillouin Zone (BZ) \cite{lester10}. The overall strength of the magnetic excitations,
corresponding to the local fluctuating moment $\left\langle{\bf m}^2\right\rangle$,
 can then be computed via \cite{lester10}
$$\left\langle{\bf m}^2\right\rangle={3\hbar\over\pi}\int_{-\infty}^\infty{\chi^{\prime\prime}(\omega)d\omega\over{1-\exp{(-\hbar\omega/k_BT)}}}.$$ One of the central purposes of inelastic neutron scattering experiments is to determine the energy and wave vector dependence of $\chi^{\prime\prime}({\bf Q},\omega)$ in absolute units for various
iron pnictides, and compare the outcome with those in copper oxide and heavy fermion superconductors.

In this article, we present a comprehensive
review of recent neutron scattering results on iron-based superconductors, mainly focusing on the evolution of the
static AF order and spin dynamics of iron pnictides and chalcogenides.
In Section II, we summarize the static AF order for various iron pnictides/chalcogenides and its doping evolution.
This includes the effects of electron and hole-doping on the static AF order and tetragonal-to-orthorhombic structural transitions (Section II. A.);
how impurity (Section II. B.) and isoelectronic substitution (Section II. C.) affect the magnetic and structural phase transitions.
Section III summarizes spin excitations and their relationship with superconductivity, including spin waves in the parent compounds
of iron-based superconductors (Section III. A.); as well as neutron spin resonance and its relationship with superconductivity (Section III. B.);
the electron and hole-doping evolution of the spin excitations in the BaFe$_2$As$_2$ family of iron pnictides (Section III. C.);
evolution of spin excitations in iron chalcogenides and alkali iron selenides (Section III. D.); impurity effects on spin excitations of iron pnictide and chalcogenide superconductors (Section III. E.); neutron polarization analysis of spin excitation anisotropy in iron pnictides (Section III. F.);
electronic nematic phase and neutron scattering
experiments under uniaxial strain (Section III. G.);
comparison with $\mu$SR, NMR,
and RIXS measurements (Section III. H.); and
comparison of spin excitations in iron-based superconductors with those in
copper oxide and heavy fermion superconductors (Section III. I.). Section IV provides a brief account of
current theoretical understanding of spin excitations in iron-based superconductors.  Finally, we
summarize the results and discuss possible future directions for the field.

\begin{figure}[t]
\includegraphics[scale=.35]{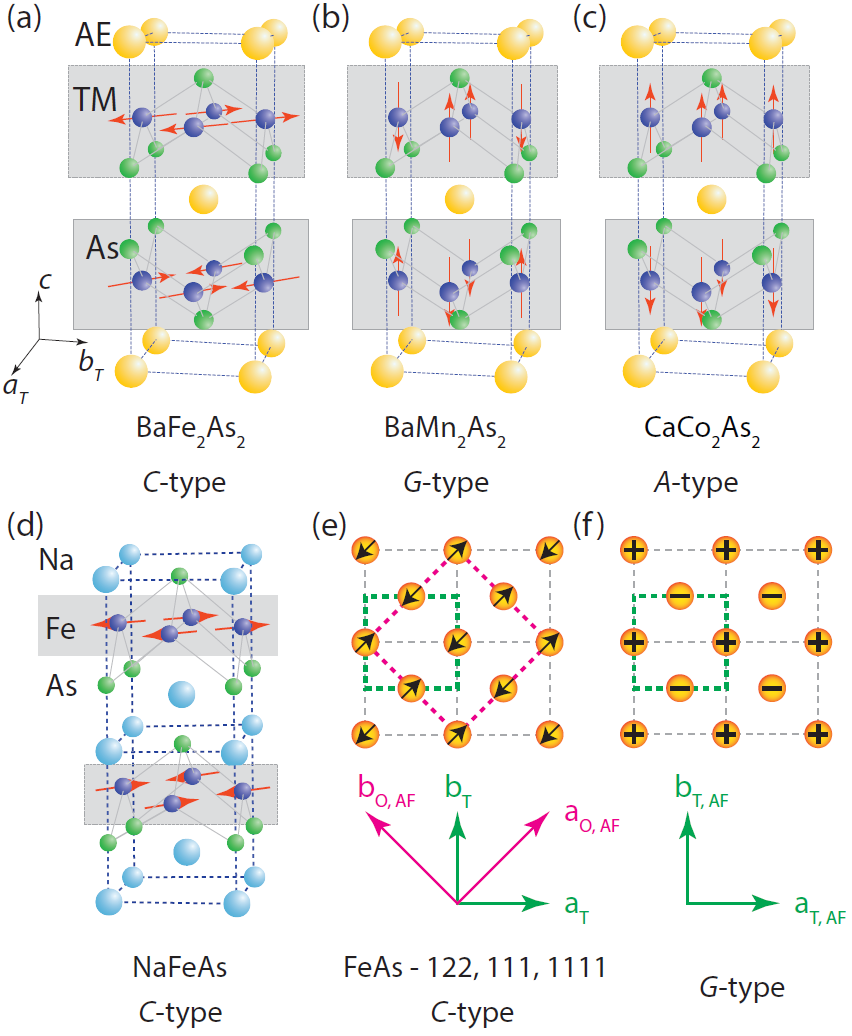}
\caption{
(Color online) Crystal structure and magnetic order in different families of pnictides.
(a) The crystal and magnetic structures of the BaFe$_2$As$_2$ in the AF ordered phase.
 The yellow, green, and blue balls indicate Ba, As, and Fe
positions, respectively.  The red arrows mark the ordered moment directions of
Fe in the AF ordered state (the $C$-type).  The $a_T$, $b_T$, and $c$ show the
Cartesian coordinate system suitable for the paramagnetic tetragonal
phase of BaFe$_2$As$_2$ \cite{qhuang}.
(b) The AF structure of BaMn$_2$As$_2$, where the ordered moments on Mn are along the
$c$-axis direction (the $G$-type) \cite{Yogesh}.
(c) The crystal structure of CaCo$_2$As$_2$, where the ordered moments
on Co form the $A$-type AF structure \cite{Quirinale}.
(d) The collinear AF order in NaFeAs doubles
the crystalline unit cell along the $c$-axis \cite{SLLi1}. (e)
The collinear $C$-type AF structure in the Fe plane, where the green dashed box
marks the tetragonal crystalline unit cell in the paramagnetic state and the magenta
dashed box indicates the orthorhombic magnetic unit cell.  The
$a_{\rm o,AF}$ and $b_{\rm o,AF}$ mark directions of the orthorhombic lattice.
(f) The in-plane moment projections for the $G$-type antiferromagnets.
}
\end{figure}

\section{Static antiferromagnetic order and its doping evolution}

\subsection{Lattice and magnetic order in the parent compounds of iron-based superconductors}

\begin{figure}[t]
\includegraphics[scale=.40]{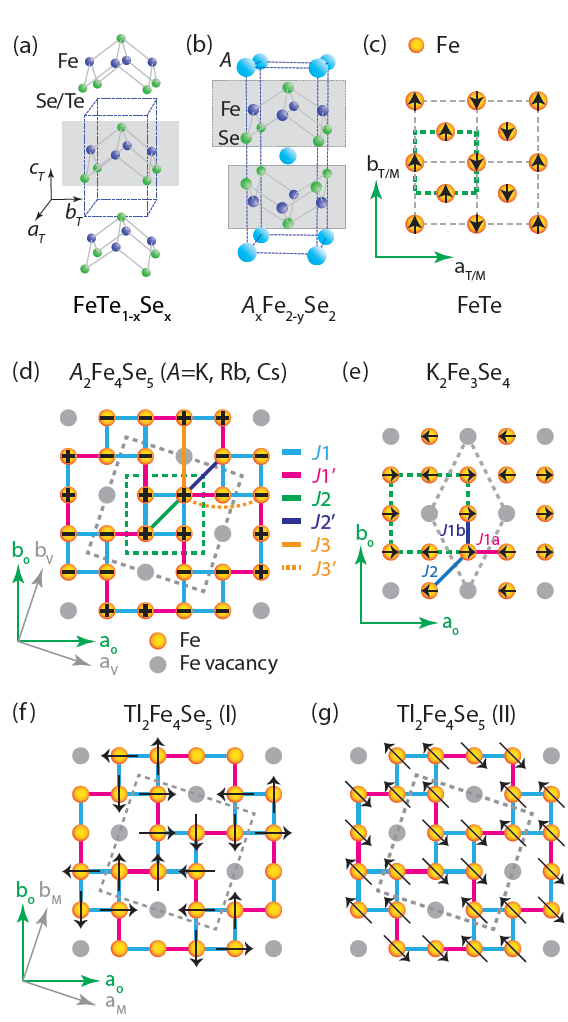}
\caption{
(Color online)
Crystal and magnetic structures of iron chalcogenides and alkali iron selenides.
(a) The crystal structure of the FeTe$_{1-x}$Se$_x$ iron chalcogenide, where Fe and
Se/Te positions are marked as blue and green, respectively. (b) The tetragonal phase
crystal structure of $A_x$Fe$_{2-x}$Se$_2$.  The positions
of $A$, Fe, and Se are marked as light blue, blue, and green, respectively.
(c) The in-plane bi-collinear magnetic structure of FeTe, where the arrows indicate the
moment directions \cite{Fruchart,WBao5,SLLi2}.
(d) The in-plane crystal and magnetic structures of
$A_2$Fe$_4$Se$_5$ in the vacancy ordered insulating phase.
Only iron positions are plotted and the grey dashed lines mark
the structural and magnetic unit cells. The light blue (magenta), green (blue), and
yellow (dashed yellow) lines represent the nearest ($J_1$, $J_1^\prime$), next nearest ($J_2$, $J_2^\prime$),
the next next nearest ($J_3$, $J_3^\prime$) neighbor exchange interactions, respectively \cite{WBao1,WBao2,mwang11,MYWang11}.
 (e) The crystal and magnetic structures of $A_2$Fe$_3$Se$_4$ in the vacancy
ordered semiconducting phase \cite{jzhao12}.  The nearest and next nearest neighbor exchange couplings are
clearly marked. (f,g) Other possible magnetic structures of Tl$_2$Fe$_4$Se$_5$ in the vacancy
ordered phase \cite{may12}.
}
\end{figure}

From a crystal structure point of view, the parent compounds of iron-based superconductors can be classified into five different families: $R$FeAsO ($R=$ La, Ce, Pr, Nd, Sm,..., the 1111 system), $A$Fe$_2$As$_2$ ($A=$ Ba, Sr, Ca, K, the 122 system), $A$FeAs ($A=$ Li, Na, the 111 system), Fe$_{1+y}$Te$_{1-x}$Se$_x$ (the 11 system), and
$A_x$Fe$_{2-y}$Se$_2$ alkali iron selenides ($A=$ K, Rb, Cs, Tl, ..., including the insulating
245 phase $A_2$Fe$_4$Se$_5$ and the semiconducting 234 phase $A_2$Fe$_3$Se$_4$)
\cite{RLGreene,johnston,GRStewart,Sadovskii,PMAswathy,dagotto,jzhao12,mwang2014}, where the 122 and 245 compounds have two FeAs(Se) layers in the unit cell and other systems have single FeAs(Se) layer.
A recent development in the field is the synthesis
of iron selenide superconductors via intercalation of molecular complexes between
layers of FeSe \cite{TPYing12,MBurrardlucas13,KrztonMaziopa12}.
The crystal structures at room temperature are all tetragonal except for the insulating
245 phase and some of them will become orthorhombic at low temperature below $T_s$.
Neutron diffraction measurements have established that
the long range AF order in the iron pnictides including the 1111, 122, 111 families is collinear with
moment aligned along the $a_o$ axis direction of the orthorhombic structure [Figs. 1(a), 1(d), and 1(e)] \cite{JWLynn}, except for the
stoichiometric LiFeAs system which is superconducting
without a magnetically ordered parent compound \cite{PChu1,XCWang,MJPitcher2008}. Although the in-plane
collinear AF structures for different classes of iron pnictides are identical [Fig. 1(e)], the ordering along the $c$ axis is material dependent.  In the
122 system, which has two magnetic irons per formula unit, the ordering is AF within a unit cell along the $c$ axis [Fig. 1(a)].
For the 111 system with one iron per unit cell, the magnetic structure doubles the chemical
unit cell along the $c$ axis  [Fig. 1(d)].
While the collinear AF structure in iron pnictide is the so-called $C$-type antiferromagnet stemming from the original work of Wollan and Koehler on perovskite manganese oxides \cite{wollan}, the related pnictide materials such as BaMn$_2$As$_2$ \cite{Yogesh} and
CaCo$_2$As$_2$ \cite{Quirinale} have the $G$- [Figs. 1(b), 1(f)] and $A$-type [Fig. 1(c)] AF structures, respectively.
Recently, another AF parent compound was found in electron-overdoped
LaFeAsO$_{1-x}$H$_x$ ($x\sim 0.5$) system in addition to the usual collinear
AF structure at $x=0$ \cite{MHiraishi14}.

\begin{table*}
\caption{\label{tab:table1} Summary of the structure transition temperature $T_s$, magnetic transition temperature $T_N$,
 and ordered magnetic moment per iron for the AF ordered parent compounds of the iron-based superconductors.
The lattice parameters in the paramagnetic tetragonal state are also listed.
}
\begin{ruledtabular}
\begin{tabular}{cccccc}
 Materials & $a_T\equiv b_T$ (\AA ) & $c$ (\AA ) & $T_s$ (K) & $T_N$ (K) & Moment/Fe ($\mu_B$) \\
\hline
LaFeAsO \footnote{\cite{cruz,MAMcGuire,QHuang2,qureshi10}.} & 4.0301 & 8.7368 & 155 & 137 & 0.36-0.6 \\
CeFeAsO \footnote{\cite{jzhao1,qzhang13}.}                  & 3.9959 & 8.6522 & 158 & 140 & 0.8      \\
PrFeAsO \footnote{\cite{JZhao2,Kimber}.}                    & 3.997  & 8.6057 & 153 & 127 & 0.48    \\
NdFeAsO \footnote{\cite{HAMook,WBao3}.}                     & 3.9611 & 8.5724 & 150 & 141 & 0.25    \\
LaFeAsO$_{0.5}$H$_{0.5}$ \footnote{\cite{MHiraishi14}.}      & 3.975  &   8.67 & 95    & 92    & 1.21 \\
CaFe$_2$As$_2$ \footnote{\cite{AIGoldman1,AIGoldman2,AIGoldman3}.} & 3.912 & 11.667  & 173    & 173    & 0.80    \\
SrFe$_2$As$_2$ \footnote{\cite{JZhao3,KKaneko,AJesche}.}     & 3.920 & 12.40 & 220    & 220   & 0.94   \\
BaFe$_2$As$_2$ \footnote{\cite{qhuang,kim11}.}              & 3.957 & 12.968 & $\sim$140    & $\sim$140  & 0.87   \\
Na$_{0.985}$FeAs \footnote{\cite{SLLi1}.}                   & 3.9448 & 6.9968 & 49     & 39      & 0.09   \\
Fe$_{1.068}$Te \footnote{\cite{WBao5,SLLi2}.}               & 3.8123 & 6.2517  & 67     & 67       & 2.25   \\
K$_2$Fe$_4$Se$_5$ \footnote{\cite{WBao1}.}                  & 8.7306 & 14.113  & 578    & 559     & 3.31    \\
Rb$_2$Fe$_4$Se$_5$ \footnote{\cite{WBao2,mwang11}.}         & 8.788  & 14.597  & 515    & 502     & 3.3     \\
Cs$_2$Fe$_4$Se$_5$ \footnote{\cite{WBao2}.}                 & 8.865  & 15.289  & 500    & 471   & 3.4     \\
TlFe$_{1.6}$Se$_2$ \footnote{\cite{may12}.}                 & $\sim$8.71 & 14.02 & 463   & 100       & $\sim$3  \\
\end{tabular}
\end{ruledtabular}
\end{table*}

For the iron chalcogenides (the 11 family) and alkali iron selenides, their crystal structures are shown in Figs. 2(a) and 2(b), respectively.
Instead of a collinear AF structure, the parent compound of the 11 family has a bi-collinear AF spin structure as shown in Fig. 2(c) \cite{Fruchart,WBao5,SLLi2}.
Compared to the collinear spin structure of the 122 family in Fig. 1(e), the iron spins are rotated 45$^\circ$ within the $a_ob_o$-plane in the 11 system. The magnetic structure in the 11 family is sensitive to the excess iron population
 in the interstitial iron site \cite{EERodriguez10,EERodriguez2011}. While the bi-collinear magnetic order is commensurate for
Fe$_{1+x}$Te with $x\leq 9\%$, it exhibits incommensurate helical magnetic order that competes with the
bi-collinear commensurate ordering close to $T_N$ for $x\geq 12\%$ \cite{EERodriguez10,EERodriguez2011,EERodriguez2013}.
The alkali iron selenides (the 245 family) \cite{XLChen,MHFang}
have two different iron vacancy structures including the insulating
$\sqrt{5}\times\sqrt{5}$ iron vacancy ordered phase [Figs. 2(d), 2(f), and 2(g)] \cite{WBao1,WBao2,mwang11} and the semiconducting
rhombus iron vacancy ordered 234 phase [Fig. 2(e)] \cite{jzhao12,mwang2014}.  While the 234 phase has a AF structure
similar to the parent compounds of the 122 family [Fig. 2(e)] \cite{jzhao12}, the
insulating 245 phase have the block AF structure with moments along the $c$ axis [Fig. 2(d)] \cite{WBao1,WBao2,mwang11} and in the
plane [Figs. 2(f) and 2(g)] \cite{may12}, respectively. Compared with the parent compounds of the 122 system,
the ordered moments of the 11 and 245 systems are much larger.  In Table I, we
summarize the lattice parameters, structure transition
temperature $T_s$, the AF phase transition temperature $T_N$, and the static
ordered moments for the parent compounds of
different iron-based superconductors.
In the 1111, 111, and 245 systems, the structural transition occurs at a temperature higher
than that of the AF phase transition \cite{johnston,GRStewart}.
In the 122 system, the structural and magnetic transitions almost occur simultaneously in the undoped parent compounds \cite{kim11}, but are well separated upon electron-doping \cite{canfield}.

\subsection{The effect of electron, hole-doping, impurity, and isoelectronic substitution on the static AF order and tetragonal-to-orthorhombic structural transitions}

\begin{figure}[t]
\includegraphics[scale=.3]{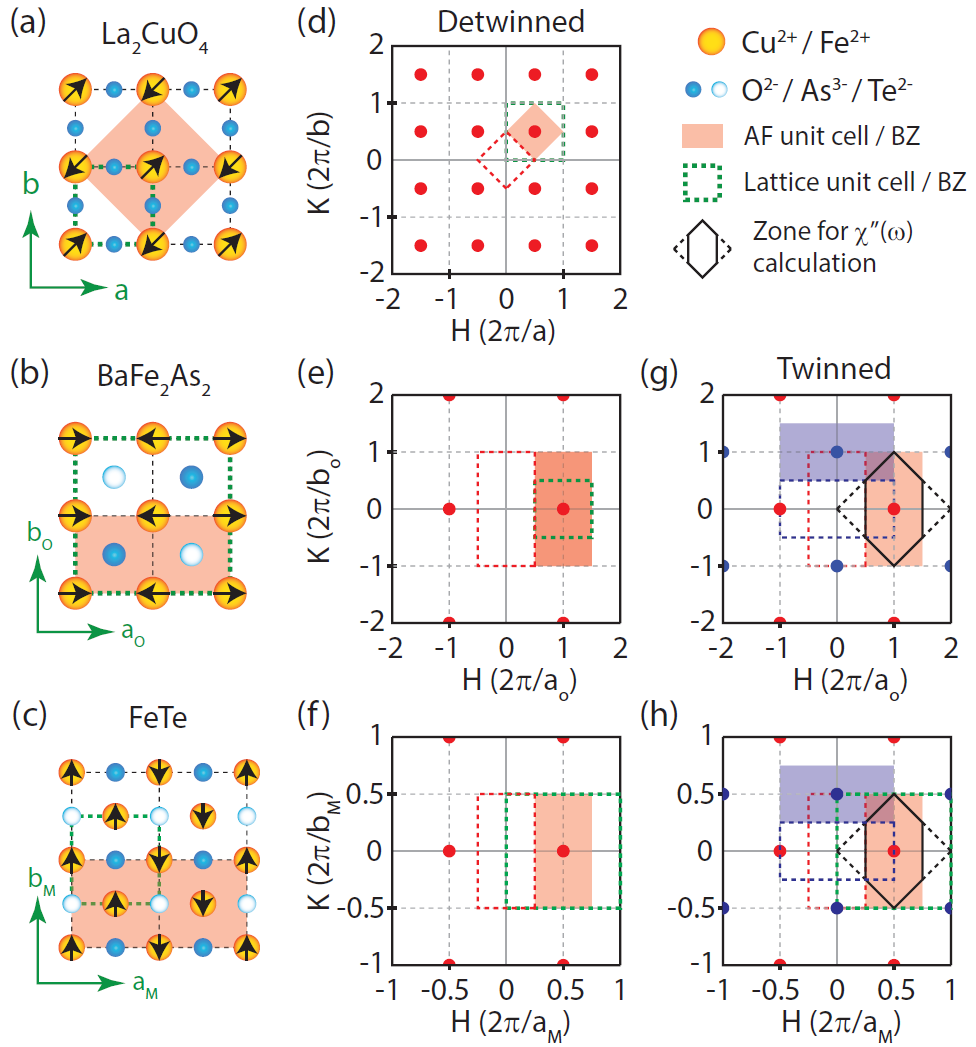}
\caption{
(Color online) Comparison of the AF structures of copper oxides and iron pnictides/chalcogenides, and the
corresponding reciprocal lattice and the twinning effect.
(a) The in-plane AF structure of the parent compound of copper oxide superconductors with
chemical and magnetic unit cells marked as dashed green
and pink area, respectively \cite{Yamada}.
(d) The reciprocal space where the solid red dots represent
the AF ordering wave vectors. The magenta area and dashed green indicate
the size of the in-plane magnetic and chemical Brillouin zone, respectively.
(b) The in-plane AF structure of BaFe$_2$As$_2$, where the open and filled blue circles
indicate As positions below and above the iron plane, respectively \cite{johnston}.
The magnetic and chemical unit cells are marked as magenta area and dashed green lines, respectively, and
(e) the corresponding reciprocal space, where red dots indicate ${\bf Q}_{\rm AF}$. (g)
The effect of twin domains for AF order and Brillouin zones.  The solid black lines mark the integration area
in reciprocal space to obtain $\chi^{\prime\prime}(\omega)$.
(c) The in-plane AF structure of FeTe, and
(f) the corresponding reciprocal space in a detwinned sample.
(h) The effect of twin domain in reciprocal space.
The magenta dashed boxes in (d), (e), and (f)
indicate the AF Brillouin zones near $\Gamma$ point
probed by RIXS.
}
\end{figure}

Before discussing the impact of electron/hole doping on the long range AF order, we shall define
momentum transfer in reciprocal space and compare the sizes of the Brillouin zones for the parent compounds of different families
of high-$T_c$ superconductors.
Figures 3(a), 3(b), and 3(c) show spin arrangements within one layer
of La$_2$CuO$_4$ \cite{vaknin}, BaFe$_2$As$_2$ \cite{qhuang}, and FeTe \cite{Fruchart,WBao5,SLLi2}, respectively. The chemical unit cells are marked as green dashed lines and the magnetic
unit cells are magenta shaded.  The positions of Cu$^{2+}$/Fe$^{2+}$ and O$^{2-}$/As$^{3-}$/Te$^{2-}$
are also marked.
The momentum transfer ${\bf Q}$ at ($q_x$, $q_y$, $q_z$) in \AA$^{-1}$ can be defined as $(H,K,L) = (q_xa/2\pi, q_yb/2\pi, q_zc/2\pi)$
in reciprocal lattice units (r.l.u.), where $a$ (or $a_o$), $b$ (or $b_o$), and $c$ are lattice parameters of the orthorhombic unit cell.
In this notation, the AF order in the parent compound of copper oxide superconductors occurs at
${\bf Q}_{\rm AF}=(H,K)=(0.5\pm m,0.5\pm n)$ where $m,n=0,1,2,\cdots$ and the first magnetic Brillouin zone is shown as
the magenta shaded box in Fig. 3(d).  Another equivalent Brillouin zone near $\Gamma$
is marked by the magenta dashed line, while the
first Brillouin zone of the chemical unit cell is the green dashed box. If the ordered moment is
entirely on the iron site in BaFe$_2$As$_2$, the chemical unit cell is twice the size of
the magnetic unit cell along the $b_o$ axis direction due to out of plane positions of the As atoms [Fig. 3(b)].
In a completely detwinned sample, the first magnetic Brillouin zone is the magenta shaded area
around ${\bf Q}_{\rm AF}=(H,K,L)=(1\pm 2m,0\pm 2n,L)$ where $L= \pm 1,3,5,\cdots$ r.l.u., larger in size
than the chemical Brillouin zone in dashed green.  Because the AF order in iron pnictides is always preceded
by a tetragonal-to-orthorhombic lattice distortion, the twinning effect in the orthorhombic state means that
AF Bragg peaks from the twinned domains appear at positions
rotated by 90$^\circ$ [Blue dots in Fig. 3(g)].  Therefore, to properly account for the twin domain
effect, one needs to carry out
wave vector integration within the region marked by solid black lines in Fig. 3(g)
to obtain the local dynamic susceptibility $\chi^{\prime\prime}(\omega)$.
Figure 3(c), 3(f), and 3(h) summarizes the bi-collinear spin structure of FeTe,
its associated magnetic Bragg peaks in reciprocal in detwinned, and twinned samples, respectively.
Depending on the size of the unit cell, the same AF Bragg peak for collinear
AF order in BaFe$_2$As$_2$ and and bi-collinear AF order in FeTe can appear with different Miller indices.
For example, if we choose unit cells of BaFe$_2$As$_2$
with one [half the magenta shaded area in Fig. 3(b) and ignoring As],
two [similar to magenta shaded
area in Fig. 3(a)], and four irons [doubling the magenta shaded area in Fig. 3(b)
along the $c$-axis], the same
AF Bragg peak would occur at in-plane wave vectors $(0.5,0)$, $(0.5,0.5)$, and $(1,0)$, respectively.
For the bi-collinear AF order FeTe, one iron and two irons per unit cell
would have the same
AF Bragg peak at the in-plane wave vectors $(0.25,0.25)$ and $(0.5,0)$, respectively.

\begin{figure}[t]
\includegraphics[scale=.40]{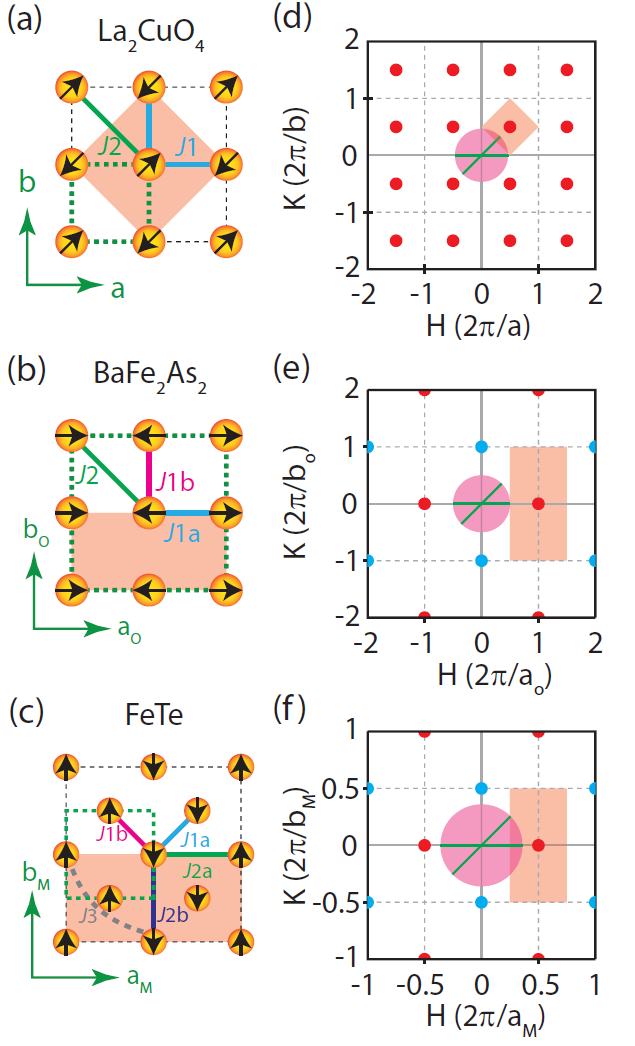}
\caption{
(Color online) Real and reciprocal space of La$_2$CuO$_4$, BaFe$_2$As$_2$, and FeTe
with magnetic exchange couplings and regions of reciprocal space probed by neutron scattering
and RIXS.
(a) The in-plane AF structure of La$_2$CuO$_4$, where nearest and next nearest
neighbor magnetic exchange couplings are marked as $J_1$ (solid blue line)
and $J_2$ (solid green line), respectively.
(d) The reciprocal space where the solid red dots represent
the AF ordering wave vectors. The purple circular area at $\Gamma$ indicates the region of reciprocal space
that can be probed by RIXS using Cu $L$-edge, while the AF Brillouin zone probed by neutron scattering
is marked by magenta square \cite{Tacon11}.
(b) The in-plane AF structure of BaFe$_2$As$_2$ with the
nearest neighbors ($J_{1a},J_{1b}$) and second nearest neighbor $J_2$ magnetic
exchange couplings.
(e) The corresponding reciprocal space, where the purple circle indicate
reciprocal space area covered by RIXS using Fe $L_3$-edge \cite{kjzhou}.
(c) The in-plane AF structure of FeTe with the
nearest neighbors ($J_{1a},J_{1b}$) and second nearest
neighbors ($J_{2a},J_{2b}$) magnetic
exchange couplings.
(f) The corresponding reciprocal space with the purple circle showing the
reciprocal space covered by RIXS.
}
\end{figure}

Figure 4(a), 4(b), and 4(c) summarizes the effective nearest neighbor and next nearest neighbor magnetic exchange couplings
for La$_2$CuO$_4$, BaFe$_2$As$_2$, and FeTe, respectively.
Figure 4(d), 4(e), and 4(f) shows the corresponding reciprocal space with ${\bf Q}_{\rm AF}$ positions marked as
red and blue dots for the two different twin domains.
While neutron scattering typically studies the magenta region of the reciprocal space within the
first Brillouin zone near ${\bf Q}_{\rm AF}$, RIXS can only probe spin excitations within the purple circles near the origin
$\Gamma$ due to kinematic constraints from energies of the incident and outgoing photons
used to enhance the magnetic scattering at Cu $L_3$-edge and Fe $L_3$-edge
\cite{Tacon11,kjzhou}.

\begin{figure}[t]
\includegraphics[scale=.3]{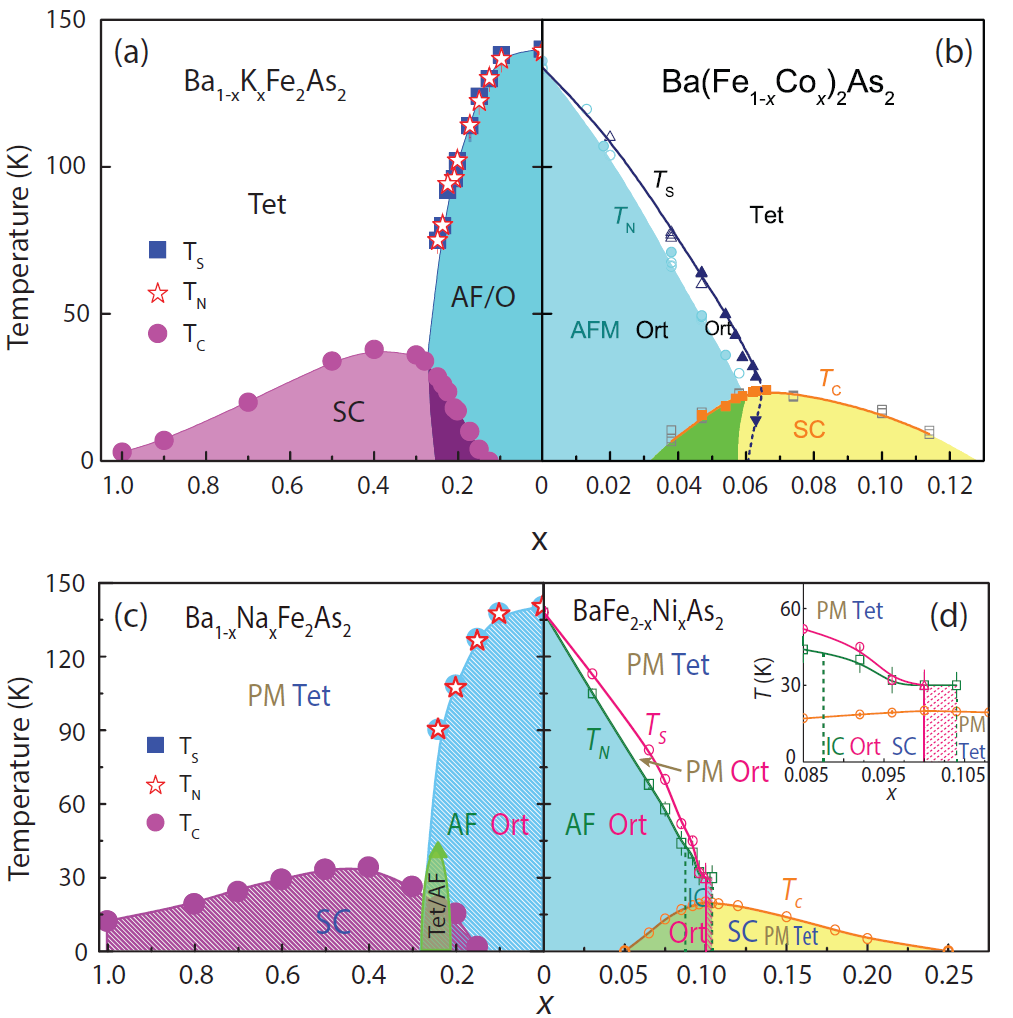}
\caption{
(Color online) The structural and magnetic phase diagrams of electron and hole-doped BaFe$_2$As$_2$.
(a) The coupled
structural and AF phase transitions in hole-doped Ba$_{1-x}$K$_x$Fe$_2$As$_2$
as determined from neutron powder diffraction experiments \cite{savci12}. The structural
and AF phase transitions are denoted as $T_s$ and $T_N$, respectively.
(b) The phase diagram of Ba(Fe$_{1-x}$Co$_x$)$_2$As$_2$ determined from X-ray
and neutron diffraction experiments \cite{snandi10}.
(c) The structural and magnetic phase diagram of
hole-doped Ba$_{1-x}$Na$_x$Fe$_2$As$_2$ from neutron diffraction experiments \cite{savci13,savci14}. The green shaded region
denotes the presence of a tetragonal AF phase. (d) Similar phase diagram for electron-doped
BaFe$_{2-x}$Ni$_x$As$_2$ \cite{hqluo12,xylu13}. Here the incommensurate (IC) AF order is a spin-glass phase coexisting and
competing with the superconducting phase \cite{xylu14}.
}
\end{figure}

\begin{figure}[t]
\includegraphics[scale=.3]{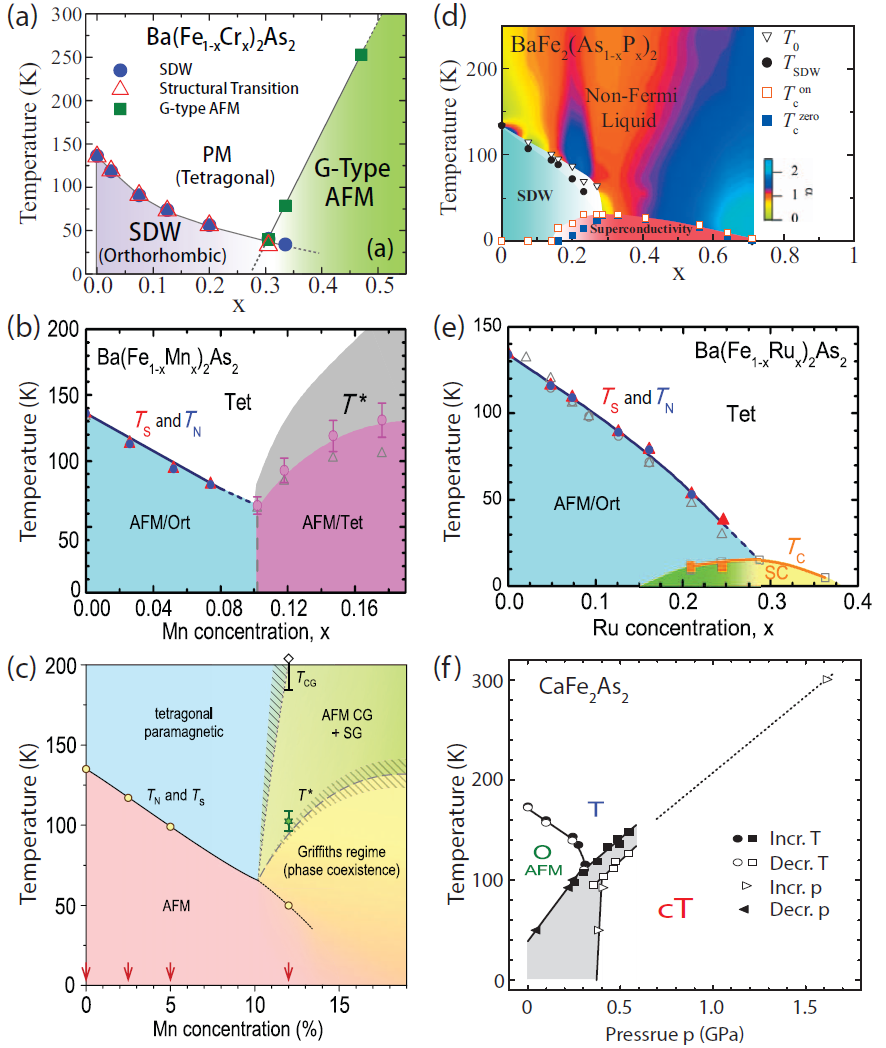}
\caption{
(Color online) The electronic phase diagrams of various other doped 122 family of iron pnictides.
(a) The structural and magnetic phase diagram of Cr-doped Ba(Fe$_{1-x}$Cr$_x$)$_2$As$_2$ \cite{marty11}.
For $x\le 0.2$, Cr doping suppresses the coupled structural and magnetic phase transition without inducing
superconductivity.  For $x>0.3$, the system becomes a $G$-type antiferromagnet (see Fig. 1).
(b) The structural and magnetic phase diagram for Mn-doped Ba(Fe$_{1-x}$Mn$_x$)$_2$As$_2$ \cite{mgkim10}.
(c) Similar phase diagram for Ba(Fe$_{1-x}$Mn$_x$)$_2$As$_2$
obtained by another group \cite{inosov13}.  Here the system is believed to
form a disordered spin glass (Griffiths) phase for $x>0.1$. (d) The structural and magnetic phase diagram of
isoelectronic doped BaFe$_2$(As$_{1-x}$P$_x$)$_2$ determined from transport and NMR measurement \cite{tshibauchi14}.
Recent neutron powder diffraction experiments indicate a coupled structural and magnetic phase transition \cite{JMAllred2014}.
(e) The structural and magnetic phase diagram of the isoelectronically doped Ba(Fe$_{1-x}$Ru$_x$)$_2$As$_2$ \cite{kim11}.  There
is no evidence of a quantum critical point near optimal superconductivity. (f) The pressure dependence of the structural and magnetic phase transitions in CaFe$_2$As$_2$.  The system enters into a collapsed tetragonal (cT) phase above
$\sim$0.4 GPa where magnetism disappears \cite{AIGoldman3}.
}
\end{figure}

Although the field of iron-based superconductors started with the discovery of the 1111 family
of materials \cite{kamihara}, majority of recent neutron scattering work
has been focused on the 122 family due to the availability of high quality single crystals \cite{canfield}.
In the undoped state, the prototypical 122 such as BaFe$_2$As$_2$ exhibits tetragonal-to-orthorhombic lattice
distortion and AF order below $T_s\approx T_N\approx 138$ K \cite{qhuang}.
Figure 5 summarizes evolution of the structural and magnetic phase transitions for the electron and hole doped BaFe$_2$As$_2$.
From transport, neutron diffraction, and X-ray diffraction measurements \cite{canfield,jhchu09,clester09,dkbratt09,adchristianson09,snandi10},
the phase diagram of
electron-doped Ba(Fe$_{1-x}$Co$_x$)$_2$As$_2$ as shown in Fig. 5(b) has been established. Upon electron-doping
via Co substitution for Fe
to suppress the static AF order and induce superconductivity, the structural and AF phase transitions
are gradually separated with the structural transition occurring at higher temperatures than the magnetic one.  The collinear
static AF order coexists and competes
with superconductivity in
the underdoped regime marked as green shaded area in Fig. 5(b) \cite{dkbratt09,adchristianson09}.
For electron-doping levels
near optimal superconductivity, the orthorhombic lattice distortion $\delta=(a-b)/(a+b)$
 initially increases with decreasing temperature below $T_N$, but then decreases dramatically below $T_c$.
For BaFe$_{1.874}$Co$_{0.126}$As$_2$, the orthorhombic structure evolves smoothly back to a tetragonal structure below $T_c$ and the
system is believed to enter into a ``reentrant'' tetragonal phase as shown in Fig. 5(b) \cite{snandi10}.
Subsequent neutron diffraction experiments revealed that the static AF order in the underdoped regime changes from commensurate to
transversely incommensurate for Ba(Fe$_{1-x}$Co$_x$)$_2$As$_2$ with $0.056\le x\le 0.062$ \cite{dkpratt11}. These results have
been hailed as direct evidence for spin-density-wave order in iron pinctides \cite{dkpratt11}, where the static AF order arises from
Fermi surface nesting  between the hole and electron pockets at the $\Gamma$ and $M$ points of the reciprocal space,
respectively \cite{jdong,ABVorontsov,JFink}.

\begin{figure}[t]
\includegraphics[scale=.35]{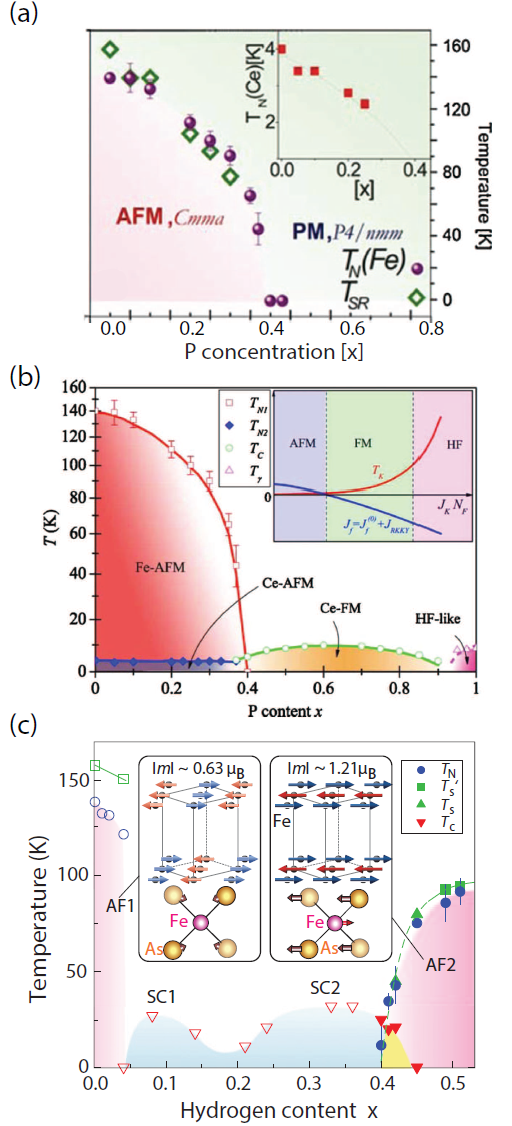}
\caption{
(Color online) The doping evolution of the electronic phase diagrams for P-doped
CeFeAs$_{1-x}$P$_x$O and H-doped LaFeAsO$_{1-x}$H$_x$.
(a) The structural and magnetic phase diagram of CeFeAs$_{1-x}$P$_x$O determined
from neutron powder diffraction experiments \cite{CCruz10}. A quantum critical point is found
near $x=0.4$ without the presence of superconductivity.  The inset shows the $T_N$ from
Ce magnetic ordering.  (b) Phase diagram of  CeFeAs$_{1-x}$P$_x$O determined by
transport measurements.
In the underdoped regime, the data is consistent with the results of neutron powder diffraction \cite{CCruz10}.
The system becomes a Ce-ordered ferromagnetic metal for $0.4<x<0.9$.  For samples with $x>0.9$, it becomes
heavy fermion-like metal \cite{YKLuo10}.  (c) The structural, magnetic, and superconducting phase transitions
in H-doped LaFeAsO$_{1-x}$H$_x$ \cite{MHiraishi14}.  There are two AF phases with different
magnetic structures near two superconducting domes.
}
\end{figure}

Figure 5(d) shows the phase diagram of electron-doped BaFe$_{2-x}$Ni$_x$As$_2$ obtained from X-ray and neutron scattering experiments \cite{hqluo12,xylu13}.  Here, the AF order decreases with increasing Ni-doping and disappears near optimal
superconductivity in a first order like fashion with an avoided quantum critical point \cite{xylu13}.
Similar to Ba(Fe$_{1-x}$Co$_x$)$_2$As$_2$ with $0.056\le x\le 0.062$ \cite{dkpratt11}, there is
short-range (60 \AA) incommensurate AF order
for samples near optimal superconductivity \cite{hqluo12}.
Although these results indicate an avoided quantum critical point in BaFe$_{2-x}$Ni$_x$As$_2$, they are in direct conflict with a recent
NMR work suggesting the presence of
two quantum critical points \cite{GQZheng1}.  However, these NMR results are inconsistent with systematic NMR and neutron scattering results on nearly optimally Co and Ni-doped BaFe$_2$As$_2$ samples revealing a cluster spin
glass state for the magnetic order \cite{apdioguardi,xylu14}.
The spin glass picture of the magnetic order near optimal superconductivity in electron-doped iron pnictides is clearly inconsistent with
the spin-density-wave explanation of the transverse incommensurate magnetic order \cite{dkpratt11}.
 These results suggest that the incommensurate AF order
in electron-doped iron pnictides
arises from a localized moment \cite{qsi,cfang,ckxu}, instead of spin-density-wave order induced from nested Fermi surfaces like incommensurate AF order in pure chromium metal \cite{EFawcett}.

The electronic phase diagrams of hole-doped Ba$_{1-x}$K$_x$Fe$_2$As$_2$ and Ba$_{1-x}$Na$_x$Fe$_2$As$_2$ determined from neutron scattering experiments on powder samples are
summarized in Figures 5(a) and 5(c), respectively \cite{savci12,savci13,savci14}.
Compared with electron-doped BaFe$_2$As$_2$, hole-doping does not separate the structural and magnetic phase transitions
and the AF and superconducting coexistence region is also present.  In particular, for
Ba$_{1-x}$Na$_x$Fe$_2$As$_2$ near $x=0.28$, a new magnetic ordered phase in the $C_4$ tetragonal symmetry of the
underlying lattice has been found \cite{savci14}.  In addition, superconductivity appears in heavily
hole-doped regimes, much different from the electron-doped case.

In copper oxide superconductors, superconductivity can only be induced by electron and hole doping
into the nearly
perfect CuO$_2$ plane, and
impurity substitution at the Cu sites by other elements dramatically suppresses superconductivity \cite{birgeneau,kivelson,armitage,Yamada,Tranquada2}.  The situation is much different for iron pnictides. While impurities such as Cr and Mn substituted for Fe in BaFe$_2$As$_2$ suppress the static AF order in the parent compound without inducing superconductivity [Figs. 6(a), 6(b) and 6(c)] \cite{marty11,mgkim10,inosov13},  isoelectronic substitution by replacing As with P [Fig. 6(d)] \cite{ZAXu2,tshibauchi14} or Fe with Ru [Fig. 6(e)] in BaFe$_2$As$_2$ \cite{kim11} can induce superconductivity.
For Cr-doped Ba(Fe$_{1-x}$Cr$_x$)$_2$As$_2$, neutron diffraction experiments on single crystals have established the structural and
magnetic phase diagram, showing a suppression of the collinear AF order for samples with $x<0.3$.
For $x>0.3$, the system becomes a $G$-type antiferromagnet with a tetragonal structure as shown in Fig. 6(a) \cite{marty11}.  The situation in Mn-doped Ba(Fe$_{1-x}$Mn$_x$)$_2$As$_2$ is somewhat similar.  With increasing Mn doping in BaFe$_2$As$_2$, the structural and AF phase transitions are gradually suppressed as shown
in Figs. 6(b) and 6(c) \cite{mgkim10,inosov13}.  For Mn concentration $x\ge 0.1$, the system goes into
a mixed phase, possibly in the Griffiths regime, with coexisting short-range spin excitations
at AF wave vectors similar to those in BaFe$_2$As$_2$ [${\bf Q}_{\rm AF}={\bf Q}_{\rm stripe}$]
and BaMn$_2$As$_2$ [${\bf Q}={\bf Q}_{N{\rm \acute{e}}el}$ rotated 45$^\circ$ from ${\bf Q}_{\rm AF}$] \cite{inosov13}.

\begin{figure}[t]
\includegraphics[scale=.43]{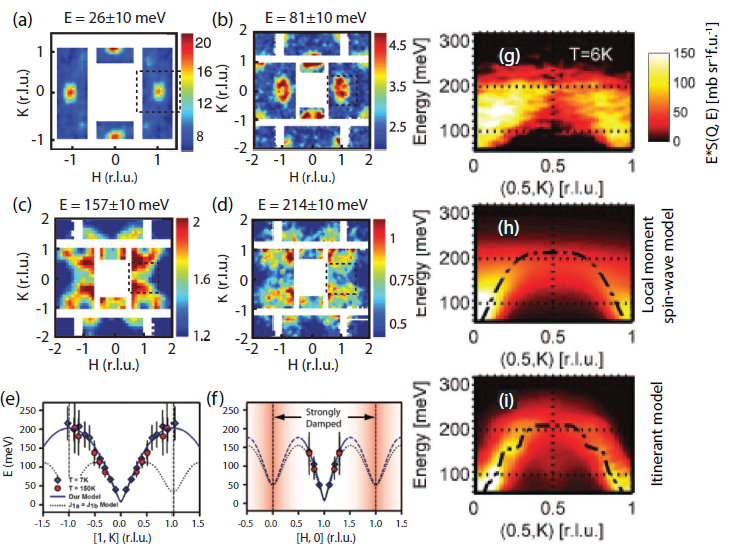}
\caption{
(Color online) Spin waves in BaFe$_2$As$_2$ and SrFe$_2$As$_2$ determined from neutron time-of-flight spectrometry.
(a) Spin waves at $E=26\pm 10$ meV.  The presence of peaks at wave vectors
${\bf Q}_{\rm AF}=(\pm 1,0)$ and $(0,\pm 1)$ is due to twinning effect.  The white regions
are detector gaps. Similar spin waves at (b) $E=81\pm 10$, (c) $157\pm 10$, and
(d) $214\pm 10$ meV \cite{LWHarriger11}.
The color bars indicate magnetic scattering in absolute units of mbarn$\cdot$sr$^{-1}$$\cdot$meV$^{-1}$$\cdot$f.u.$^{-1}$.
(e) Spin-wave dispersion curves and fits using a
Heisenberg Hamiltonian with different exchange couplings along the $[1,K]$ direction.
(f) Similar Heisenberg Hamiltonian fits along the $[H,0]$ direction \cite{LWHarriger11}.
(g) Spin waves in the energy and wave vector cuts
along the $[0.5,K]$ direction for SrFe$_2$As$_2$ \cite{raewings11}.
(h) The dashed line shows fit of a Heisenberg Hamiltonian assuming one
set of exchange coupling constants. (i) A RPA
calculation of $\chi^{\prime\prime}({\bf Q},\omega)$ based
on a 5-band model \cite{raewings11}. The reciprocal space notation in (g,h,i) is tetragonal
where ${\bf Q}_{\rm AF}=(0.5,0.5)$, different from those in (a-f).
}
\end{figure}

In contrast to Cr and Mn doping, isoelectronic doping by replacing As with P in BaFe$_2$(As$_{1-x}$P$_x$)$_2$
induces optimal superconductivity near $x=0.3$ \cite{ZAXu2}.  From the systematic transport and London penetration depth measurements
on BaFe$_2$(As$_{1-x}$P$_x$)$_2$,
a quantum critical point has been identified near optimal superconductivity at $x=0.3$ [Fig. 6(d)] \cite{tshibauchi14}.
For isoelectronic Ba(Fe$_{1-x}$Ru$_x$)$_2$As$_2$, optimal superconductivity again appears
near Ru-doping level of $x=0.3$ [Fig. 6(e)] \cite{kim11}.  However, there are no report for the presence of a quantum critical point in this system.  Figure 6(f) shows the pressure dependence of the CaFe$_2$As$_2$ phase diagram \cite{AIGoldman3}.
While superconductivity can be induced directly via applying hydrostatic pressure in BaFe$_2$As$_2$ and
SrFe$_2$As$_2$ \cite{johnston,GRStewart}, external pressure exerted on
CaFe$_2$As$_2$ results in a nonmagnetic collapsed tetragonal (cT) phase, eliminating the static AF ordered moment
and spin excitations
 without inducing superconductivity \cite{AIGoldman3}.

Although a majority of neutron scattering work has been focused on the 122 family of materials
because of the availability of high-quality single crystals, there are also important phase diagram results in the 1111 family.  For example, P-doping in the CeFeAsO family of
materials suppresses static AF order, but does not induce superconductivity \cite{CCruz10}.
Systematic neutron scattering studies of the structural and magnetic phase transitions in powder samples of
CeFeAs$_{1-x}$P$_x$O suggest that the pnictogen height [the average Fe-As(P) distance] and orthorhombicity of the
CeFeAs$_{1-x}$P$_x$O unit cell critically control the iron AF ordered moment and
N$\rm \acute{e}$el temperature of the system. Figure 7(a) shows the P-doping dependence of the structural and AF phase transition
temperatures in CeFeAs$_{1-x}$P$_x$O, suggesting the presence of a magnetic quantum critical point near $x=0.4$ \cite{CCruz10}.
A complete mapping of the CeFeAs$_{1-x}$P$_x$O phase diagram shown in Fig. 7(b) was obtained via transport and susceptibility measurements,
which reveal that superconductivity does not appear in the  entire phase diagram, possibly
due to heavy fermion properties of the rare earth element
Ce \cite{YKLuo10}.
Another recent advance is the discovery of
bipartite magnetic parent phases in the H-doped LaFeAsO$_{1-x}$H$_x$ family of materials
\cite{MHiraishi14}.  In contrast to the general phase diagram of iron pnictides, superconductivity
in LaFeAsO$_{1-x}$H$_x$ appears in two domes adjacent to two different AF phases with different
magnetic structures and N$\rm \acute{e}$el temperatures [Fig. 7(c)] \cite{MHiraishi14}.
These results again confirm the notion that superconductivity in iron-based superconductors is intimately
connected with the magnetic interactions.

\section{Spin excitations and their relationship with superconductivity}

The rapid development of neutron time-of-flight chopper spectrometers in recent years has allowed
measurements of spin excitations in high-$T_c$ superconductors
throughout the Brillouin zone for energy transfers up to
1 eV.  In the case of copper oxides, spin waves in La$_2$CuO$_4$ have been mapped out
throughout the Brillouin zone \cite{RColdea,NSHeadings}. While the low energy spin excitations are well described
by theory based on the Heisenberg Hamiltonian,
high-energy spin waves are damped near the ${\bf Q}=(1/2,0)$ position in reciprocal space and
merge into a momentum dependent continuum suggesting the decay of spin waves into other excitations \cite{RColdea,NSHeadings}.  The effective magnetic exchange couplings of La$_2$CuO$_4$ determined from the Heisenberg model are summarized in Table II.
The doping evolution of spin excitations
as a function of electron and hole doping and their coupling to superconductivity
are reviewed recently \cite{armitage,Yamada,Tranquada2}.  In the case of iron-based superconductors, the situation is more complicated. Of
the 11, 111, 122, 1111, 245 families of materials, spin waves in the AF parent compounds
throughout the Brillouin zone were mapped out for the 11 \cite{OJLipscombe,IAZaliznyak}, 111 \cite{clzhang14}, 122 \cite{sodiallo09,jzhao09,LWHarriger11,raewings11}, and 245 \cite{MYWang11,yxiao13,schi13} families of materials
due to the availability of large single crystals needed for inelastic neutron scattering experiments.
Although single crystals of the 1111 family of materials are still not large enough to allow
a determination of the entire spin wave spectra, measurements of low-energy spin waves reveal that the system is highly two-dimensional with weak magnetic exchange coupling along the $c$ axis \cite{MRamazanoglu13}.
In the Sections III A and B, we describe spin wave measurements in the parent compounds of different families of
iron-based superconductors and discuss their relationship with superconductivity.

\subsection{Spin waves in the parent compounds of iron-based superconductors}

Inelastic neutron scattering studies of spin waves in the parent compounds of iron-based superconductors
began soon after the availability of single crystals of the 122 family \cite{canfield}.
For a magnetically ordered system, spin waves occur
when the magnetic moments precess around their ordered configuration.
Regardless of the microscopic origin of the magnetic order,
spin waves of an ordered
system should exhibit
sharp excitations in the long wavelength limit (small ${\bf Q}$) and can be described by a suitable Hamiltonian using
perturbation theory.  For a spin Hamiltonian, one can starts with a
Heisenberg model where the energy of spin waves depends only on the relative
orientation of neighboring spins.
In the initial neutron scattering experiments on low-energy spin waves in SrFe$_2$As$_2$ \cite{jzhao08},
CaFe$_2$As$_2$ \cite{RJMcQueeney08}, and BaFe$_2$As$_2$ \cite{KMatan09}, a spin gap
due to magnetic iron anisotropy
was identified.  In addition, the low-energy spin waves were described by either a local moment
 Heisenberg Hamiltonian \cite{jzhao08,RJMcQueeney08} or the spin excitation continuum from itinerant electrons \cite{sodiallo09,KMatan09}. However, these measurements were unable to reach spin waves near the zone boundary and thus did not allow a conclusive determination of the effective nearest and next nearest neighbor magnetic exchange couplings denoted
as $J_{1a}$/$J_{1b}$ and $J_2$, respectively [Fig. 4(b)].  In the itinerant picture of the magnetism in iron pnictides \cite{mazin2011n,hirschfeld}, spin waves from the AF ordered phase should arise from quasiparticle excitations between the electron and hole Fermi surfaces and form a spin excitation continuum at high-energies \cite{Kaneshita10}. In the initial neutron time-of-flight experiments on CaFe$_2$As$_2$, spin waves up
to an energy of $\sim$100 meV were measured and found to fit a Heisenberg Hamiltonian \cite{sodiallo09}.
However, no spin wave signals were found for energies above 100 meV consistent with
$ab$ initio calculations of the dynamic magnetic susceptibility, indicating that the high energy spin excitations are dominated by the damping of spin waves by particle-hole excitations \cite{sodiallo09}.

\begin{figure}[t]
\includegraphics[scale=.45]{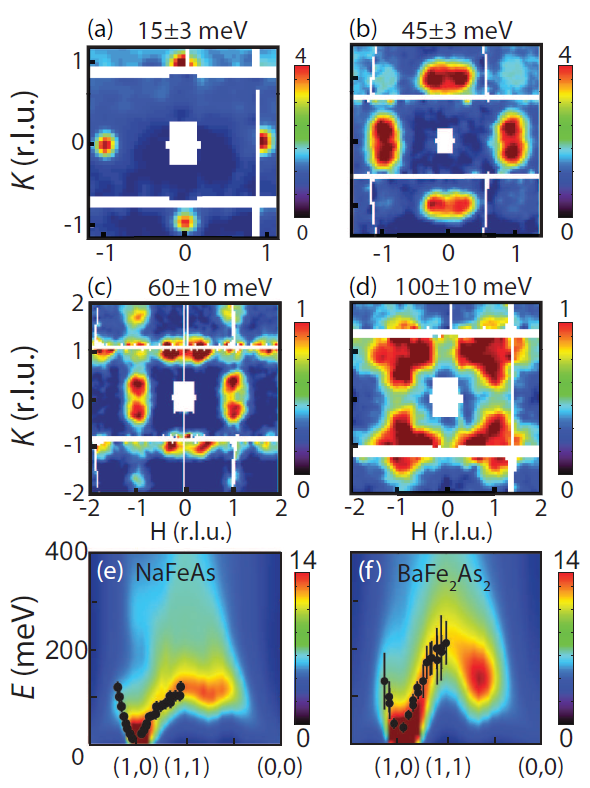}
\caption{
(Color online) Comparison of spin waves in NaFeAs and combined DFT and DMFT calculations.
(a) Spin waves of NaFeAs at $E=15\pm 3$ meV. Similar spin waves at (b) $E=45\pm 3$;
(c) $60\pm 10$; (d) $100\pm 10$ meV.  The magnetic intensity are in absolute units of mbarn$\cdot$sr$^{-1}$$\cdot$meV$^{-1}$$\cdot$f.u.$^{-1}$.
(e) The solid circles show dispersion of spin waves, while the color plots are calculations
from a combined DFT and DMFT theory.
(f) Similar comparison for BaFe$_2$As$_2$ \cite{clzhang14}.
}
\end{figure}

In subsequent neutron scattering experiments on CaFe$_2$As$_2$ \cite{jzhao09}, BaFe$_2$As$_2$ \cite{LWHarriger11}, and SrFe$_2$As$_2$
using more sample mass \cite{raewings11}, spin waves were mapped out throughout the Brillouin zone and the zone boundary energy scales were found to be around 220 meV.
Figures 8(a)-8(d) show images of spin waves in BaFe$_2$As$_2$ in the AF ordered state at energies of $E=26\pm 10$, $81\pm 10$, $157\pm 10$, and $214\pm10$ meV, respectively \cite{LWHarriger11}. With increasing energy, spin waves become diffusive but one can still see clear excitations near the zone boundary at $E=214$ meV, different from the earlier experiment \cite{sodiallo09}.  Figures 8(e) and 8(f) show spin wave dispersions of BaFe$_2$As$_2$ along the in-plane $[1,K]$ and $[H,0]$ directions.  Using a Heisenberg Hamiltonian with anisotropic
spin wave damping, one can fit the entire spin wave spectrum with a large in-plane nearest neighbor magnetic exchange anisotropy
($J_{1a}>0$, $J_{1b}<0$) and finite next nearest neighbor exchange coupling ($J_2>0$) \cite{jzhao09,LWHarriger11}.
The details of Heisenberg Hamiltonian for spin waves have been discussed in \cite{sodiallo09,jzhao09,LWHarriger11}.
The outcomes of the fits with anisotropic in-plane magnetic exchanges are shown as solid lines in Fig. 8(e), while the dashed lines
in the Figure are calculations assuming isotropic in-plane magnetic exchange couplings.
The discovery of large in-plane exchange anisotropy is surprising
given the small orthorhombic lattice distortion in the AF ordered state \cite{ALWysocki11}.  Only by probing spin waves at high energies near the zone boundary, one can conclusively determine the effective magnetic exchange couplings in the system.
Different magnetic structures and spin exchange couplings in iron-based materials has been
studied using a localized moment model with different nearest and next nearest
neighbor exchange couplings \cite{JPHu2012a}.

Although spin waves in CaFe$_2$As$_2$ \cite{jzhao09} and BaFe$_2$As$_2$ \cite{LWHarriger11} can be modeled by a local moment
Heisenberg Hamiltonian, one still has to use anisotropic spin wave damping characteristic of an itinerant electron
system.  In the neutron scattering work on spin waves of SrFe$_2$As$_2$ [Fig. 8(g)] \cite{raewings11}, the authors report
that a Heisenberg Hamiltonian that can fit the low-energy spin wave data fails to describe the spectrum
near the zone boundary [Fig. 8(h)].  The overall spin wave spectrum is instead best described by an itinerant model with large spin wave damping near the zone boundary [Fig. 8(i)] \cite{raewings11}.

\begin{figure}[t]
\includegraphics[scale=.30]{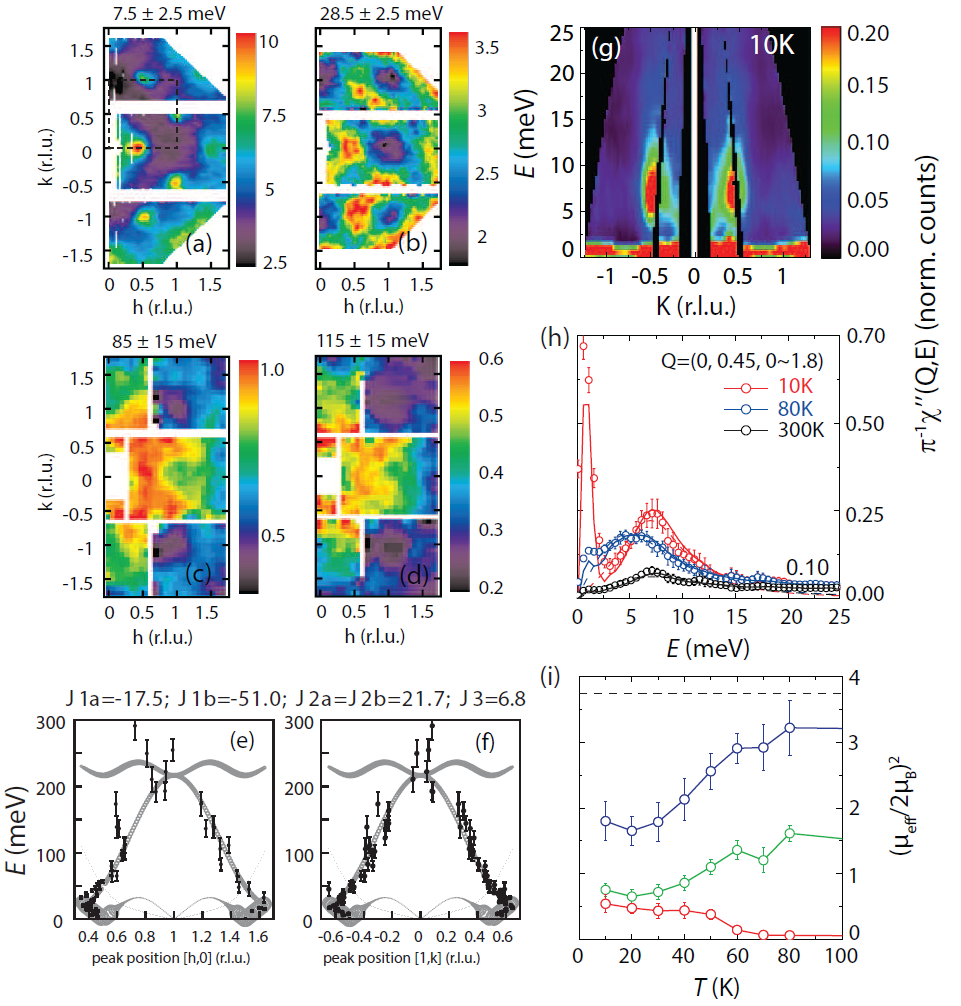}
\caption{
(Color online) Spin waves in Fe$_{1.05}$Te and Fe$_{1.1}$Te.
(a) Wave vector dependence of spin waves in Fe$_{1.05}$Te at
$E=7.5\pm 2.5$ meV.  Similar spin waves at (b) $E=28.5\pm 2.5$,
(c) $85\pm 15$, (d) $115\pm 15$ meV \cite{OJLipscombe}.
The diffusive nature of spin waves is clearly seen at high energies.
(e,f)  Dispersion curves of spin waves along the $[H,0]$ and
 $[1,K]$ directions and the Heisenberg Hamiltonian fits using
nearest, next nearest, and next next nearest neighbor exchange couplings \cite{OJLipscombe}.
(g) Energy and wave vector cut of spin waves in Fe$_{1.1}$Te.
(h) Energy dependence of spin waves at different temperatures.
(i) Temperature dependence of the integrated moments for
Fe$_{1.1}$Te.  The data suggests an increased total integrated moment
on warming from 10 K to 100 K across $T_N$ and $T_s$
\cite{IAZaliznyak}.
}
\end{figure}

Similar to the 122 family of materials, NaFeAs, the parent compound of the 111 family of iron pnictides,
has the collinear AF structure albeit with a greatly reduced ordered moment size \cite{SLLi1}.  Triple-axis neutron
scattering experiments on single crystals of NaFeAs studied low-energy spin waves and
found a small gap in the excitation spectrum \cite{JTPark12,ysong13}.
Figure 9 summarizes the evolution of spin waves to the zone boundary as a function of increasing energy \cite{clzhang14}.
Compared with the spin wave zone boundary energy of $\sim$220 meV in
BaFe$_2$As$_2$ as shown in Fig. 8, spin waves in NaFeAs reach the zone boundary at the in-plane
wave vector ${\bf Q}=(1,1)$
around $\sim$110 meV [Fig. 9(d)].
This means that the overall magnetic excitation bandwidth in the 111 family is considerably
lower than that of the 122 family of iron pnictides. Figures 9(e) and 9(f) compare the
experimental and the
combined density functional theory and dynamical mean field theory (DFT+DMFT) calculations of
spin wave dispersion of NaFeAs and BaFe$_2$As$_2$, respectively. The outcome suggests that
pnictogen height is correlated with the strength of electron-electron correlations and consequently the effective
bandwidth of magnetic excitations in iron pnictides \cite{clzhang14,ZPYin2014}.

\begin{figure}[t]
\includegraphics[scale=.5]{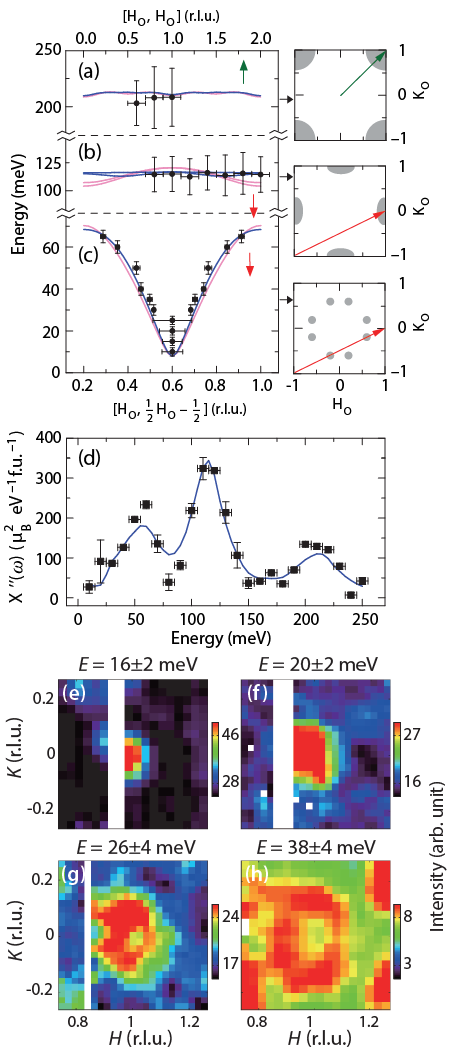}
\caption{
(Color online) Spin waves in the insulating 245 and semiconducting 234 phases.
(a,b,c) Spin-wave dispersions in the insulating Rb$_{0.89}$Fe$_{1.58}$Se$_2$ \cite{MYWang11}.
The solid circles are data from the cut directions marked by the arrows in the right panels
of (a,b,c).  The solid lines are fits from a Heisenberg Hamiltonian considering nearest,
the next nearest, and next next nearest neighbor exchange couplings \cite{MYWang11}. (d) The energy
dependence of the local dynamic susceptibility, where the solid line is the calculated value
from the Heisenberg Hamiltonian. The total moment sum rule appears to be satisfied \cite{MYWang11}.
(e,f,g,h) The wave vector dependence of spin waves from the semiconducting K$_{0.85}$Fe$_{1.54}$Se$_2$ \cite{JZhao14}.
}
\end{figure}

Figure 10 summarizes spin wave measurements for the iron chalcogenide Fe$_{1+x}$Te \cite{Fruchart,WBao5,SLLi2}, the parent compound of the 11 family of iron-based superconductors \cite{OJLipscombe,IAZaliznyak}.  The static magnetic order and spin excitations of Fe$_{1+x}$Te are sensitive to the excess iron
in the interstitial sites \cite{EERodriguez10,EERodriguez2011,cstock11,JSWen1}. This is rather different from the iron pnictides, which cannot
accommodate any excess iron in the crystal structure.
For Fe$_{1.05}$Te and Fe$_{1.1}$Te, the AF structure is commensurate bi-collinear \cite{EERodriguez2011}.
 Figures 10(a), 10(b), 10(c), and 10(d) show the two-dimensional images of spin waves in Fe$_{1.05}$Te at $E=7.5\pm 2.5$,
$28.5\pm 2.5$, $85\pm 15$, and $115\pm 15$ meV, respectively \cite{OJLipscombe}. The dispersion
of spin waves is different from that of the 122 and 111 families, and becomes
diffusive for energies
above 85 meV without well-defined spin waves [Figs. 10(c) and 10(d)].
The solid lines in Figs. 10(e) and 10(f) show fits of the
dispersion using a Heisenberg
Hamiltonian assuming exchange couplings $J_{1a}$, $J_{1b}$, $J_{2}$, and $J_{3}$ \cite{OJLipscombe}.
In a separate neutron scattering experiment \cite{IAZaliznyak}, the authors find that the low-energy spin excitations can be well-described by liquid-like spin plaquette correlations [Fig. 10(g)].
Furthermore, the integrated magnetic excitation intensity increases on warming [Fig. 10(h)].
The effective spin per Fe $S\approx 1$ at $T\approx 10$ K in the AF ordered phase
grows to $S\approx 3/2$ at $T=80$ K in the paramagnetic
phase, suggesting that the local magnetic moments are
entangled with the itinerant electrons in the system [Fig. 10(i)] \cite{IAZaliznyak}.

Of all the iron-based superconductors, alkali iron selenides $A_x$Fe$_{2-y}$Se$_2$ \cite{XLChen,MHFang} are unique
in that superconductivity in this class of materials always
coexists with a static long-range AF order
with a large moment and high N$\rm\acute{e}$el temperature \cite{WBao1,WBao2,mwang11,jzhao12,may12}.
Although there is ample evidence indicating that the superconducting alkali iron selenides are mesoscopically phase
separated from
the insulating $A_2$Fe$_4$Se$_5$ phase with the $\sqrt{5}\times\sqrt{5}$ block AF structure as shown in Fig. 2(d) \cite{ZWWang12,SCSpeller12,ARicci11,DPShoemaker12,WLi11,VKsenofontov11,ZShermadini12,ACharnukha12,YTexier12,SVCarr14}, there is still no consensus on the
chemical and magnetic structures of their parent compounds \cite{WBao2,mwang11,jzhao12,may12}.
Assuming that the insulating $A_2$Fe$_4$Se$_5$ phase is the parent compound of the superconducting
$A_x$Fe$_{2-y}$Se$_2$, its spin waves have been mapped out by several groups \cite{MYWang11,yxiao13,schi13}.
Compared with spin waves in iron pnictides and iron chalcogenides (Figs. 7-10), the dispersion of
the spin waves in insulating $A_2$Fe$_4$Se$_5$ has two optical branches at high energies
and one acoustic branch at low energy, where the arrows are wave vector scales and
the thin dashed line separates the vertical energy scale for the acoustic and low-energy optical spin waves from the high-energy
optical spin waves [Figs. 11(a)-11(c)] \cite{MYWang11}. By integrating the energy dependence of the
local dynamic susceptibility in Rb$_{0.89}$Fe$_{1.58}$Se$_2$ [Fig. 11(d)], it was found that the total moment sum
rule is exhausted for magnetic scattering at energies below 250 meV.
Therefore, spin waves in insulating Rb$_{0.89}$Fe$_{1.58}$Se$_2$ can be regarded as
a classic local moment system where a Heisenberg Hamiltonian is
an appropriate description of the spin wave spectrum.

\begin{figure}[t]
\includegraphics[scale=.28]{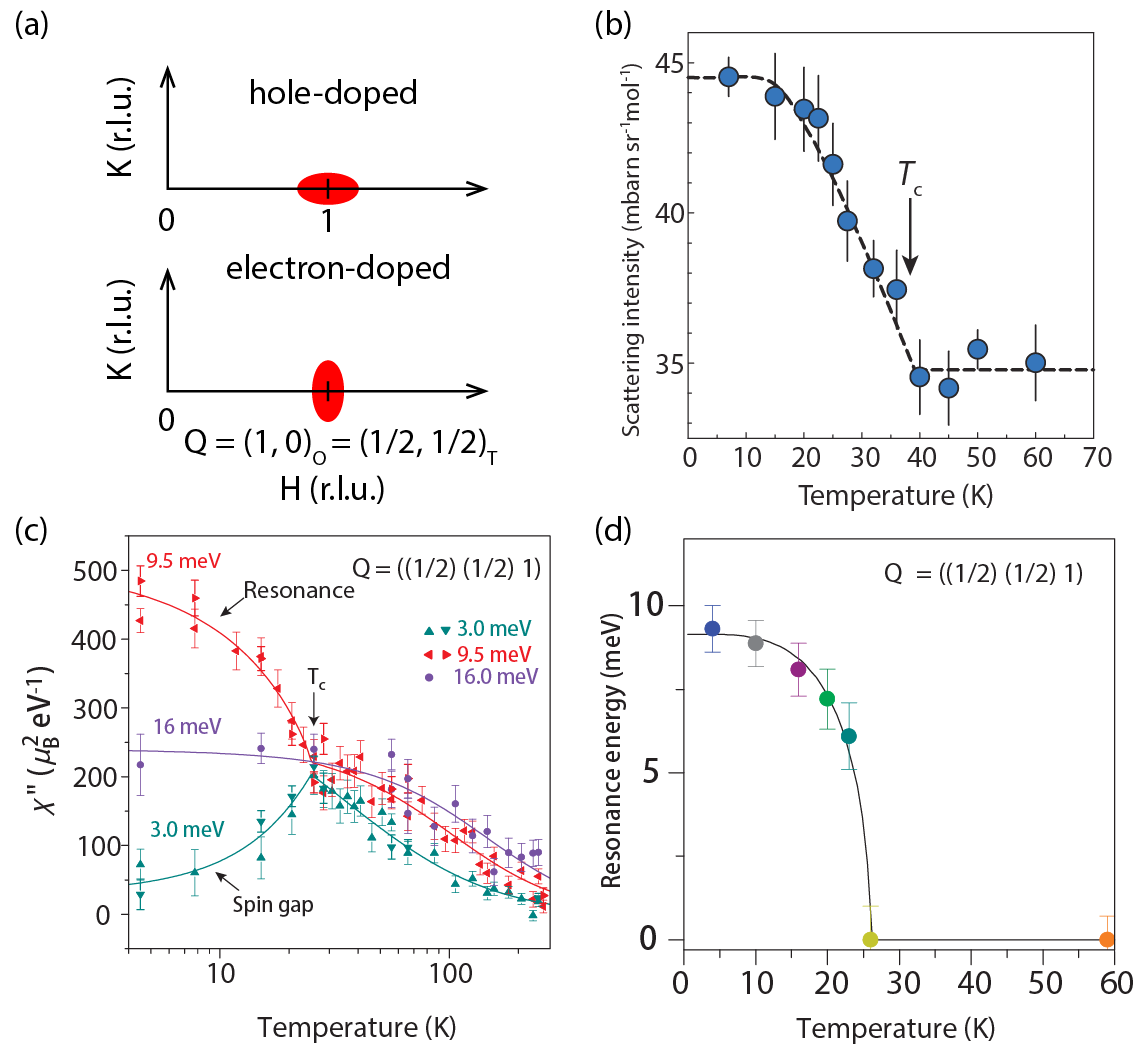}
\caption{
(Color online) Neutron spin resonance in electron- and hole-doped iron pnictides.
(a) The schematic drawings of the wave vector dependence of the low-energy spin excitations
in optimally hole- (upper panel) and electron-doped (lower panel) superconducting
iron pnictides. (b) Temperature dependence of the resonance at $E=16$ meV, showing clear
superconducting order parameter-like enhancement below $T_c$ for a powder sample of
Ba$_{0.6}$K$_{0.4}$Fe$_2$As$_2$ \cite{ADChristianson}.
(c) Temperature dependence of the magnetic scattering at energies $E=3.0$, 9.5, and 16 meV
for optimally electron-doped superconducting BaFe$_{1.85}$Co$_{0.15}$As$_2$ ($T_c=25$ K) \cite{DSInosov10}.
While magnetic intensity at the resonance energy ($E=9.5$ meV) shows a clear enhancement below $T_c$
at the expense of opening a spin gap at $E=3.0$ meV, the scattering at $E=16$ meV is not sensitive to superconductivity.
(d) Temperature dependence of the resonance energy for BaFe$_{1.85}$Co$_{0.15}$As$_2$ \cite{DSInosov10}.
}
\end{figure}

On the other hand, if the semiconducting AF phase with
rhombus iron vacancy order [Fig. 2(e)] is the parent compound \cite{jzhao12,mwang2014}, one
finds that spin waves of the system are rather close to those of iron pnictides.
Figure 11(e), 11(f), 11(g), 11(h) shows the evolution of spin waves as a function of increasing energy for
the semiconducting K$_{0.85}$Fe$_{1.54}$Se$_2$ with collinear AF order and $T_N=280$ K \cite{JZhao14}.
The data agrees well with calculations using a Heisenberg Hamiltonian.
A comparison of the observed spin wave spectrum in this system with those
of the CaFe$_2$As$_2$ single crystals \cite{jzhao09} reveals
remarkable similarity, and thus suggesting similar effective magnetic exchange couplings in these systems \cite{JZhao14}.

\begin{figure}[t]
\includegraphics[scale=0.4]{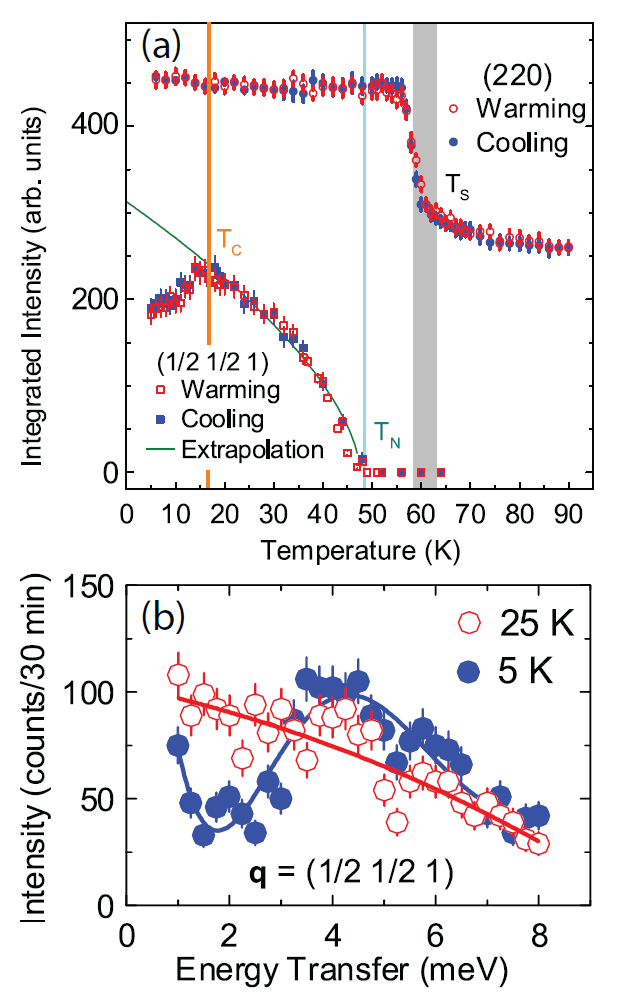}
\caption{
(Color online) Effect of electron-doping on magnetism and superconductivity in electron underdoped iron pnictides.
(a) Temperature dependence of the nuclear $(2,2,0)$ and $(1/2,1/2,1)$ (in tetragonal notation)
magnetic scattering in the electron underdoped
Ba(Fe$_{0.953}$Co$_{0.047}$)$_2$As$_2$ ($T_c=17$ K).  The structural, magnetic, and superconducting transitions are clearly marked.
(b) A weak resonance appears below $T_c$ at $E=4$ meV \cite{dkbratt09}.
}
\end{figure}

Table II summarizes the effective magnetic exchange couplings for the parent compounds of known iron-based superconductors.  We also list the effective magnetic exchange couplings for La$_2$CuO$_4$, the parent compound of copper oxide superconductors.  They are dominated by the large nearest neighbor and weak next nearest neighbor magnetic exchange couplings (Fig. 4).
For the parent compounds of iron-based superconductors, it is instructive to compare their effective magnetic exchange couplings.
In spite of their dramatically different AF structures summarized in Figs. 1-3
, they all appear to have similar next nearest neighbor
magnetic exchange couplings (see Table II).  This is consistent with the idea that the next nearest neighbor coupling $J_2$
is mainly determined by a local superexchange mechanism mediated
by As or Se/Te, regardless of their metallic
or insulating ground states \cite{EAbrahams11,JPHu2012}.

\begin{table*}
\caption{\label{tab:table2} Comparison of the effective magnetic exchange couplings
for parent compounds of copper-based and iron-based superconductors. Here
the nearest, next nearest, next next nearest neighbor, and $c$ axis exchange couplings are
 $SJ_{1a}$($SJ_{1b}$), $SJ_{2a}$($SJ_{2b}$), $SJ_{3}$, and $SJ_{c}$, respectively, where $S$ is the spin of the system.
\\
}
\begin{ruledtabular}
\begin{tabular}{ccccccc}
 Materials & $SJ_{1a}$ (meV) & $SJ_{1b}$ (meV) & $SJ_{2a}$ (meV) & $SJ_{2b}$ (meV) & $SJ_{3}$ (meV) & $SJ_{c}$ (meV) \\
\hline
La$_2$CuO$_4$ \footnote{\cite{RColdea}.} & $55.9\pm 2$ & $55.9\pm 2$ & $-5.7 \pm 1.5$ & $-5.7 \pm 1.5$ & 0 & 0 \\
NaFeAs \footnote{\cite{clzhang14}.}      & $40\pm 0.8$  & $16\pm 0.6$  & $19\pm 0.4$   & $19\pm 0.4$   & 0 & $1.8\pm 0.1$ \\
CaFe$_2$As$_2$ \footnote{\cite{jzhao09}.} & $49.9\pm 9.9$& $-5.7\pm 4.5$& $18.9\pm 3.4$ & $18.9\pm 3.4$ & 0 & $5.3\pm 1.3$ \\
BaFe$_2$As$_2$  \footnote{\cite{LWHarriger11}.} & $59.2\pm 2.0$ & $-9.2 \pm 1.2$ & $13.6 \pm 1$  & $13.6\pm 1$  & 0  & $1.8 \pm 0.3$ \\
SrFe$_2$As$_2$ (L)   \footnote{\cite{raewings11}.}  & $30.8\pm 1$ & $-5\pm 4.5$ & $21.7\pm 0.4$
&  $21.7\pm 0.4$ & 0  &  $2.3\pm 0.1$ \\
SrFe$_2$As$_2$ (H) \footnote{The L and H are fits using low and high-energy spin waves, respectively.}  & $38.7\pm 2$ & $-5\pm 5$ & $27.3\pm 0.3$
&  $27.3\pm 0.3$ & 0  &  $2.3\pm 0.1$ \\
Fe$_{1.05}$Te \footnote{\cite{OJLipscombe}.}  & $-17.5\pm 5.7$ & $-51.0\pm 3.4$ & $21.7 \pm 3.5$   & $21.7\pm 3.5$ & $6.8\pm 2.8$  & $\sim$1 \\
Rb$_{0.89}$Fe$_{1.58}$Se$_2$ \footnote{\cite{MYWang11}.}  & $-36\pm 2$ & $15\pm 8$     & $12\pm 2$ & $16\pm 5$ & $9\pm 5$ &  $1.4\pm 0.2$   \\
(Tl,Rb)$_2$Fe$_4$Se$_5$ \footnote{\cite{schi13}.} & $-30\pm 1$ & $31 \pm 13$ & 10 $\pm$ 2 &29 $\pm 6$  & 0   &  $0.8\pm 1$   \\
K$_{0.85}$Fe$_{1.54}$Se$_2$ \footnote{\cite{JZhao14}.} & $-37.9\pm 7.3$ & $-11.2\pm 4.8$ & $19.0\pm 2.4$ & $19.0\pm 2.4$ & 0 & $0.29\pm 0.06$  \\
\end{tabular}
\end{ruledtabular}
\end{table*}

\begin{figure}[t]
\includegraphics[scale=.20]{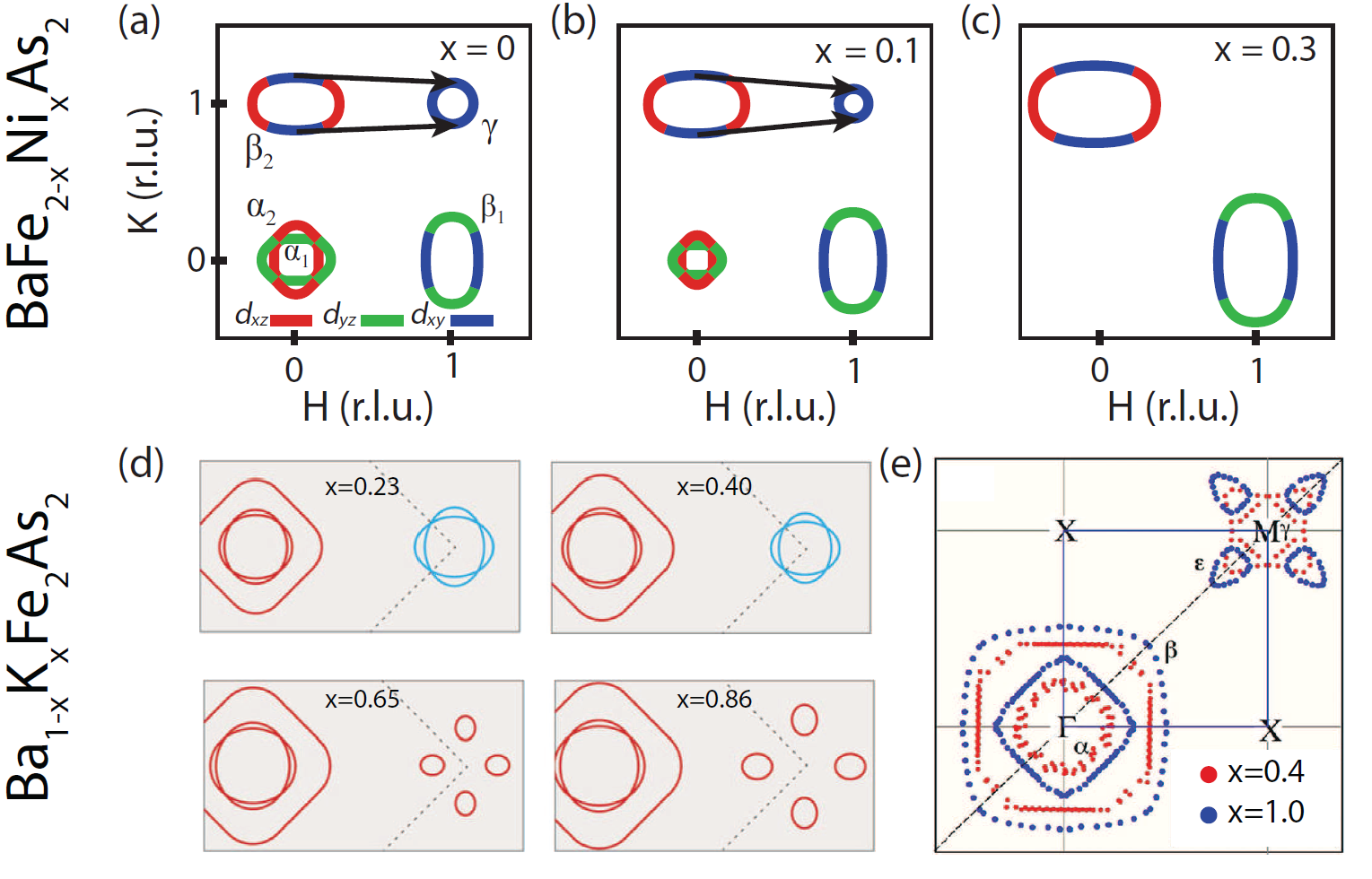}
\caption{
(Color online) The evolution of Fermi surfaces in electron and hole-doped BaFe$_2$As$_2$.
(a) Schematics of Fermi surfaces corresponding to BaFe$_2$As$_2$
with possible nesting vectors marked with
arrows \cite{PCDai1}. The $d_{xz}$, $d_{yz}$, and $d_{xy}$ orbitals for different Fermi surfaces
 are colored as red, green and blue, respectively. (b) Fermi surfaces when 10\% electrons are doped into
BaFe$_2$As$_2$ to form optimal superconductivity. (c) Fermi surfaces with 30\% electron doping when superconductivity is
suppressed \cite{MWang13}. (d) Schematics of Fermi surfaces for hole-doped
Ba$_{1-x}$K$_x$Fe$_2$As$_2$ with increasing K-doping to $x=0.23$ (upper left panel),
0.40 (upper right panel), 0.65 (lower left), and 0.86 (lower right panel) \cite{PRichard11}.
(e) A comparison of the Fermi surfaces for $x=0.4$ and 1 in the folded Brillouin zone \cite{PRichard11}.
}
\end{figure}

\subsection{Neutron spin resonance and its relationship with superconductivity}

The neutron spin resonance is a collective magnetic excitation
occurring below $T_c$ with a temperature dependence similar to the superconducting order parameter \cite{eschrig}.
First discovered in hole doped YBa$_2$Cu$_3$O$_{6+x}$
copper oxide superconductors \cite{RossatMignod}, the resonance is located near the AF ordering wave vector ${\bf Q}_{\rm AF}$
of the nonsuperconducting parent compound
and occurs at an energy related to the superconducting $T_c$ \cite{PCDai2,SDWilson06} or gap energy \cite{GYu09}.
It has been argued that the mode is a signature of the $d$-wave pairing as a result of quasiparticle excitations between
the sign reversed $d$-wave
superconducting gaps \cite{eschrig}.  Soon after the discovery of iron pnictide superconductors \cite{kamihara}, a
 neutron spin resonance was found in powder samples of Ba$_{0.6}$K$_{0.4}$Fe$_2$As$_2$
 \cite{ADChristianson}.  Since the resonance occurs below $T_c$ at the momentum transfer ($Q=1.15\ \mathrm{\AA}^{-1}$)
close to ${\bf Q}_{\rm AF}$ in BaFe$_2$As$_2$ [Fig. 12 (a) and 12(b)] \cite{ADChristianson},
the mode is believed to arise from
the sign reversed quasiparticle excitations
 between the hole and electron Fermi surfaces near the $\Gamma$ and $M$ points in
reciprocal space, respectively (Figs. 12-14) \cite{mazin2011n,hirschfeld}. In subsequent inelastic neutron scattering
experiments on single crystals of
electron doped Ba(Fe$_{1-x}$Co$_x$)$_2$As$_2$ \cite{MDLumsden09,DSInosov10} and
BaFe$_{2-x}$Ni$_x$As$_2$ superconductors \cite{SXChi,SLLi09}, the resonance was indeed found
at the in-plane AF ordering wave vector ${\bf Q}_{\rm AF}=(1,0)$ [Fig. 12(a)].
Similar measurements on powder samples of LaFeAsO$_{1-x}$F$_x$ \cite{Ishikodo09,Wakimoto10,Shamoto10}
and molecular-intercalated FeSe \cite{AETaylor2013} also revealed resonance like spin excitations below $T_c$.
Figure 12(c) shows temperature dependence of the imaginary part of the dynamic susceptibility
$\chi^{\prime\prime}({\bf Q}_{\rm AF},\omega)$ for energies below
($E=\hbar\omega=3$ meV), at ($E=9.5$ meV), and above ($E=16$ meV) the resonance
in superconducting BaFe$_{1.85}$Co$_{0.15}$As$_2$ ($T_c=25$ K).
It is clear that the intensity gain of the resonance below $T_c$ is at the expense of opening a spin gap at energies below it.  By carefully monitoring the temperature dependence of the resonance, the authors of \cite{DSInosov10} suggest that the
energy of the mode decreases with increasing temperature and may be directly correlated
with the temperature dependence of the superconducting gap energy [Fig. 12(d)].
However, recent experiments on the nearly
optimally doped BaFe$_{1.904}$Ni$_{0.096}$As$_2$ superconductor found that the resonance energy is essentially temperature independent on warming \cite{HQLuo13}, different from that of BaFe$_{1.85}$Co$_{0.15}$As$_2$ \cite{DSInosov10}.

In the electron underdoped regime where static AF order coexists and competes with superconductivity [Figs. 5(b) and 5(d)], the static AF order occurs at a lower temperature than $T_s$.
Figure 13(a) shows the temperature dependence of the nuclear peak intensity at $(2,2,0)$
and magnetic
Bragg scattering at ${\bf Q}_{\rm AF}$ for underdoped Ba(Fe$_{0.953}$Co$_{0.047}$)$_2$As$_2$ ($T_c=17$ K) \cite{dkbratt09}.
In the high temperature tetragonal state, the observed neutron scattering intensity from a strong nuclear Bragg peak $(2,2,0)$ is lower than that expected from the structure factor calculation due to multiple scattering effect, termed as the
neutron extinction effect \cite{Hamilton1957}.
When the symmetry of the system is reduced from tegragonal to orthorhombic, there is a
dramatic intensity gain  below $\sim$60 K arising from the release of the neutron extinction
effect and the AF order occurs below $T_N\approx 48$ K. Upon entering into the superconducting
state, the intensity of the static AF order decreases. Simultaneously, a weak neutron spin resonance
appears at $E=4$ meV [Fig. 13(b)], suggesting that the intensity gain of the mode arises from suppression
of the static AF order and spin excitations at energies below the resonance \cite{dkbratt09,adchristianson09}.
Application of a magnetic field which partially suppresses superconductivity will enhance the intensity of the static AF order and
suppress the resonance \cite{MYWang11b}.  These results further suggest that the static AF order
coexists and competes with superconductivity in electron underdoped iron pnictides.

\begin{figure}[t]
\includegraphics[scale=.36]{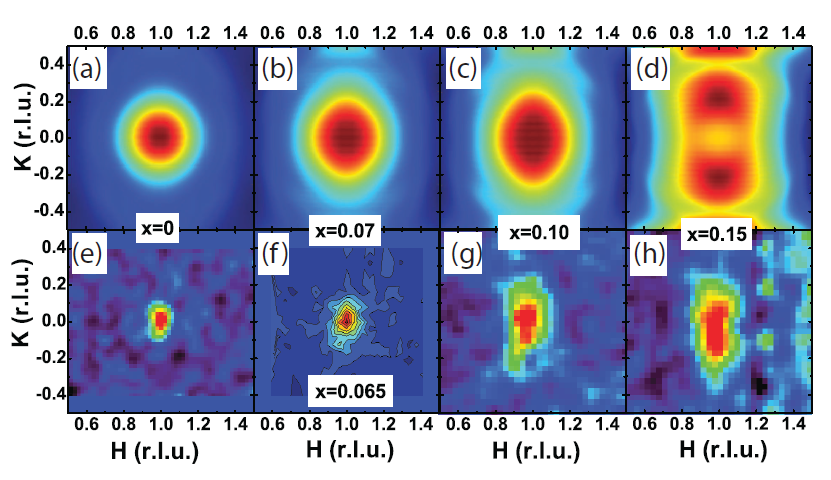}
\caption{
(Color online) Comparison of wave vector evolution of the low-energy spin excitations in electron-doped
BaFe$_{2-x}$Ni$_x$As$_2$ with the RPA calculation based on a rigid band shift.
(a,b,c,d) RPA calculation results obtained for an energy $E=8$ meV for electron dopings of $x=0, 0.07, 0.1$, and
0.15, respectively.  As the doping increases from (a) to (d), one clearly sees an
enhancement of the anisotropy in spin excitations (transverse
elongation).  (e,f,g,h) In-plane wave vector dependence of the spin excitations at $E=8$ meV for
$x=0, 0.065, 0.1$, and 0.15, respectively \cite{hqluo12}. For the electron overdoped $x=0.15$ sample,
two transverse incommensurate peaks are expected from the RPA calculation.  This is indeed
observed in neutron scattering experiments \cite{HQLuo13}.
}
\end{figure}

From density functional theory calculations \cite{IIMazin08,KKuroki08} and
angle resolved photoemission spectroscopy (ARPES) experiments on electron/hole doped iron pnitides
\cite{PRichard11,XHChen14}, we know that Fermi surfaces in most of these
materials are composed of hole-like pockets near $\Gamma$ and electron-like pockets near $M$ point at
${\bf Q}_{\rm AF}=(1,0)$.
The neutron spin resonance in iron pnictides at
${\bf Q}_{\rm AF}=(1,0)$
can arise from the sign reversed quasiparticle excitations
 between the hole and electron Fermi surfaces in an $s^{+-}$-wave superconductor as shown in Fig. 14 \cite{mazin2011n,hirschfeld},
exhibiting the same signature as the sign changed superconducting gap function in the $d$-wave copper oxides \cite{eschrig}.
With increasing electron-doping, the hole and electron Fermi surfaces decrease
and increase in size, respectively [Figure 14(a)-14(c)].  Similarly, the hole Fermi pockets
at the $\Gamma$ point
increase in size with increasing hole-doping, while the electron Fermi surfaces exhibit
a Lifshitz transition at $M$ point before becoming hole overdoped KFe$_2$As$_2$ [Fig. 14(d)-14(f)] \cite{XHChen14}.

\begin{figure}[t]
\includegraphics[scale=.4]{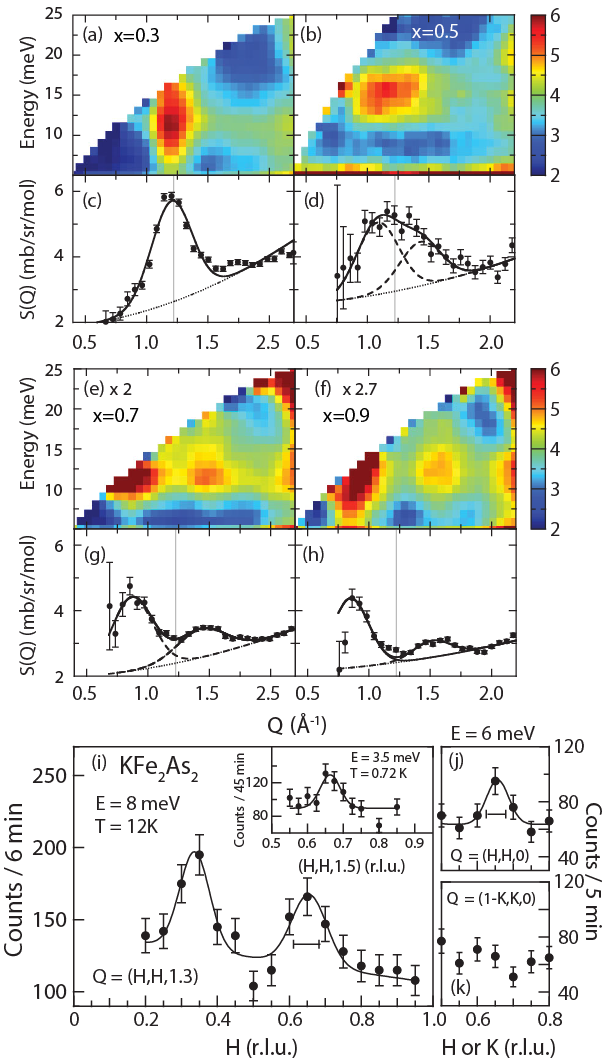}
\caption{
(Color online) The K-doping evolution of the neutron spin resonance and low-energy spin excitations in
Ba$_{1-x}$K$_x$Fe$_2$As$_2$.
(a)  Inelastic neutron scattering experiments on superconductivity-induced
low-energy spin excitations of powder samples in
Ba$_{1-x}$K$_x$Fe$_2$As$_2$ at $x=0.3$.  A clear resonance is seen around 12 meV and $Q=1.25$ \AA$^{-1}$
as shown in (c).
(b,d) The resonance becomes broader in $Q$ at $x=0.5$, and splits into two peaks at
different wave vectors at (e,g) $x=0.7$ and (f,h) $x=0.9$ due to the changing
Fermi surfaces \cite{JPCastellan}.  (i) Longitudinal scans along the $[H,H,1.3]$ direction
above $T_c$ at $E=8$ meV for single crystals of KFe$_2$As$_2$.  Two incommensurate peaks
are seen.  The inset shows similar scan below $T_c$ at $E=3.5$ meV.
Longitudinal (j) and transverse (k) scans at $E=6$ meV \cite{CHLee11}.
}
\end{figure}

Using the random phase approximation (RPA) based on a
three-dimensional tight-binding model in the local density
approximation (LDA) \cite{SGraser10}, calculations can predict the momentum anisotropy of the low-energy spin excitations
and the resonance \cite{JTPark10}. For electron-doped BaFe$_{2-x}T_x$As$_2$, low-energy spin excitations
become progressively elongated ellipses
along the transverse direction
relative to the spin waves in BaFe$_2$As$_2$ due to the enhancement of the intra-orbital, but inter-band,
pair scattering process between the $d_{xy}$ orbitals [Figs. 14(a) and 14(b)] \cite{JHZhang10}. Figure 15 shows
the comparison of the RPA calculations and experimentally measured in-plane spin excitation anisotropy in
BaFe$_{2-x}$Ni$_x$As$_2$ superconductors \cite{HQLuo5}, confirming that
the quasiparticle excitations between the hole and electron Fermi surfaces are consistent with the wave vector evolution of the
low-energy spin excitations \cite{HQLuo13}.

\begin{figure}[t]
\includegraphics[scale=0.2]{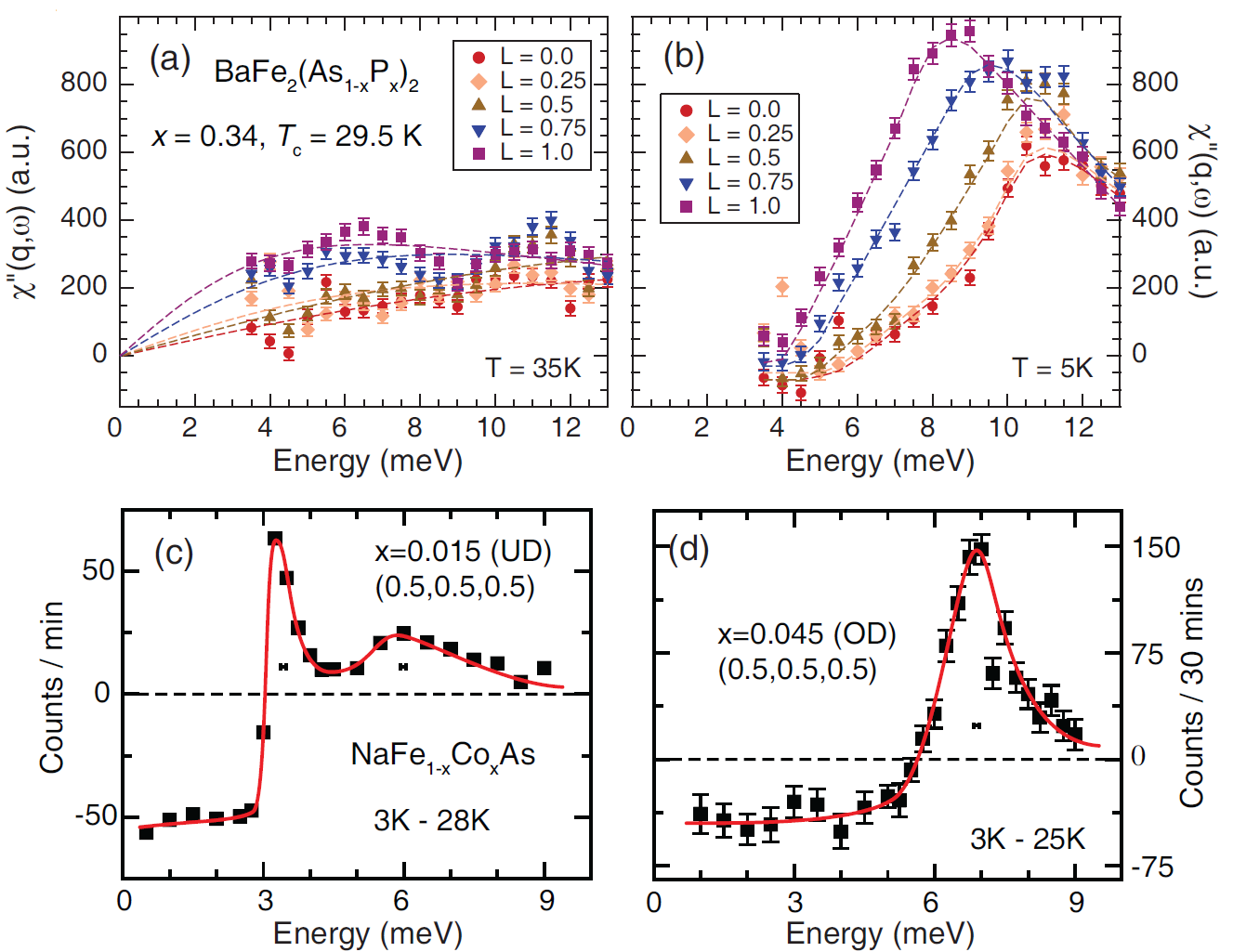}
\caption{
(Color online) The wave vector and energy dependence of the neutron spin resonance for BaFe$_2$(As$_{1-x}$P$_x$)$_2$ and
NaFe$_{1-x}$Co$_x$As.
(a) The energy dependence of the dynamic susceptibility at the in-plane AF wave vector position and different $L$ values
above $T_c$ for BaFe$_2$(As$_{1-x}$P$_x$)$_2$ with $x=0.34$. (b) Identical scans at temperature $T=5$ K well below $T_c$.
The resonance shows clear dispersion for different values of $L$ \cite{CHLee13}. (c)
The energy dependence of the superconductivity-induced double resonance for an underdoped (UD)
NaFe$_{1-x}$Co$_x$As with $x=0.015$.  There are two peaks in the energy scan at
$E=3.5$ and 6 meV. (d) The double resonance in the underdoped sample becomes a single resonance
for electron overdoped NaFe$_{1-x}$Co$_x$As with $x=0.045$ \cite{CLZhangL13}.
}
\end{figure}

In the case of hole-doped materials,
RPA calculations have predicted that spin excitations should be longitudinally elongated, and thus rotated 90$^\circ$ from those of the
electron-doped BaFe$_{2-x}T_x$As$_2$ \cite{JTPark10}.  Inelastic neutron scattering experiments on hole-doped
single crystals of superconducting Ba$_{0.67}$Ka$_{0.33}$Fe$_2$As$_2$ ($T_c=38$ K) reveal longitudinally elongated spin excitations for energies near the resonance, consistent with RPA calculations \cite{CLZhang11}.
Figure 16(a)-16(h) plots the hole-doping dependence of the resonance obtained using high-quality powder samples of
Ba$_{1-x}$K$_x$Fe$_2$As$_2$ \cite{SAvci,JPCastellan}. Although these measurements do not provide precise
information concerning the wave vector dependence of the spin excitations, they do give the
hole-doping evolution of the total momentum transfer of the mode.  With increasing
hole-doping, the sharp resonance centered at ${Q}\approx 1.25$ \AA$^{-1}$ for $x=0.3$ [Fig. 16(a) and 16(c)]
becomes broader in ${Q}$ and splits into two peaks for $x=0.7$ and 0.9 [Fig. 16(e)-16(h)] \cite{JPCastellan}.
This is consistent with the RPA result that hole-doping induces longitudinal incommensurate spin excitations \cite{JPCastellan}.
Indeed, neutron scattering experiments on hole overdoped KFe$_2$As$_2$ found two incommensurate spin
excitation peaks located longitudinally away from ${\bf Q}_{\rm AF}$ [Fig. 16(i)-16(k)], again confirming the notion that low-energy spin excitations
in hole and electron-doped iron pnictides are controlled by quasiparticle excitations
between the hole and electron Fermi surfaces \cite{CHLee11}.

In addition to electron or hole-doping to BaFe$_2$As$_2$, isoelectronic substitution to
BaFe$_2$As$_2$ by replacing Fe with Ru \cite{AThaler10} or As with P \cite{ZAXu2} can also induce superconductivity. Compared with the electron-doped BaFe$_{2-x}T_x$As$_2$, isoelectronic substitution is much less effective in suppressing AF order and inducing superconductivity. Inelastic neutron scattering experiments on BaFe$_{2-x}$Ru$_x$As$_2$ near optimal superconductivity reveal a neutron spin resonance
similar to electron-doped BaFe$_{2-x}T_x$As$_2$, but with greatly damped intensity, possibly due to the weakening of the
electron-electron correlations by Ru doping \cite{JZhao13}.  In the case of BaFe$_2$(As$_{1-x}$P$_x$)$_2$, initial neutron
scattering experiments on powder samples with $T_c=30$ K have revealed the presence of a resonance
at $E\approx 12$ meV \cite{MIshikado11}.  Figure 17(a) and 17(b) shows the energy dependence of $\chi^{\prime\prime}({\bf Q},\omega)$ above
and below $T_c$, respectively, obtained for single crystals of BaFe$_2$As$_{1.32}$P$_{0.68}$ ($T_c=29.5$ K) \cite{CHLee13}.
In the normal state, $\chi^{\prime\prime}({\bf Q},\omega)$  is featureless and changes only slightly at different momentum transfers along
the $c$ axis ($L=0,0.25,0.5,0.75,1$). Upon entering into the superconducting state, a neutron spin resonance is formed and
its energy is significantly dispersive along the $c$ axis [Fig. 17(b)] \cite{CHLee13}.  Since the bandwidth of the dispersion becomes larger
on approaching the AF ordered phase, the dispersive feature may arise from the three-dimensional AF spin correlations in the undoped parent \cite{CHLee13}.

So far, most of the neutron scattering work has been focused on single crystals of electron/hole-doped BaFe$_2$As$_2$ family of materials. For the NaFe$_{1-x}$Co$_x$As family of materials \cite{DRParker13,GTTan13}, the air sensitive nature of these materials makes it very difficult
to perform inelastic neutron scattering experiments \cite{MATanatar12}.  By using hydrogen free glue to coat the samples, neutron scattering experiments can be carried out to study the evolution of spin excitations in NaFe$_{1-x}$Co$_x$As \cite{JTPark12,ysong13}. From ARPES experiments \cite{ZHLiu11,QQGe13}, it was found that the superconducting gap in the electron Fermi pockets of the
underdoped regime near $x=0.0175$ has a large anisotropy, which is absent in the hole Fermi pocket.  The superconducting gap anisotropy disappears upon increasing $x$ to 0.045. Figure 17(c) and 17(d) shows the intensity gain of the
resonance below $T_c$ for underdoped
$x=0.015$ \cite{CLZhangL13} and overdoped $x=0.045$ \cite{CLZhangB13}, respectively.  Instead of a single resonance peak, superconductivity induces a sharp resonance at $E_{r1}=3.25$ meV and a broad resonance at $E_{r2}=6$ meV \cite{CLZhangL13}.  Similar
measurements on electron overdoped $x=0.045$ reveal only one sharp resonance \cite{CLZhangB13}.   The appearance of the double resonance
and the superconducting gap anisotropy
in the underdoped sample was interpreted as originating from either the orbital dependence of
the superconducting pairing \cite{CLZhangL13,RYu14} or superconductivity coexisting with static AF order
in the iron pnictides \cite{WRowe12,WCLv14}.
\clearpage
\begin{figure}[t]
\includegraphics[scale=0.32]{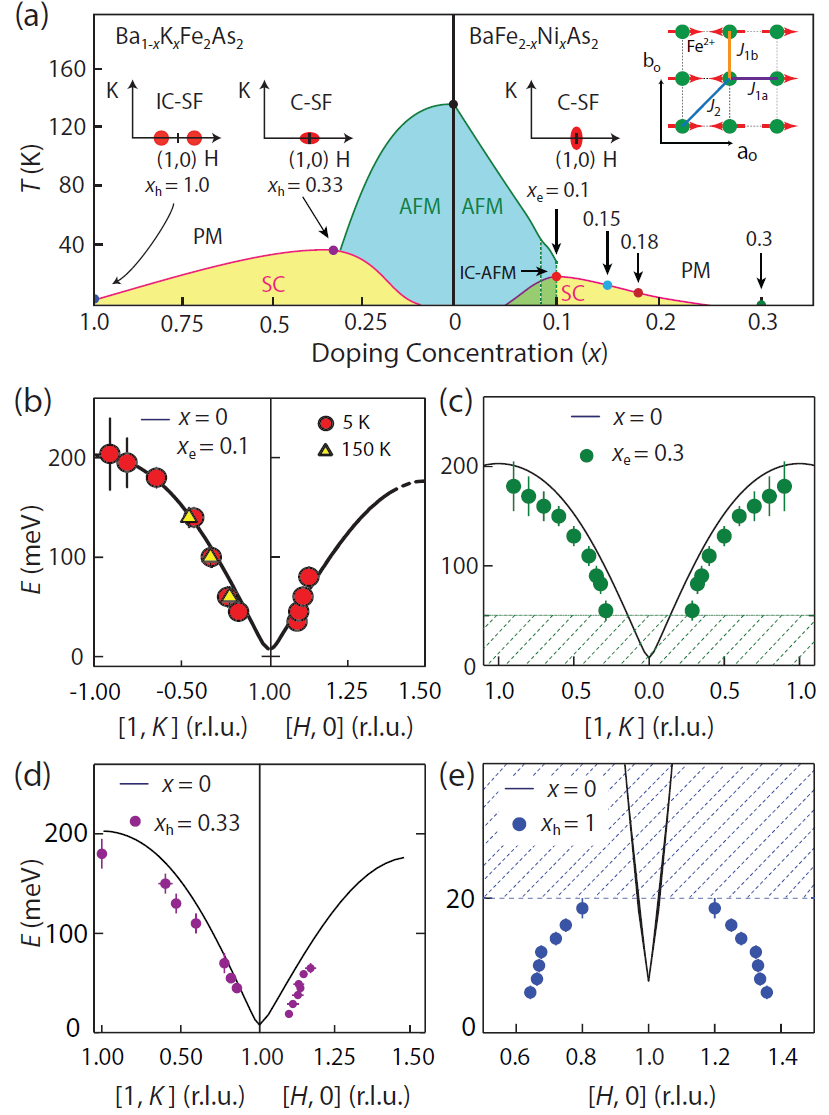}
\caption{
(Color online)
The evolution of spin excitation dispersions for hole and electron-doped BaFe$_2$As$_2$.
(a) The electronic phase diagram of electron and hole-doped BaFe$_2$As$_2$, where the arrows indicate the doping levels of inelastic neutron scattering
experiments. The right
inset shows the crystal and AF spin structures of BaFe$_2$As$_2$. The inset above $x_e=0.1$
shows the transversely elongated ellipse representing the low-energy
spin excitations
in electron-doped BaFe$_{2-x}$Ni$_x$As$_2$ in the $(H,K)$ plane of reciprocal space.
The left insets show the evolution of low-energy spin excitations in hole-doped Ba$_{1-x}$K$_x$Fe$_2$As$_2$ in the $(H,K)$ plane.
C-SF and IC-SF indicate commensurate and incommensurate spin fluctuations, respectively.
(b-e) The solid lines in the figure are spin wave dispersions of the undoped BaFe$_2$As$_2$ along the two high-symmetry directions.
The symbols in (b), (c), (d), and (e) are dispersions of spin excitations for BaFe$_{1.9}$Ni$_{0.1}$As$_2$, BaFe$_{1.7}$Ni$_{0.3}$As$_2$,
Ba$_{0.67}$K$_{0.33}$Fe$_2$As$_2$, and KFe$_2$As$_2$, respectively.
The shaded areas indicate vanishing spin excitations \cite{MWang13}.
}
\end{figure}

\subsection{The electron and hole-doping evolution of the spin excitations in the BaFe$_2$As$_2$ family of iron pnictides}

To understand the interplay between magnetism and superconductivity in iron pnictides, one must first
determine the electron and hole-doping evolution
of the spin excitation spectra throughout the Brillouin zone. Since single crystals of electron and hole-doped BaFe$_2$As$_2$
are available, one can systematically map out the evolution of spin excitations
at different electron/hole-doping levels marked with arrows in the phase diagram [Fig. 18(a)] \cite{XHChen14,LWHarriger11,CHLee11,MSLiu12,GSTucker12,HQLuo13,MWang13}. The solid lines in Figs. 18(b)-18(e) show the dispersion of spin waves in BaFe$_2$As$_2$
along the $[1,K]$ and $[H,0]$ directions \cite{LWHarriger11}.
Upon electron-doping to induce optimal superconductivity, spin excitations become broader at low-energies ($E\le 80$ meV)
and couple to superconductivity via the resonance
 while remaining almost unchanged at high energies ($E>80$ meV) \cite{MSLiu12,GSTucker12}.
The red circles and yellow upper triangles in Fig. 18(b)
show spin excitation dispersions of the optimally electron doped BaFe$_{1.9}$Ni$_{0.1}$As$_2$ at
$T=5$ K and 150 K, respectively \cite{MSLiu12}.  Figure 18(c) shows the dispersions of spin excitations of the electron-overdoped
nonsuperconducting
BaFe$_{1.7}$Ni$_{0.3}$As$_2$, where a large spin gap forms for energies below $\sim$50 meV \cite{MWang13}.
Figure 18(d) and 18(e) shows the dispersions of spin excitations
for optimally hole-doped Ba$_{0.67}$K$_{0.33}$Fe$_2$As$_2$ and hole-overdoped KFe$_2$As$_2$, respectively \cite{MWang13}.
While electron doping does not much affect the high-energy spin excitations and dispersion, hole-doping suppresses the high-energy
spin excitations.

\begin{figure}[t]
\includegraphics[scale=.4]{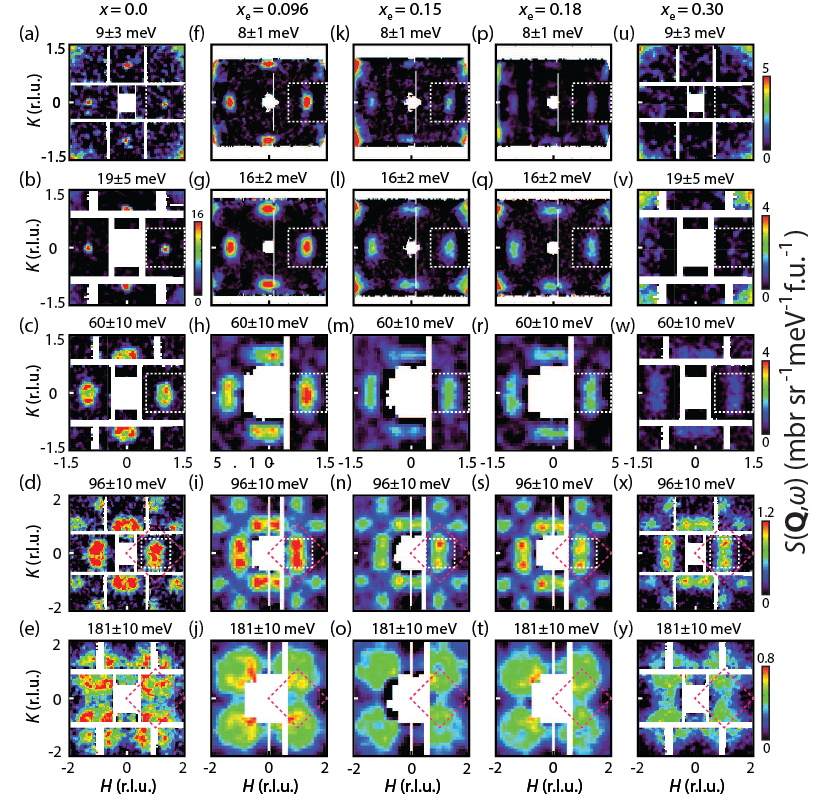}
\caption{
(Color online)
Constant-energy slices through magnetic excitations of electron-doped
BaFe$_{2-x}$Ni$_x$As$_2$ iron pnictides at different energies.
The color bars represent the vanadium
normalized absolute spin excitation intensity in the units of mbarn$\cdot$sr$^{-1}$meV$^{-1}$f.u.$^{-1}$.
(a-e) Spin waves in BaFe$_2$As$_2$ at excitation energies of $E=9\pm 3$, $19\pm 5$, $60\pm10$, $96\pm 10$, and $180\pm 10$ meV \cite{LWHarriger11}.  Spin waves peak at
the AF ordering wave vectors ${\bf Q}_{\rm AF}=(\pm 1,0)$ in the orthorhombic notation.  Spin waves are also seen at ${\bf Q}_{\rm AF}\approx (0,\pm 1)$ due to the twin domains of the orthorhombic structure.
(f-j) Two-dimensional images of spin excitations for BaFe$_{1.904}$Ni$_{0.096}$As$_2$ at $E=8\pm 1$, $16\pm 2$, $60\pm 10$, $96\pm 10$,
$181\pm 10$ meV.  Identical slices as that of (f-j) for (k-o) BaFe$_{1.85}$Ni$_{0.15}$As$_2$ and (p-t) BaFe$_{1.82}$Ni$_{0.18}$As$_2$ \cite{HQLuo5}.
(u-y) Constant-energy slices through magnetic excitations of electron overdoped doped nonsuperconducting
 BaFe$_{1.7}$Ni$_{0.3}$As$_2$ at $E=9\pm 3$, $19\pm 5$, $60\pm 10$, $96\pm 10$,
$181\pm 10$ meV \cite{MWang13}. The white dashed box indicate wave vector integration range at low-energies, while the purple
dashed boxes in (d-h) mark the integration range for high-energy spin excitations.
}
\end{figure}

Figure 19 reveals the evolution of the two-dimensional constant-energy images of spin excitations in the $(H, K)$ plane at different energies
 as a function of electron-doping for BaFe$_{2-x}$Ni$_x$As$_2$ \cite{LWHarriger11,HQLuo13,MWang13}. In
undoped BaFe$_2$As$_2$, there is an anisotropy spin gap
below $\sim$15 meV, thus there is essentially no signal at $E=9\pm 3$ meV [Fig. 19(a)] \cite{KMatan09}. For nearly
optimally electron doped $x=0.096$, the spin gap is suppressed and low-energy spin excitations
are dominated by the resonance [Fig. 19(f)] \cite{MDLumsden09,DSInosov10,SXChi,SLLi09,HQLuo5}.
In electron overdoped BaFe$_{2-x}$Ni$_x$As$_2$ with $x=0.15$ ($T_c=14$ K) and 0.18 ($T_c=8$ K),
spin excitations at $E=8\pm 1$ meV become weaker and more transversely elongated [Figs. 19(k) and 19(p)] \cite{HQLuo13}.
For the nonsuperconducting $x=0.3$ sample,
a large spin gap forms in the low-energy excitation spectra [Fig. 19(u)].
Figures 19(b)-19(e), 19(g)-19(j), 19(q)-19(t), 19(v)-19(y)
show the evolution of spin excitations at different energies for BaFe$_{2-x}$Ni$_x$As$_2$ with $x=0,0.096,0.15,0.18,$ and 0.30, respectively.  While
electron doping modifies
spin excitations at energies below $E=96\pm 10$ meV, high energy spin excitations remain similar and only soften slightly.

\begin{figure}[t]
\includegraphics[scale=.25]{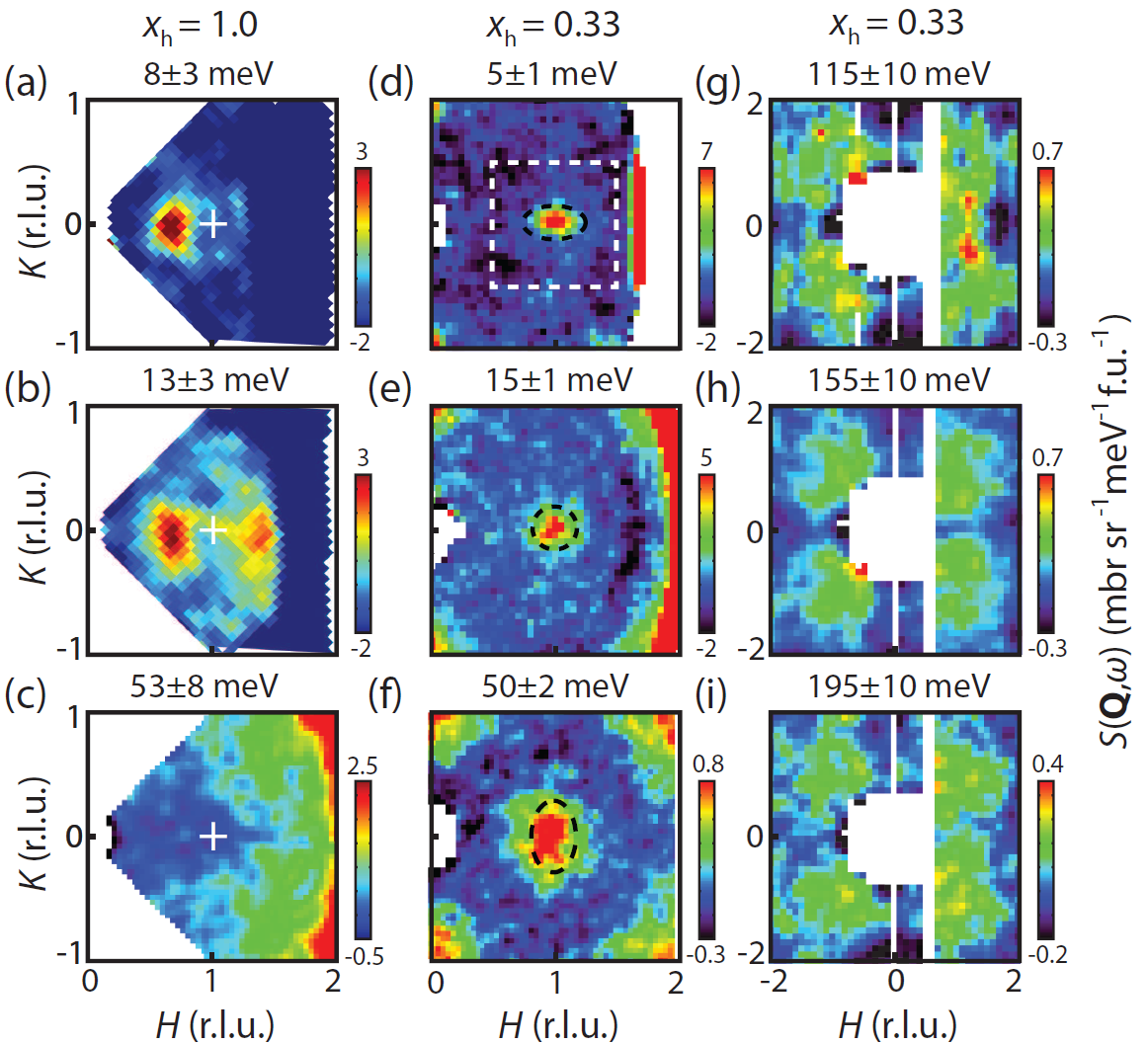}
\caption{
(Color online)
Wave vector dependence of spin excitations in hole doped
Ba$_{1-x}$K$_x$Fe$_2$As$_2$ from single crystal measurements.
Two-dimensional images of spin excitations at different energies for hole-doped KFe$_2$As$_2$ at 5 K.
(a) $E=8\pm 3$ meV obtained with $E_i=20$ meV along the $c$-axis.
The right side incommensurate peak is obscured by background scattering.
 (b) $13\pm3$ meV with $E_i=35$ meV,  and (c) $53\pm 8$ meV with $E_i=80$ meV.
For  Ba$_{0.67}$K$_{0.33}$Fe$_2$As$_2$ at $T=45$ K, images of spin excitations at (d) $E=5\pm 1$ meV obtained with $E_i=20$ meV, (e) $15\pm 1$ meV with
$E_i=35$ meV, and (f) $50\pm 2$ meV obtained with $E_i=80$ meV.
Spin excitations in Ba$_{0.67}$K$_{0.33}$Fe$_2$As$_2$ at energy transfers
(g) $115\pm 10$ meV; (h) $155\pm 10$ meV (i) $195\pm 10$ meV obtained
with $E_i=450$ meV, all at 9 K. Wave vector dependent backgrounds have been subtracted from the images \cite{MWang13}.
}
\end{figure}

Figure 20 shows the constant-energy images
of spin excitations as a function of hole-doping. For pure KFe$_2$As$_2$, incommensurate spin excitations along the longitudinal direction are seen at $E=8\pm 3$ meV [Fig. 20(a)] and $13\pm 3$ meV [Fig. 20(b)] \cite{CHLee11}.
However, spin excitations become much weaker at $E=53\pm 8$ meV \cite{MWang13}.
For optimally hole-doped Ba$_{0.67}$K$_{0.33}$Fe$_2$As$_2$,
the low-energy spin excitations change from
longitudinally elongated ellipses at $E=5\pm1$ meV [Fig. 20(d)] to transversely
elongated ellipses at $E=50\pm 2$ meV [Fig. 20(f)].
At the neutron spin resonance energy of $E=15\pm 1$ meV, spin
excitations change from longitudinally elongated ellipses above $T_c$ (not shown) to
isotropic circles below $T_c$ in reciprocal space [Fig. 20(e)].
For energies above 100 meV, spin excitations in hole-doped Ba$_{0.67}$K$_{0.33}$Fe$_2$As$_2$ [Figs. 20(g)-20(i)]
behave similarly to those of
 electron-doped BaFe$_{2-x}$Ni$_x$As$_2$ (Fig. 19) \cite{MWang13}.

\begin{figure}[t]
\includegraphics[scale=.28]{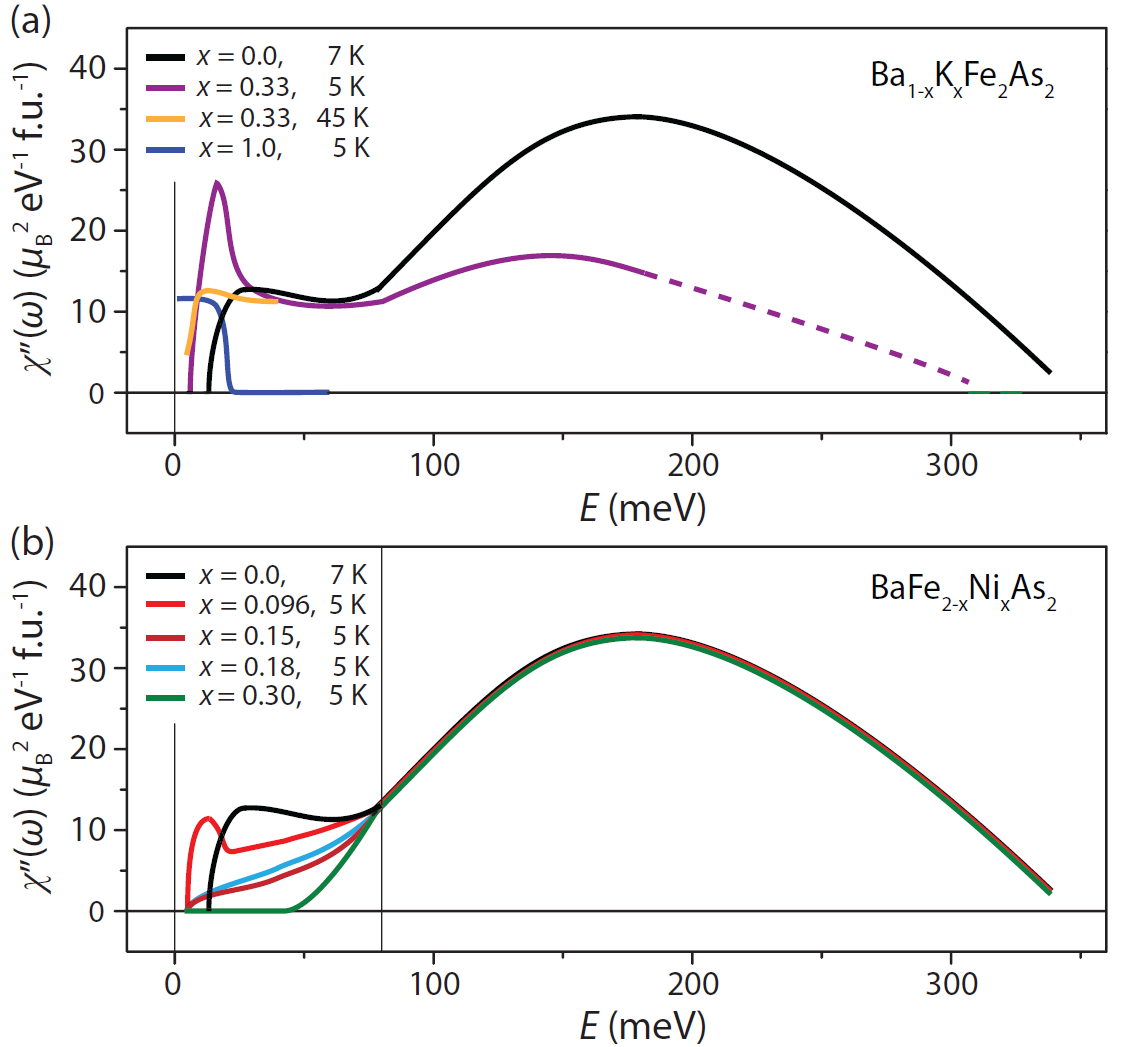}
\caption{
(Color online)
Energy and temperature dependence of the local dynamic susceptibility  $\chi^{\prime\prime}(\omega)$ for (a)
BaFe$_2$As$_2$ (black solid line), Ba$_{0.67}$K$_{0.33}$Fe$_2$As$_2$ (purple and yellow
solid lines for below and above $T_c$, respectively), KFe$_2$As$_2$ (blue solid line),
and (b) BaFe$_{2-x}$Ni$_{x}$As$_2$ with $x=0,0.096,0.15,0.18,0.3$,
corresponding to black, red, dark red, light blue, and green lines respectively.
The intensity is in absolute units of $\mu_B^2$eV$^{-1}$f.u.$^{-1}$ obtained
by integrating the $\chi^{\prime\prime}(Q,\omega)$ in the
dashed regions specified in Figs. 18 and 19 \cite{MSLiu12,MWang13}.
}
\end{figure}

To quantitatively determine the electron and hole-doping evolution of the spin excitations in iron pnictides, one can estimate the energy dependence of the
local dynamic susceptibility per formula unit $\chi^{\prime\prime}(\omega)$ \cite{lester10,MSLiu12}.
The dashed boxes in Figs. 19 and 20 show the
integration region of the
local dynamic susceptibility in reciprocal space. At low-energies, we only integrate the scattering
within the white dashed box since it includes all magnetic responses in the Brillouin zone.  Approaching to
the zone boundary, we integrate the response within the purple dashed box in
Fig. 19 as discussed in Section II A (Fig. 3).
The energy dependence of the
local dynamic susceptibility for hole and
electron doped iron pnictides are plotted in
Figs. 21(a) and 21(b), respectively.
We see that the effect of hole-doping near optimal superconductivity is to
suppress high-energy spin excitations and transfer spectral weight to low energies.
The intensity changes across $T_c$ for hole-doped Ba$_{0.67}$K$_{0.33}$Fe$_2$As$_2$ are
much larger than that of the electron-doped BaFe$_{1.9}$Ni$_{0.1}$As$_2$ \cite{MSLiu12}.  As a function of increasing electron-doping,
the local dynamic susceptibility at low energies decreases
and finally vanishes for electron-overdoped nonsuperconducting  BaFe$_{1.7}$Ni$_{0.3}$As$_2$ \cite{HQLuo13,MWang13}.

\subsection{Evolution of spin excitations in iron chalcogenides and alkali iron selenides}

Compared with iron pnictides,
iron chalcogenide (Fe$_{1+y}$Te$_{1-x}$Se$_{x}$) superconductors have a different static
AF ordered  (bi-collinear instead of collinear) parent compound \cite{Fruchart,WBao5,SLLi2},
but a similar Fermi surface topology \cite{ASubedi,KNakayama10,DLFeng10}.
If the resonance originates from the hole and electron Fermi surface nesting, one would also expect a resonance
at a wave vector similar to that of the iron pnictides.
The neutron scattering experiments on FeTe$_{0.6}$Se$_{0.4}$ reveal that this is indeed the case \cite{WBao09,HAMook10,Babkevich10}.
Figure 22(a) shows that the resonance energy is weakly temperature dependent and suddenly
vanishes above $T_c$ \cite{WBao09,LWHarriger12}.  Another interesting aspect of
Fe$_{1+y}$Te$_{1-x}$Se$_{x}$ is the presence of transverse incommensurate
spin excitations at different energies [Figs. 22(b) and 22(c)] \cite{MDLumsden10,SHLee10,DNArgyriou10,SLLi10}.
Since the parent compound of iron chalcogenide superconductors has bi-collinear spin structure,
the AF Bragg peaks and associated spin excitations in nonsuperconducting iron chalcogenides
stem from wave vector positions rotated 45$^\circ$ from those of the resonance in reciprocal space.
The enhancement of the resonance in superconducting Fe$_{1+y}$Te$_{1-x}$Se$_{x}$
occurs at the expense of the spin excitations associated with the AF nonsuperconducting parent
compound \cite{ZJXu10,TJLiu10,SXChi11}.

\begin{figure}[t]
\includegraphics[scale=.35]{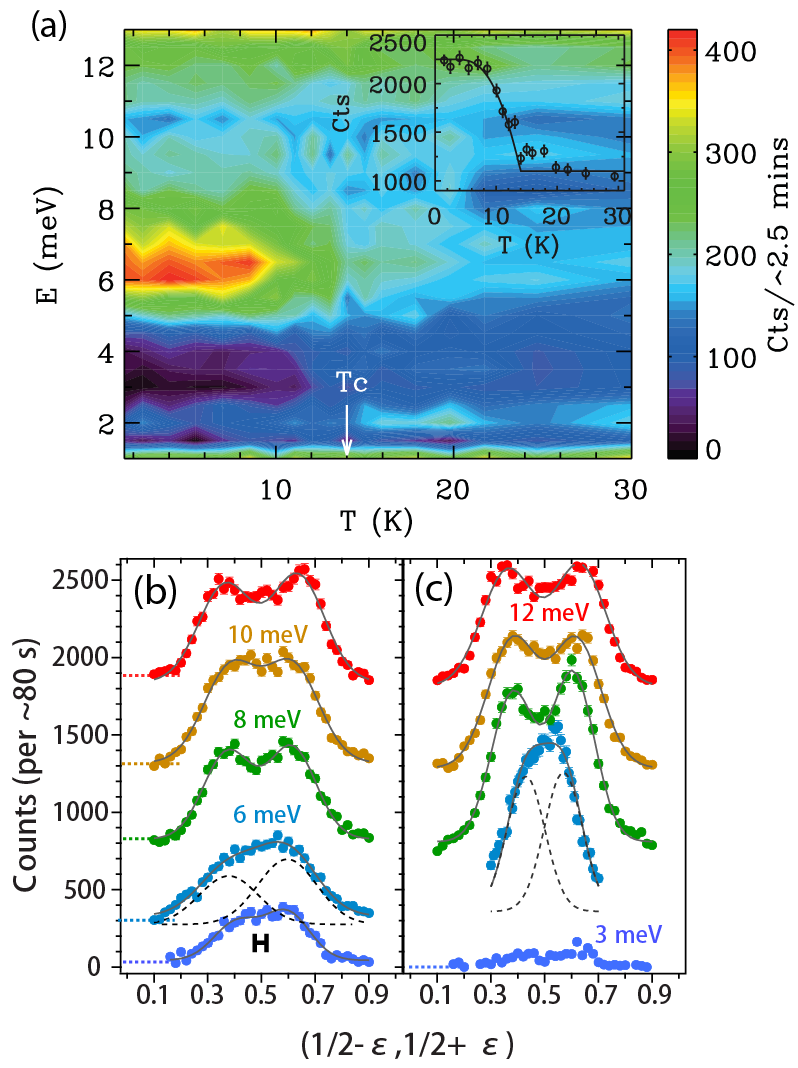}
\caption{
(Color online)
Temperature and wave vector dependence of the resonance and low-energy spin excitations in iron
chalcogenide superconductors.
(a) Temperature dependence of the resonance energy for optimally doped FeTe$_{0.6}$Se$_{0.4}$.
The mode energy is essentially temperature independent \cite{WBao09}.  The inset shows the
temperature dependence of the resonance intensity. The wave vector dependence of the spin excitations
at different energies
along the transverse direction (b) above and (c) below $T_c$ for FeTe$_{0.6}$Se$_{0.4}$ \cite{DNArgyriou10}.
}
\end{figure}

Figure 23 compares the wave vector dependence of spin excitations at different energies within the $(H,K)$ plane for
nonsuperconducting Fe$_{1+y}$Te$_{0.73}$Se$_{0.27}$
and superconducting Fe$_{1+y}$Te$_{0.51}$Se$_{0.49}$ \cite{MDLumsden10}.
Using the tetragonal crystalline lattice unit cell,
the reciprocal lattice units in Fe$_{1+y}$Te$_{1-x}$Se$_{x}$ are rotated 45$^\circ$
from that for the AF ordered orthorhombic iron pnictides (Fig. 3). In this notation, spin waves from the bi-collinear AF ordered
Fe$_{1+y}$Te stem from ${\bf Q}_{\rm AF}=(\pm 0.5,0)$ in reciprocal space while the resonance occurs at $(0.5,0.5)$ \cite{MDLumsden10}.
For the nonsuperconducting Fe$_{1+y}$Te$_{0.73}$Se$_{0.27}$, spin excitations at low energies ($E=10\pm 1$, $22\pm 3$ meV)
peak at transversely incommensurate positions from $(0.5,0.5)$ [Figs. 23(a) and 23(b)]. On increasing the
energies to $E=45\pm 5$ [Fig. 23(c)] and $120\pm 10$ meV [Fig. 23(d)], spin excitations become fourfold symmetric and move
to positions near $(\pm 1,0)$ and $(0,\pm 1)$ \cite{MDLumsden10}. For superconducting Fe$_{1+y}$Te$_{0.51}$Se$_{0.49}$,
the transverse incommensurate spin excitations in the nonsuperconducting sample at $E=10\pm 1$ and
$22\pm 3$ meV are replaced by
the resonance and transversely elongated spin excitations near $(\pm 0.5,\pm 0.5)$ [Figs. 23(e) and 23(f)].  Spin excitations at
$E=45\pm 5$ [Fig. 23(g)] and $120\pm 10$ meV [Fig. 23(h)]
are not greatly affected by superconductivity. These results are similar to spin excitations in electron-doped
iron pnictides \cite{MSLiu12}, suggesting that superconductivity in
iron chalcogenides only affects low-energy spin excitations and has commensurate
spin excitations consistent with the hole and electron Fermi surface nesting \cite{TJLiu10,SXChi11}.

\begin{figure}[t]
\includegraphics[scale=.4]{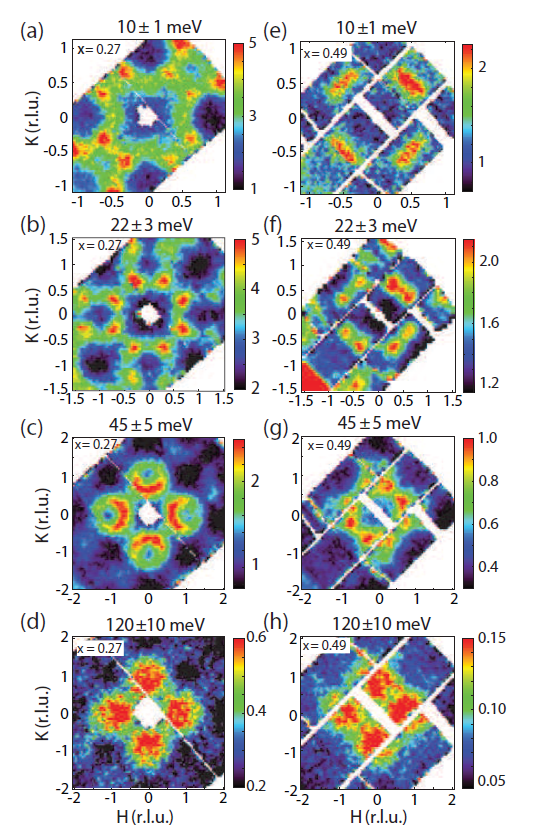}
\caption{
(Color online)
Wave vector evolution of the spin excitations in
FeTe$_{1-x}$Se$_{x}$ throughout the Brillouin zone.
The in-plane wave vector dependence of
the spin excitations in FeTe$_{0.73}$Se$_{0.27}$ at (a) $E=10\pm 1$, (b) $22\pm 3$,
(c) $45\pm 5$, (d) $120\pm 10$ meV.  Identical scans for FeTe$_{0.51}$Se$_{0.49}$
at (e) $E=10\pm 1$, (f) $22\pm 3$,
(g) $45\pm 5$, (h) $120\pm 10$ meV \cite{MDLumsden10}.
}
\end{figure}

In iron pnictide and iron chalcogenide superconductors,
the neutron spin resonance is believed to arise from quasiparticle excitations between the
hole and electron Fermi pockets near the
$\Gamma$ and $M$ points, respectively \cite{mazin2011n,hirschfeld}.
Since alkali iron selenide superconductors $A_x$Fe$_{2-y}$Se$_2$ \cite{XLChen,MHFang}
do not have hole pockets near the Fermi energy \cite{TQian11,DXMou11,YZhang11}, it is important to determine if the system also has a resonance arising from quasiparticle excitations between the two electron-like Fermi pockets near the $(\pm 1,0)$ and $(0,\pm 1)$
positions in reciprocal space \cite{TAMaier11}.  From the earlier work on copper oxide superconductors, it is generally believed that the resonance arises from sign reversed quasiparticle excitations between two different parts of the Fermi surfaces \cite{eschrig}.
As there are no hole Fermi surfaces in superconducting $A_x$Fe$_{2-y}$Se$_2$, a determination of the location of the resonance in reciprocal space will directly test the prediction from the RPA and weak coupling calculation
concerning the nature of the superconducting pairing interaction \cite{TAMaier11}. If a resonance is seen approximately at a wave vector connecting the two electron Fermi pockets, one would expect a sign change between the two Fermi pockets reminiscent of
 the $d$-wave pairing symmetry state of the copper oxide superconductors \cite{TAMaier11,FWang11,TDas11}.

\begin{figure}[t] \includegraphics[scale=.25]{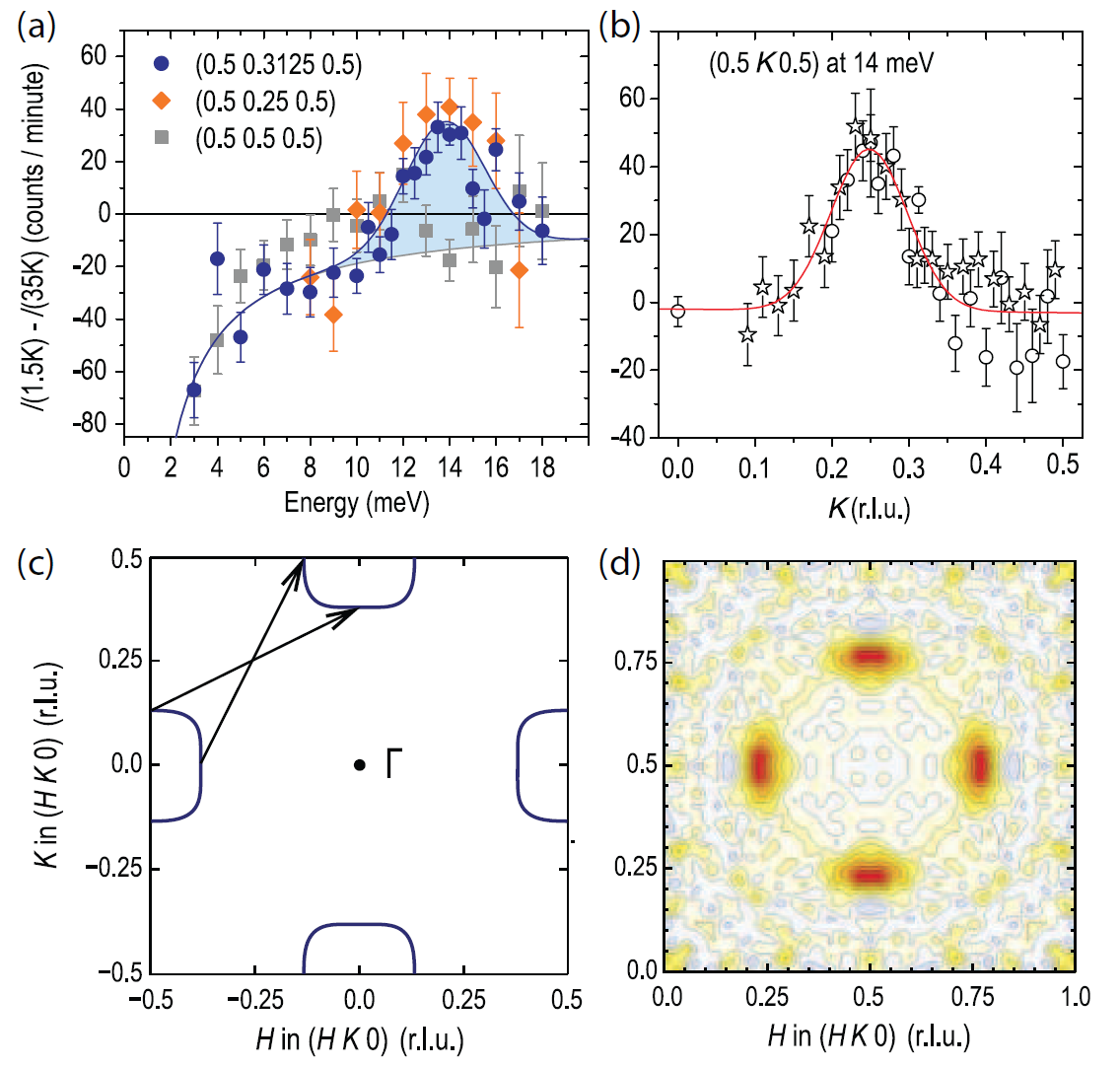}
\caption{
(Color online)
Energy and wave vector dependence of the resonance in the superconducting
alkali iron selenides. (a) Superconductivity-induced neutron scattering intensity
changes in the superconducting Rb$_2$Fe$_4$Se$_5$ with $T_c=32$ K.  A resonance like feature is seen
below $T_c$ at $E=14$ meV. (b) The temperature differences in wave vector scans across the resonance
along the $[0.5,K,0.5]$ direction. (c) Possible nesting wave vectors connecting the two
electron-like Fermi surfaces. (d) The temperature difference plot in the in-plane reciprocal
space reveals the location of the resonance in the superconducting
alkali iron selenides \cite{JTPark11,GFriemel12a,GFriemel12}.
}
\end{figure}

Experimentally, a neutron spin resonance has been observed at an energy of $E_r=14$ meV
in the superconducting Rb$_2$Fe$_4$Se$_5$ with $T_c=32$ K [Fig. 24(a)] \cite{JTPark11}. A complete mapping of the reciprocal space within the $(H,K)$ scattering plane of the system reveals that the mode occurs near the wave vector
$(0.5,0.25,0.5)$ in the tetragonal unit cell notation \cite{GFriemel12a}. Figure 24(a) and 24(b) plots the temperature difference
 between 1.5 K($<T_c$) and 35 K($>T_c$) showing the superconductivity-induced
resonance in energy and wave vector scans, respectively.
Figure 24(c) shows the Fermi surfaces in the $(H,K,0)$ plane corresponding to the
doping level of 0.18 electrons/Fe.  The arrows are the in-plane nesting wave vectors
consistent with the resonance \cite{GFriemel12a}. Figure 24(d) plots the
difference of the RPA calculated dynamic susceptibility between the superconducting
and normal states at the resonance energy \cite{GFriemel12a}. The calculated results are in qualitatively agreement with the
neutron scattering experiments, thus suggesting that the mode arises from quasiparticle excitations between the electron pockets \cite{GFriemel12a,GFriemel12}. Subsequent neutron scattering experiments on superconducting Rb$_{0.82}$Fe$_{1.68}$Se$_2$
($T_c=32$ K) \cite{MYWang12} and Cs$_x$Fe$_{2−y}$Se$_2$ \cite{AETaylor12} also found the resonance at wave vector positions
connecting the two electron Fermi surfaces, thus confirming this is a general feature of the superconducting
alkali iron selenides.
Although the resonance mode energy in molecular-intercalated FeSe superconductors \cite{TPYing12,MBurrardlucas13,KrztonMaziopa12} approximately
follows $\sim 5k_BT_c$ consistent with other iron-based superconductors \cite{DSInosov2011}, its wave vector
is better matched to those of the superconducting
component of $A_x$Fe$_{2-y}$Se$_2$ \cite{AETaylor2013}.

\subsection{Impurity effect on spin excitations of iron pnictide and chalcogenide superconductors}

As described in the earlier sections, low-energy
spin excitations in high-$T_c$ copper oxide and iron-based superconductors are coupled to superconductivity via the opening of a spin gap and
re-distributing the weight to a neutron spin resonance, both at the AF ordering wave vector of their parent compounds \cite{eschrig}.
Since superconductivity in high-$T_c$ superconductors can be altered rather dramatically with impurity doping,
it is important to determine the effect of impurities on spin excitations.
In the case of the copper oxide superconductors, the resonance and low-energy spin excitations respond to
magnetic and nonmagnetic impurity doping differently \cite{ysidis00}.
When magnetic impurities such as Ni are doped into optimally superconducting  YBa$_2$Cu$_2$O$_7$,
the resonance peak shifts to lower energy with a preserved energy-to-$T_c$ ratio \cite{ysidis00}.
In contrast, nonmagnetic
impurity Zn doping to YBa$_2$Cu$_2$O$_7$ restores normal state spin excitations but hardly changes the energy
of the resonance \cite{ysidis00}.  Similar Zn-substitution in the underdoped YBa$_2$Cu$_2$O$_{6.6}$
induces static magnetic order at low temperatures and triggers a spectral-weight redistribution from the resonance to the low-energy
incommensurate spin excitations \cite{ASuchaneck10}.

\begin{figure}[t]
\includegraphics[scale=.3]{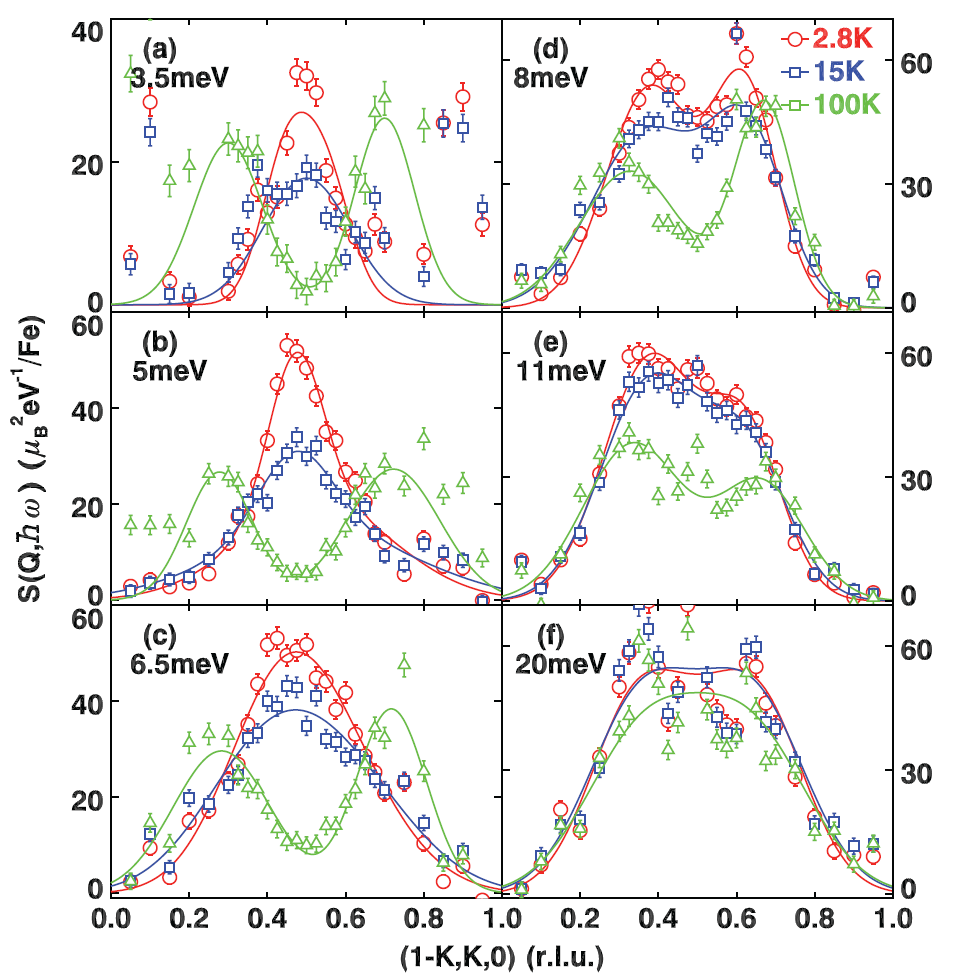}
\caption{
(Color online)
Temperature dependence of the low-energy
spin excitations in Fe$_{1+y-x}$(Ni/Cu)$_x$Te$_{0.5}$Se$_{0.5}$ family of iron chalcogenides.
Wave vector dependence of the spin excitations along the transverse direction through
the AF ordering wave vector ${\bf Q}_{\rm AF}$
 for the Ni-doped sample with $x=0.04$ at $T=2.8$ K (red circles),
15 K (blue squares), and 100 K (green triangles), obtained at (a) $E=3.5$, (b) 5, (c) 6.5,
(d) 8, (e) 11, and (f ) 20 meV [which was measured
in a higher zone, near ${\bf Q}=(1.5,0.5,0)$].
The low-energy spin excitations change from commensurate at low-temperature ($T=2, 15$ K)
to transversely incommensurate at 100 K.
Solid lines are guides to
the eye \cite{ZJXu12}.
}
\end{figure}

To see the impurity effect on the resonance and low-energy spin excitations in iron-based superconductors,
inelastic neutron scattering experiments were carried out on Ni and Cu-doped superconducting  Fe$_{1+y}$Te$_{0.5}$Se$_{0.5}$ \cite{ZJXu12}. Figure 25 shows the temperature dependence
of the spin excitations at different energies
for Fe$_{1+y-0.04}$Ni$_{0.04}$Te$_{0.5}$Se$_{0.5}$ \cite{ZJXu12}.  In addition to reducing the energy of the resonance,
the spin excitations at $E=3.5$ [Fig. 25(a)], 5 [Fig. 25(b)], and 6.5 meV [Fig. 25(c)]
change from commensurate below $T_c$ to transversely incommensurate around 100 K.  Wave vector scans at $E=8$ [Fig. 25(d)], 11 [Fig. 25(e)], and 20 meV [Fig. 25(f)] have similar lineshapes on warming from 2.8 K to 100 K. Such a dramatic spectral reconstruction
for temperatures up to $\sim 3T_c$ is not seen in copper oxide and iron pnictide superconductors, and may indicate the presence of strong electron correlations in iron chalcogenide superconductors \cite{ZJXu12}.  In subsequent transport and neutron scattering
experiments on Cu-doped Fe$_{0.98-z}$Cu$_{z}$Te$_{0.5}$Se$_{0.5}$ with $z=0,0.02$, and 0.1 \cite{JSWen13}, a metal-insulator transition was found for $z>0.02$.  In addition, low-energy spin excitations of the system are enhanced with increasing Cu-doping.  These results suggest that the localization of the conducting states and electron correlations
induced by the Cu-doping may play an important role \cite{JSWen13}.

\begin{figure}[t]
\includegraphics[scale=.4]{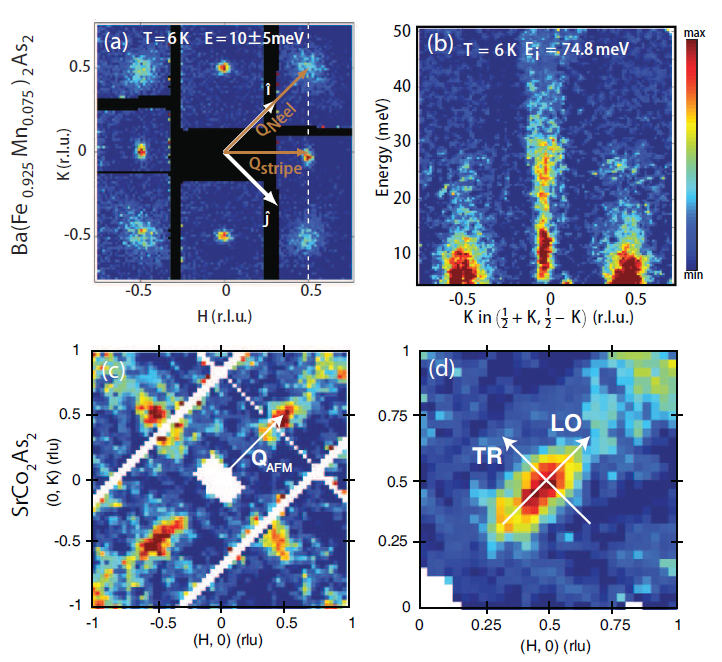}
\caption{
(Color online)
Spin excitations in Mn impurity doped BaFe$_2$As$_2$ and SrCo$_2$As$_2$ pnictides. (a) Spin excitations in
Ba(Fe$_{0.925}$Mn$_{0.075}$)$_2$As$_2$ with incident
beam ($E_i = 74.8$ meV) parallel to the crystallographic $c$ axis.
Data are displayed in the $(H+K,H-K)$ plane
and averaged over an energy transfer of $E = 5$-15 meV. In addition to the usual spin excitations from the
collinear AF ordered phase at ${\bf Q}_{\rm stripe}$, there are spin excitations at ${\bf Q}_{\rm N\acute{e}el}$.
(b) Spin excitations emanating from ${\bf Q}_{\rm N\acute{e}el}$ and
${\bf Q}_{\rm stripe}$ after averaging over
the range $H = 0.50\pm 0.05$ \cite{GSTucker12b}.
(c) Wave vector dependence of the spin excitations in SrCo$_2$As$_2$ measured with incident beam along $c$ axis
and $E_i=75$ meV at $T=5$ K.  The energy integration range is between $E = 15$-25 meV, highlighting anisotropic spin excitations centered at ${\bf Q}_{\rm AF}$.
The panel (d) plots the same data as in panel (c),
but symmetry-equivalent quadrants have been averaged together.  The wave vector anisotropy becomes even more apparent \cite{Jayasekara13}.
}
\end{figure}

While it is well known that hole-doping via K substitution for Ba in BaFe$_2$As$_2$ induces high-$T_c$
superconductivity \cite{Johrendt1}, substitution of Mn and Cr for Fe in BaFe$_2$As$_2$ never
induces superconductivity \cite{AThaler11,ASSefat09}.  In the case of Cr-doping, the system adopts
a checkerboard $G$-type AF structure for Ba(Fe$_{1−x}$Cr$_x$)$_2$As$_2$ with $x>0.3$ [Fig. 6(a)] \cite{marty11}.
How spin excitations in the parent compound BaFe$_2$As$_2$ are modified by Cr-doping is unclear.
On the other hand, Mn-doped BaFe$_2$As$_2$ represents a more complicated situation: while BaMn$_2$As$_2$
forms a simple AF structure with the ordered moment along the
$c$ axis \cite{Yogesh}, Mn-doping of BaFe$_2$As$_2$ may induce a
Griffiths regime associated with the suppression of the collinear AF order in BaFe$_2$As$_2$ by the randomly introduced localized Mn moments acting as strong magnetic impurities \cite{inosov13}.
Inelastic neutron scattering experiments were carried out on
single crystals of Ba(Fe$_{1-x}$Mn$_x$)$_2$As$_2$ with $x = 0.075$, which has a
tetragonal-to-orthorhombic lattice distortion and
orders into a collinear AF structure simultaneously below $T_s=T_N=80$ K \cite{GSTucker12b}.
Figure 26(a) shows spin excitations of the system measured with the crystallographic $c$ axis
parallel to the incident neutron beam at $E_i=74.8$ meV.  In addition to spin excitations associated with
the collinear AF structure denoted as ${\bf Q}_{\rm stripe}={\bf Q}_{\rm AF}$, there are excitations at the
AF wave vector positions of BaMn$_2$As$_2$ (${\bf Q}_{\rm N\acute{e}el}$) \cite{GSTucker12b}.
At present, it is unclear if this is an intrinsic effect of the system or there is real space
phase separation between Mn and Fe.  Figure 26(b) shows energy dependence of the scattering
at ${\bf Q}_{\rm stripe}$ and ${\bf Q}_{\rm N\acute{e}el}$.
While spin excitations at ${\bf Q}_{\rm stripe}$ extend well above 50 meV, they are limited to below $\sim$30 meV
at ${\bf Q}_{\rm N\acute{e}el}$ [Fig. 26(b)].

\begin{figure}[t]
\includegraphics[scale=.35]{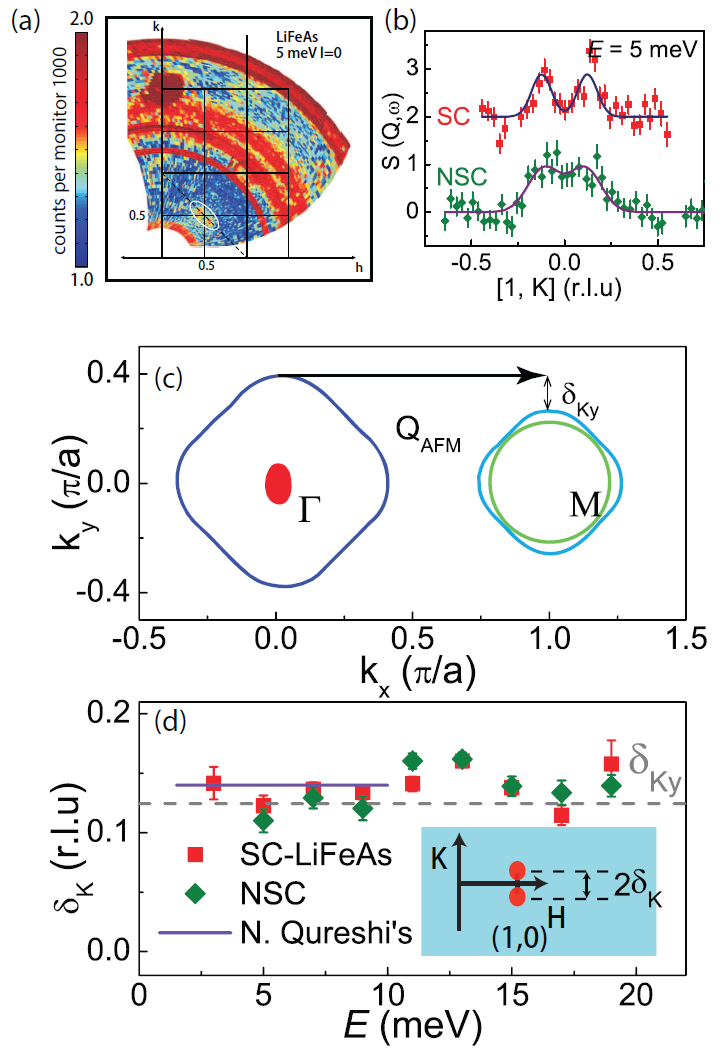}
\caption{
(Color online)
Spin excitations in superconducting and nonsuperconducting LiFeAs without static AF order.
(a) Mapping of inelastic neutron scattering intensity at $E=5$ meV in
the $(H,K)$ reciprocal space of LiFeAs in the tetragonal notation.
Note the transverse incommensurate peaks away from the ${\bf Q}_{\rm AF}=(0.5,0.5)$ position \cite{Qureshi12b}.
(b) Comparison of incommensurate spin excitations for
the superconducting (SC) and nonsuperconducting (NSC) LiFeAs at $E=5$ meV \cite{MWang2012d}.
(c) Hole and electron Fermi surfaces in LiFeAs from ARPES \cite{SVBborisenko10}.  The arrow
indicates possible nesting condition between hole and electron Fermi surfaces.
The $\delta_{Ky}$ indicates the expected transverse incommensurability.
 (d) The experimentally observed transverse incommensurate spin fluctuations and its energy dependence
for SC and NSC LiFeAs \cite{Qureshi12b,MWang2012d,Qureshi2014b}. The $\delta_K$ is the observed
transverse incommensurability.
}
\end{figure}

In the study of electron-doping evolution of the spin excitations in iron pnictides, it was found that electron-doping via
Co or Ni substitution for Fe in BaFe$_2$As$_2$
induces transversely elongated spin excitations near ${\bf Q}_{\rm stripe}$ due to the mismatched hole and electron Fermi
surfaces (Figs. 15 and 16) \cite{JHZhang10}.  If this scenario is correct for all electron-doping levels, one would expect transversely elongated spin excitations in heavily Co-doped BaFe$_2$As$_2$ or SrFe$_2$As$_2$.
Figure 26(c) and 26(d) shows
wave vector dependence of spin excitations in SrCo$_2$As$_2$ \cite{Jayasekara13}.  Although spin excitations still appear at the same wave vector positions as those of BaFe$_2$As$_2$, they are longitudinally elongated.  As SrCo$_2$As$_2$
may have complicated Fermi surfaces like that of BaCo$_2$As$_2$ \cite{nxu2013},
Fermi surface nesting could potentially explain the line-shape of the spin excitations.
It will be interesting to sort out how spin excitations evolve from transversely elongated to longitudinally elongated
in reciprocal space as a function of Co-doping for BaFe$_{2-x}$Co$_x$As$_2$ and SrFe$_{2-x}$Co$_x$As$_2$.

\begin{figure}[t]
\includegraphics[scale=.18]{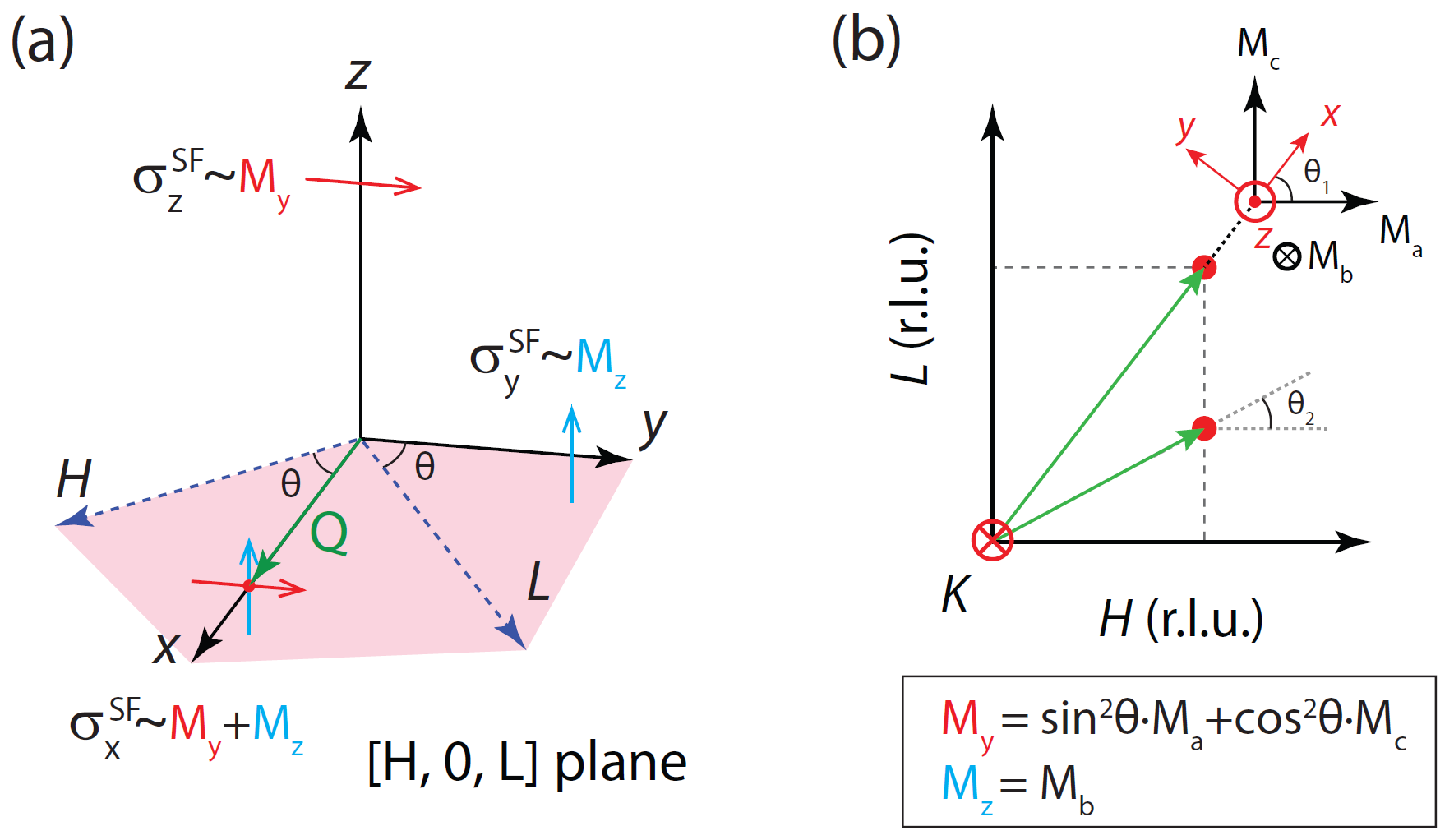}
\caption{
(Color online)
Neutron polarization analysis used to determine the magnitude of spin excitations along the crystallographic $a$ ($M_a$),
$b$ ($M_b$), and $c$ ($M_c$) axis directions.
(a) Incident beam neutrons are polarized along the $x$, $y$, and $z$ directions, corresponding to directions along the momentum transfer ${\bf Q}$, perpendicular to ${\bf Q}$ but in the horizontal scattering plane,
and perpendicular to ${\bf Q}$ and the horizontal scattering plane, respectively. In this geometry,
neutron spin-flip scattering $\sigma_{x}^{\rm SF}\sim M_y+M_z$, where $M_y$ and $M_z$ are magnitude of
spin excitations along the $y$ and $z$ directions, respectively. Similarly,
$\sigma_{y}^{\rm SF}\sim M_z$ and $\sigma_{z}^{\rm SF}\sim M_y$. If the
angle between $x$ direction and the $H$ axis is $\theta$,
we have $M_y=\sin^2\theta M_a+\cos^2\theta M_c$ and $M_z=M_b$. (b) Since $M_a$, $M_b$, and $M_c$ should be the same
at equivalent wave vectors in reciprocal space except for the magnetic form factor, we can conclusively determine $M_a$, $M_b$, and $M_c$ by measuring $\sigma_{\alpha}^{\rm SF}$ at two or more equivalent reciprocal lattice vectors \cite{HQLuo2013}.
}
\end{figure}

Another iron pnictide worthy of mention is LiFeAs \cite{XCWang,PChu1,MJPitcher2008}.  Although this material has the same
crystal structure as that of NaFeAs [Fig. 1(d)], it is a superconductor without static AF order in stoichiometric LiFeAs and
Li deficiency tends to suppress superconductivity.
Initial inelastic neutron scattering measurements on powder samples indicate the presence of superconductivity-induced
resonance near the usual AF ordering wave vector ${\bf Q}_{\rm AF}$ \cite{AETaylor2011}.
Subsequent neutron scattering experiments on single crystals showed that spin excitations in this system
are transversely incommensurate away from the ${\bf Q}_{\rm AF}$ for both the superconducting LiFeAs [Fig. 27(a)] \cite{Qureshi12b}
and nonsuperconducting Li$_{1-x}$FeAs [Fig. 27(b)] \cite{MWang2012d}.
 The absence of the static AF order has been interpreted as
due to poor Fermi surface nesting between $\Gamma$ and $M$ consistent with
ARPES measurements [Fig. 27(c)] \cite{Brydon2011,SVBborisenko10}.  However, the
quasiparticle scattering between the hole pockets near $\Gamma$ and
electron pocket $M$ should give rise to transverse incommensurate spin fluctuations and this is indeed
the case [Fig. 27(a)] \cite{Qureshi12b,MWang2012d,Qureshi2014b}.
Furthermore, the incommensurate spin excitations are weakly energy dependent and only broaden slightly
for nonsuperconducting Li$_{1-x}$FeAs [Fig. 27(d)] \cite{MWang2012d}.
These results suggest that spin excitations in LiFeAs have the same origin as other iron pnictides, and the low-energy spin
excitations in the system follow the nested Fermi surface scenario.

\subsection{Neutron polarization analysis of spin excitation anisotropy in iron pnictides}

Although most neutron scattering experiments are carried out with unpolarized neutrons, neutron polarization analysis can provide some unique information concerning the nature of the ordered moment and the anisotropy of spin excitations.  The neutron polarization analysis was first developed at Oak Ridge National Laboratory in the pioneering work of Moon, Riste, and Koehler \cite{Moon69}.
This technique was used to unambiguously identify the magnetic nature of the neutron spin resonance in
optimally doped high-$T_c$ copper oxide YBa$_2$Cu$_3$O$_7$ \cite{HAMook93}.
In modern polarized neutron scattering experiments on high-$T_c$ superconductors at the Institut Laue-Langevin,
the Cryopad capability \cite{Berna2005} is typically used
to ensure that the sample is in a strictly zero magnetic field environment, thus avoiding errors due to flux inclusion and field expulsion in the superconducting phase of the sample \cite{lester10}.  The Cryopad device can also be
used for neutron polarimetry in which an arbitrary incident and scattered neutron beam polarization can
be measured \cite{Berna2005}.

Polarized neutrons were produced using a focusing Heusler monochromator and analyzed using a focusing Heusler analyzer. Polarization analysis can be used to separate magnetic (e.g. spin excitations) and nuclear (e.g. phonon) scattering because the former has a tendency to flip the spin of the neutron, whereas the latter leaves the neutron spin unchanged. More specifically, the spin of the neutron is always flipped in a magnetic interaction
where the neutron polarization is
parallel to the wave vector transfer $\bf{Q}$ and
the magnetic moment or excitation polarization in the sample is transverse to ${\bf Q}$. We therefore describe the neutron polarization in a coordinate system where $x$ is parallel to $\bf{Q}$. For convenience, we then define the other orthogonal directions
with $y$ in the scattering plane, and $z$ out of plane [see Fig. 28(a)].  There are then six independent channels in which the instrument can be configured at a specific wave vector and energy point: three neutron polarization directions $x$, $y$, $z$, each
of which can be measured to detect neutrons that flip or do not flip their spins when scattering at the sample. The measured neutron cross-sections are labeled by the experimental configuration in which they were measured, and are written $\sigma_{\alpha}^{\rm SF}$, $\sigma_{\alpha}^{\rm NSF}$, where $\alpha$ is the neutron polarization direction ($x$, $y$ or $z$) and the superscript represents either spin-flip (SF) or non-spin flip (NSF) scattering.

\begin{figure}[t]
\includegraphics[scale=.42]{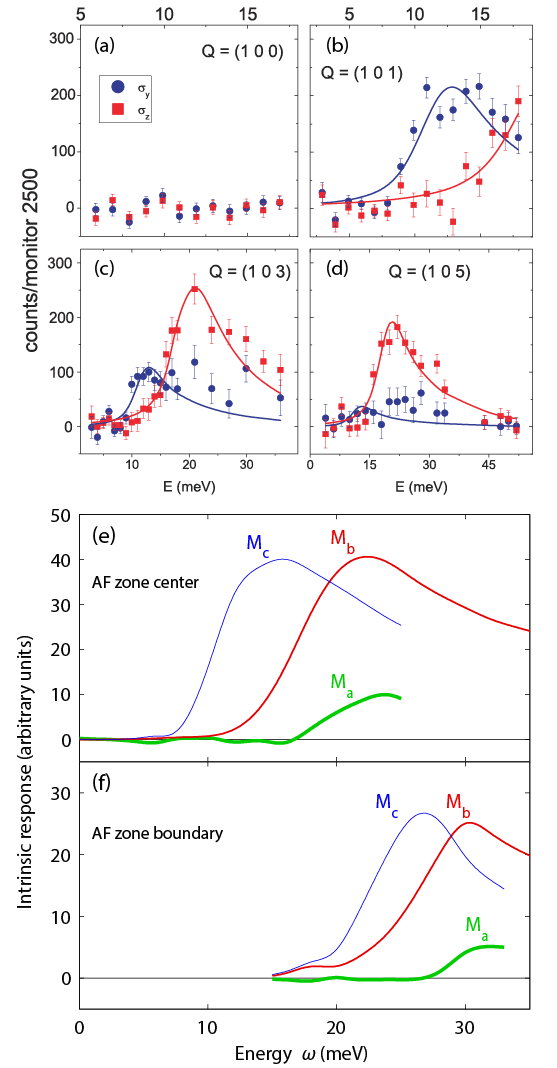}
\caption{
(Color online) Polarized neutron scattering study of spin waves in BaFe$_2$As$_2$.
The AF ordering Brillouin-zone centers are at ${\bf Q}_{\rm AF}=(1,0,L)$ with $L=1,3,\cdots$.
The magnetic zone boundaries are at ${\bf Q}=(1,0,L)$ with $L=0,2,\cdots$.
(a) Inelastic constant-${\bf Q}$
scans at the AF zone boundary ${\bf Q}=(1,0,0)$.  Here the blue dots are
 the magnetic scattering along the $y$ direction, or
$\sigma_y\sim \sigma_{x}^{\rm SF}-\sigma_{y}^{\rm SF} \sim M_y$, while
red squares depict $\sigma_z\sim \sigma_{x}^{\rm SF}-\sigma_{z}^{\rm SF} \sim M_z$.
(b,c,d) Similar scans at ${\bf Q}_{\rm AF}=(1,0,L)$ with $L=1,3,$ and 5, respectively. The solid lines
are spline-interpolated spin-wave theory calculations
folded with the experimental resolution \cite{Qureshi12}.  Using identical experimental
setup as that of \cite{Qureshi12} but with much more sample mass, the energy dependence of $M_a$, $M_b$, and
$M_c$ is determined at the magnetic Brillouin (e) zone center and (f) zone boundary.  The presence of longitudinal
spin excitations, or nonzero $M_a$, is seen above 20 meV at ${\bf Q}_{\rm AF}=(1,0,1)$ and above 30 meV
at ${\bf Q}=(1,0,0)$ \cite{CWang2013}.
}
\end{figure}

Magnetic neutron scattering only probes the magnetic moment perpendicular to $\mathbf{Q}$. The cross-sections can therefore be written in terms of $M_y$ and $M_z$ [see Fig. 28(a)], the two spatial components (perpendicular to $\mathbf{Q}$) of the spin direction of the magnetic excitations, and the nuclear scattering strength $N$. However, a measured cross-section component on an imperfect instrument contains a leakage between SF and NSF channels due to imperfect neutron polarization. This leakage can be quantified by measuring the nuclear Bragg peak contamination into the spin flip channel, the ``instrumental flipping ratio'' $R={\rm NSF_N/SF_N}$ (for an unpolarized neutron scattering experiment $R$=1, and $R\to\infty$ in an ideal polarized neutron scattering experiment).
The measured cross-section components can then be written \cite{Moon69,OJLipscombe2010}
\begin{equation}
\begin{split}
\left( \begin{array}{c}
\sigma_x^{\rm SF}\\
\sigma_y^{\rm SF}\\
\sigma_z^{\rm SF}\\
\sigma_x^{\rm NSF}\\
\sigma_y^{\rm NSF}\\
\sigma_z^{\rm NSF}
\end{array} \right)&=
\frac{1}{(R+1)}  \\
&\left( \begin{array}{ccccc}
R & R & 1 & 2R/3+1/3 & (R+1)\\
1 & R & 1 & 2R/3+1/3 & (R+1)\\
R & 1 & 1 & 2R/3+1/3 & (R+1)\\
1 & 1 & R & R/3+2/3 & (R+1)\\
R & 1 & R & R/3+2/3 & (R+1)\\
1 & R & R & R/3+2/3 & (R+1)
\end{array} \right)
\left( \begin{array}{c}
M_y\\
M_z\\
N\\
\rm NSI\\
\rm B
\end{array} \right),
\end{split}
\end{equation}
where B is a background term to take account of instrumental background
and the non-magnetic nuclear incoherent scattering from the sample, which is assumed to be equal in all six cross-sections when measured at the same wave vector and energy. NSI is the nuclear spin incoherent scattering caused by moments within the nuclei of the isotopes in the sample. NSI is independent of $\mathbf{Q}$ and typically is negligible in magnitude compared with the nuclear coherent cross-section $N$.
Furthermore, in the case where only SF (or only NSF) cross-section components are collected, NSI would be absorbed into the B term.

\begin{figure}[t]
\includegraphics[scale=.3]{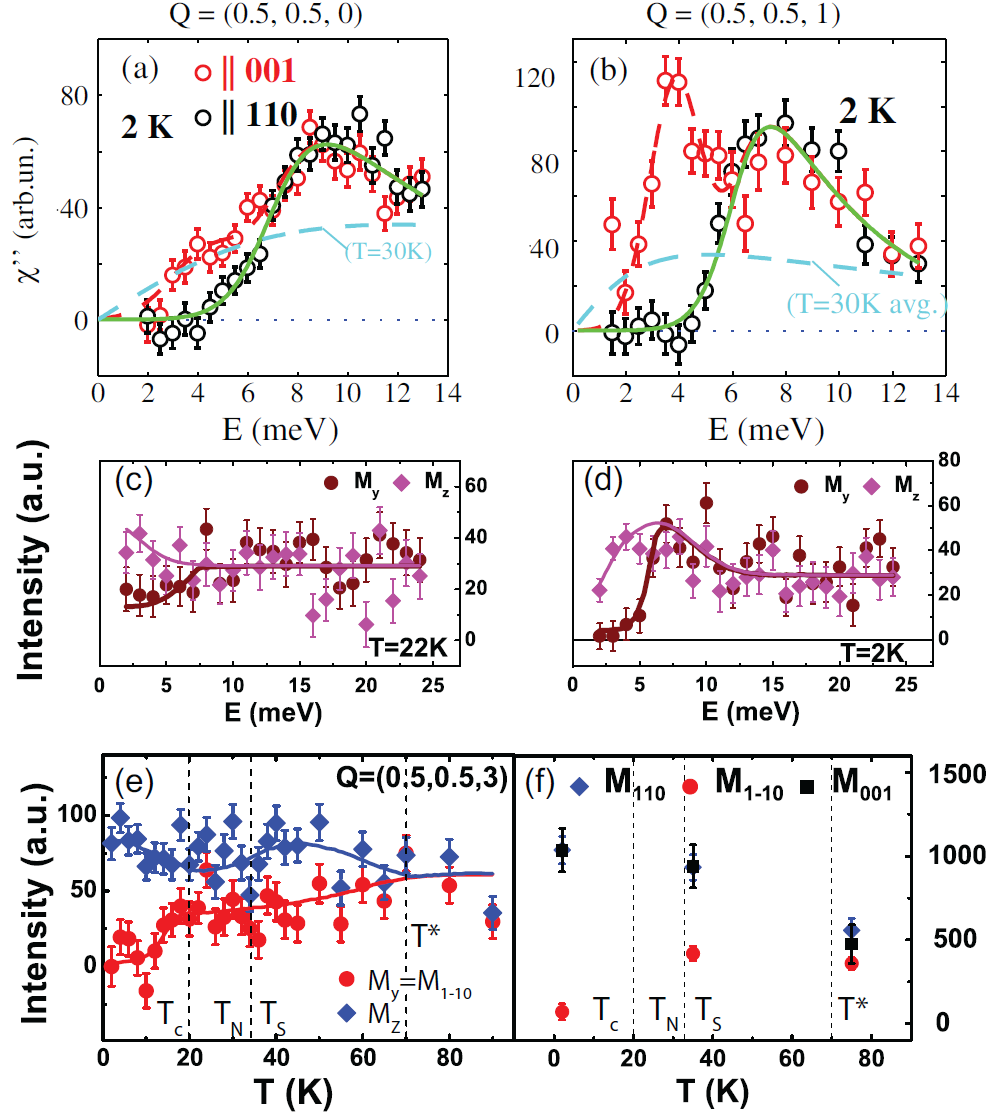}
\caption{
(Color online)
Neutron polarization analysis of the spin excitation anisotropy in
electron-doped iron pnictide superconductors.
(a) Energy dependence of the imaginary
part of the out-of-plane ($M_c$) and in-plane ($M_b$) generalized magnetic susceptibility
at ${\bf Q}=(0.5,0.5,0)$ in tetragonal
notation for Ba(Fe$_{0.94}$Co$_{0.06}$)$_2$As$_2$ at $T=2$ K.  The dashed line shows
isotropic paramagnetic scattering at $T=30$ K. (b)  Similar data at
${\bf Q}_{\rm AF}=(0.5,0.5,1)$, where the mostly $c$ axis polarized susceptibility
exhibits a peak in the superconducting phase \cite{Psteffens2013}. (c)
Energy dependence of $M_y$ and $M_z$ in the normal state
at ${\bf Q}_{\rm AF}=(0.5,0.5,3)$ for BaFe$_{1.904}$Ni$_{0.096}$As$_2$ ($T_c=19.8$ K, $T_N\approx T_s=33\pm2$ K).
(d) Similar data at $T=2$ K.  The data show
clear low-energy spin excitation anisotropy in both the normal
and superconducting states.
(e) Temperature dependence of $M_y$ and $M_z$, where $T^\ast$ marks
the temperature of the in-plane resistivity anisotropy. (f) Temperature
dependence of $M_a=M_{110}$, $M_b=M_{1-10}$,
and $M_c=M_{001}$. Vertical dashed lines mark temperatures for $T^\ast$,
$T_s$, $T_N$, and $T_c$ \cite{HQLuo2013}.
}
\end{figure}

For SF neutron scattering measurements at $\mathbf{Q}$, one can conclusively
determine the magnetic components $M_y$ and $M_z$.
If the magnetic components of the system along the $x$, $y$, and $z$
are $M_x$, $M_y$, and $M_z$, respectively, we would have
$M_z=M_b$ and $M_y=M_a\sin^2\theta+M_c\cos^2\theta$ if the sample is aligned in the
$[H,0,L]$ scattering plane, where $M_a$, $M_b$, and $M_c$ are magnitudes of spin excitations along
the orthorhombic $a$, $b$, and $c$-axis directions of the lattice, respectively, and
$\theta$ is the angle between $M_a$ and $x$-axis [Fig. 28(b)] \cite{OJLipscombe2010}.
Since there are three unknowns ($M_a,M_b,M_c$) and only two equations with known $M_y$ and $M_z$,
one can only determine the values of $M_a$, $M_b$, and $M_c$ by measuring at least two equivalent reciprocal lattice vectors with different $\theta$ angle as illustrated in Fig. 28(b).  In the initial polarized neutron scattering experiments on optimally electron-doped superconducting
BaFe$_{1.9}$Ni$_{0.1}$As$_2$, spin excitation anisotropy near the resonance energy was observed \cite{OJLipscombe2010}.
Similar results were also found for the resonance in optimal superconducting Fe(Se,Te) \cite{Babkevich11}. For electron-overdoped
BaFe$_{1.85}$Ni$_{0.15}$As$_2$, the resonance and spin excitations at all energies probed are
isotropic with $M_y=M_z$ \cite{MSLiu2012}.

Figure 29 summarizes the outcome from polarized neutron scattering experiments on BaFe$_2$As$_2$.
From unpolarized neutron scattering
experiments, it is well known that spin waves in the AF ordered state
are gapped below about $\sim$15 meV at the magnetic Brillouin zone center ${\bf Q}_{\rm AF}$ \cite{KMatan09,JTPark12}.
Figure 29(a)-29(d) shows neutron SF inelastic constant-${\bf Q}$ scans at the zone center
$(1,0,L)$ for different $L$ values and at $(1,0,0)$. The blue dots represent
the magnitude of magnetic scattering along the $y$ axis direction, or $M_y$,
while red squares depict $M_z$, where
$\sigma_{y,z}=(R+1)[\sigma_x^{\rm SF}-\sigma_{y,z}^{\rm SF}]/(R-1)\approx M_{y,z}$ \cite{Qureshi12}.
Consistent with unpolarized measurements \cite{KMatan09,JTPark12}, there are large spin gaps at $(1,0,0)$ and $(1,0,1)$.
However, the gap value for $\sigma_z^{\rm SF}$ is significantly larger than
that for $\sigma_y^{\rm SF}$ [Figs. 29(b)-29(d)].  These results indicate strong single-iron anisotropy within the layer,
suggesting that it costs more energy to rotate a
spin within the orthorhombic $a$-$b$ plane than to rotate it perpendicular to the FeAs layers \cite{Qureshi12}.
In addition, there is no evidence for longitudinal spin fluctuations typically associated with itinerant electrons and nested
Fermi surfaces like in the spin-density-wave state of pure chromium metal \cite{EFawcett,Qureshi12}.

\begin{figure}[t]
\includegraphics[scale=.4]{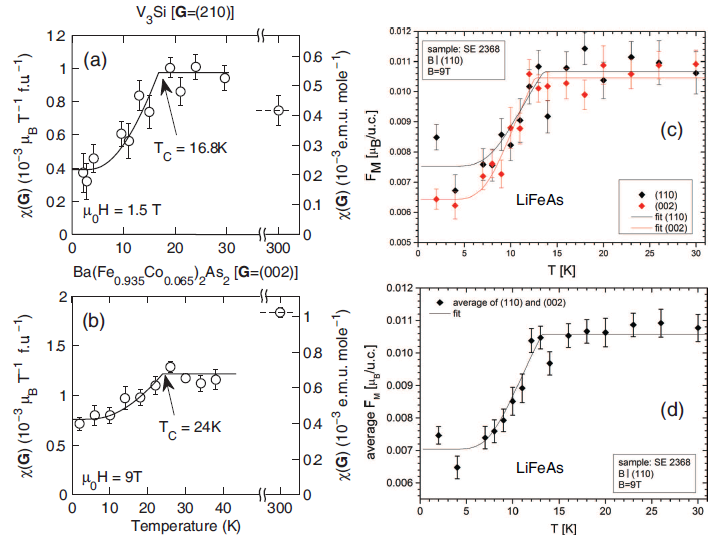}
\caption{
(Color online)
Polarized neutron diffraction studies of the induced magnetization density for different iron pnictide superconductors.
The temperature dependence of the susceptibility and
induced moment for (a) the conventional BCS superconductor V$_3$Si,
and (b) iron pnictide superconductor
Ba(Fe$_{0.935}$Co$_{0.065}$)$_2$As$_2$. The solid lines are the Yosida behavior expected for a singlet order
parameter \cite{CLester2011}. (c)
The temperature dependence of the field-induced magnetization for superconducting LiFeAs obtained using
the $(1,1,0)$ and $(0,0,2)$ nuclear Bragg peaks under a 9-T magnetic field. (d) The average of the $(1,1,0)$ and $(0,0,2)$ shows a clear drop below $T_c$, suggesting spin singlet pairing for LiFeAs
\cite{JBrand2014}.
}
\end{figure}

In subsequent polarized neutron scattering experiments on BaFe$_2$As$_2$ with greater sample mass \cite{CWang2013}, three
distinct spin-excitation components, $M_a$, $M_b$, and $M_c$, with magnetic moments
fluctuating along the three crystallographic axes are identified at the AF Brillouin zone center [Fig. 29(e)]
and zone boundary [Fig. 29(f)].  The data reveals the presence of finite $M_a$ at the AF
zone center for energies above $\sim$20 meV [Fig. 29(e)].  Similar measurements at the AF zone boundary suggest nonzero values
of $M_a$ above $\sim$30 meV [Fig. 28(f)].  While $M_b$ and $M_c$, the two transverse components of spin waves,
can be described by a linear spin-wave theory with
magnetic anisotropy and inter-layer coupling,
the presence of $M_a$, the longitudinal component of spin waves, is generically incompatible
with transverse spin waves at low-${\bf Q}$ from a
local-moment Heisenberg Hamiltonian.
These results suggest a contribution of itinerant electrons to the
magnetism that is already in the parent compound of this family of Fe-based superconductors.
This means that one cannot account for spin waves in the parents of iron pnictides with a purely local moment picture, and must take the contribution from itinerant electrons into account to understand the magnetism in these materials \cite{CWang2013}.

In a polarized inelastic neutron scattering experiment on optimally
electron-doped BaFe$_{1.88}$Co$_{0.12}$ ($T_c=24$ K), two resonance-like excitations were found in the superconducting state \cite{Psteffens2013}.  While the high-energy mode occurring at $E=8$ meV is an isotropic resonance with weak dispersion along the $c$ axis, there is a 4 meV spin excitation that appears only in the $c$ axis polarized
channel and whose intensity modulates along the $c$ axis similar to spin waves in the
undoped BaFe$_2$As$_2$ [Fig. 30(a) and 30(b)].  These results suggest that spin excitations in undoped and
optimally electron doped
BaFe$_2$As$_2$ have similar features, different from what one
might expect for superconducting and AF phases of iron pnictides \cite{Psteffens2013}.

In a separate polarized inelastic neutron scattering experiment on electron underdoped BaFe$_{1.904}$Ni$_{0.096}$As$_2$, where the
system exhibits an AF order and tetragonal-to-orthorhombic lattice distortion
temperatures near $T_N\approx T_s=33\pm2$ K, and superconductivity below $T_c=19.8$ K, neutron SF cross sections have been measured at various energies and wave vectors. Figure 30(c) and 30(d) shows energy dependence of $M_y$ and $M_z$ in the normal and superconducting states at the
AF wave vector \cite{HQLuo2013}.  In addition to confirming that the low-energy spin excitations are
highly anisotropic below $\sim$5 meV in the superconducting state [Fig. 30(d)], the magnetic scattering appears to be anisotropic in the normal state with $M_z>M_y$
[Fig. 30(c)].  Figure 30(e) shows that the magnitudes of $M_y$ and $M_z$
become different below $T^\ast$, illustrating that the magnetic
anisotropy first appears below the temperature where transport measurements on uniaxial strain
detwinned samples display in-plane resistivity anisotropy \cite{JHCHu2010,MATanatar2010,IRFisher2011}.  To quantitatively determine if the spin excitation anisotropy is indeed within the $a$-$b$ plane, neutron SF cross sections were measured at multiple equivalent wave vectors.  The outcome suggests that the presence of in-plane spin excitation anisotropy is associated with resistivity anisotropy in strain-induced sample [Fig. 30(f)].  Therefore, spin excitation anisotropy in iron pnictides is a direct probe of the spin-orbit coupling in these materials \cite{HQLuo2013}.
Recent polarized inelastic neutron scattering experiments on superconducting Ba$_{0.67}$K$_{0.33}$Fe$_2$As$_2$ \cite{CLZhang2013b} and
Ba$_{0.5}$K$_{0.5}$Fe$_2$As$_2$ \cite{NQureshi2014} reveal that the low-energy spin excitation anisotropy
persists to hole overdoped iron pnictides far away from the AF ordered phase. Similar polarized neutron scattering experiments on
underdoped NaFe$_{0.985}$Co$_{0.015}$As with double resonances [Fig. 17(c)] suggest that
the first resonance is highly anisotropic
and polarized along the $a$ and $c$ axes, while the second mode is isotropic similar to that
of electron overdoped NaFe$_{0.935}$Co$_{0.045}$As. Since the
$a$ axis polarized spin excitations of the first resonance appear below $T_c$, the itinerant
electrons contributing to the magnetism may also be coupled to the superconductivity \cite{CLZhang2014d}.

Polarized neutron scattering is not only useful for determining the spin excitation anisotropy, it can also be used to
to measure the susceptibility and induced magnetization in the normal and superconducting states
of a superconductor.  The technique of using polarized neutron diffraction to study the magnetization
of the paramagnetic crystal by an externally
applied magnetic field was developed by Shull and Wedgewood in their study of electron spin pairing
of a BCS superconductor V$_3$Si \cite{shull66}. Instead of the full neutron polarization analysis as described above,
the magnetization measurements are performed under a magnetic field
using a polarized incident beam of neutrons \cite{PJBrown10,CLester2011}. The flipping ratio $R$, defined as the
ratio of the neutron scattering cross sections with neutrons parallel and antiparallel
to the applied magnetic field, is associated with the nuclear structure factors $F_N({\bf G})$ and the Fourier transform of the
real-space magnetization density ${\bf M}({\bf G})$ via
$R\approx 1-2\gamma r_0 M({\bf G})/[\mu_B F_N({\bf G})]$, where
$\gamma r_0 = 5.36 \times 10^{-15}$ m and
$\bf G$ is the reciprocal lattice vector \cite{CLester2011}. In a conventional BCS superconductor such as V$_3$Si,
where electrons below $T_c$ form
singlet Cooper pairs, the temperature dependence of the field induced magnetization shows the characteristic Yosida drop below $T_c$ expected
for singlet pairing [Fig. 31(a)] \cite{shull66}.  For a spin-triplet superconductor such as Sr$_2$RuO$_4$ \cite{APMackenzie03}, there are no change in the field induced magnetization across $T_c$ \cite{JADuffy2000}.  The temperature dependence of the field-induced magnetization shows a clear drop below $T_c$ in nearly optimally electron-doped BaFe$_{1.87}$Co$_{0.13}$As$_2$ [Fig. 31(b)] \cite{CLester2011}, consistent
with measurements of the NMR Knight shift in the same compound \cite{FNing2008,SOh2011}.  The large residual contribution
of the field induced magnetization below $T_c$ seen in both V$_3$Si \cite{shull66} and BaFe$_{1.87}$Co$_{0.13}$As$_2$ \cite{CLester2011}
has been attributed to the van Vleck or orbital contribution to the susceptibility.

\begin{figure}[t]
\includegraphics[scale=.25]{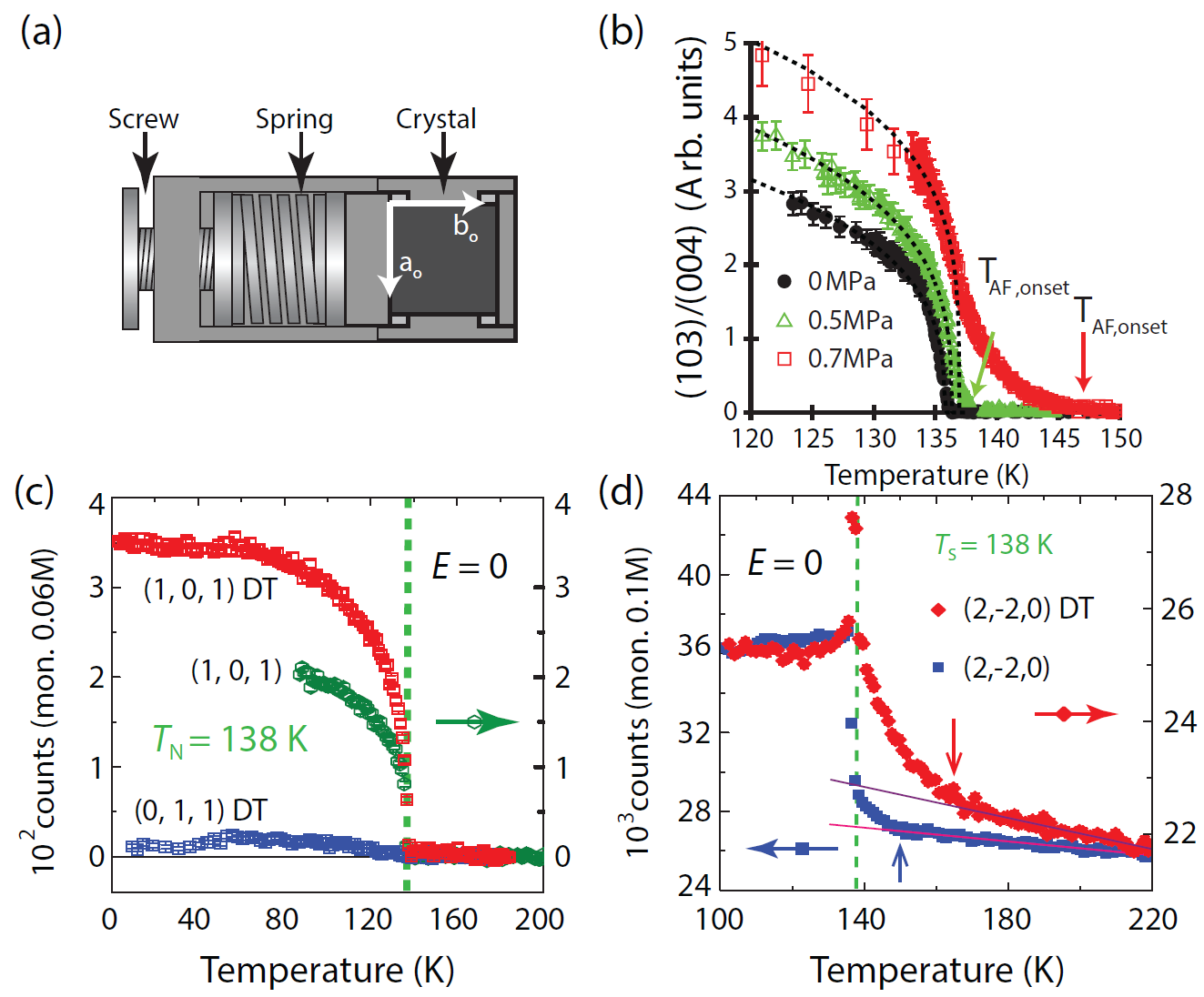}
\caption{
(Color online)
The effect of uniaxial strain on structural and magnetic phase transitions in the as-grown and annealed BaFe$_2$As$_2$.
(a) Schematic drawing of the device used to apply uniaxial strain to detwin single crystals of BaFe$_2$As$_2$.
The sample is cut into a square shape with $a_o$/$b_o$ parallel to the applied pressure direction.
(b) The temperature dependence of the magnetic order parameters under different
applied unaixial strain for the as-grown BaFe$_2$As$_2$.  The onset of the AF ordering temperature increases with increasing
uniaxial pressure \cite{CDhital2012}.  (c)
Magnetic Bragg peak intensity at the $(1,0,1)$ and $(0,1,1)$ positions for the annealed
BaFe$_2$As$_2$ at zero pressure (green) and $P\sim 15$ MPa uniaxial pressure along the $b_o$ axis.
No shift in $T_N$ is seen under uniaxial pressure. (d) The blue squares show the temperature dependence of the $(2,-2,0)$
nuclear Bragg peak at zero pressure. The sharp step at $T_s$ is caused by releasing of the
neutron extinction due to tetragonal to orthorhombic lattice distortion. The identical
scan under $P\sim 15$ MPa uniaxial pressure is shown as
red diamonds \cite{XYLu2014S}.
}
\end{figure}

Similar polarized neutron diffraction experiments have also been carried out on superconducting LiFeAs \cite{JBrand2014},
which does not have a static AF ordered parent compound \cite{PChu1,XCWang,MJPitcher2008} and may have triplet electron pairing due to a large density of states near the Fermi level favoring a ferromagnetic instability \cite{Brydon2011}.  Figure 31(c) shows temperature dependence of the field-induced magnetization at wave vectors $(1,1,0)$ and $(0,0,2)$.  The average of the $(1,1,0)$ and $(0,0,2)$ is shown in Fig. 31(d).
Different from the spin triplet superconductor Sr$_2$RuO$_4$ \cite{JADuffy2000}, the field-induced magnetization clearly decreases at the onset of $T_c$, consistent with the spin singlet electron pairing \cite{JBrand2014}.  Therefore, the mechanism of superconductivity in
LiFeAs is likely the same as all other iron-based superconductors.

\subsection{Electronic nematic phase and neutron scattering experiments under uniaxial strain}

As mentioned in Section F, transport measurements on uniaxial strain detwinned
electron-doped BaFe$_2$As$_2$
 reveal clear evidence for the in-plane resistivity anisotropy first occurring at a temperature above the zero pressure $T_N$ and $T_s$
\cite{JHCHu2010,MATanatar2010,IRFisher2011}. As a function of increasing electron-doping,
the resistivity anisotropy first increases and then vanishes near optimal superconductivity \cite{IRFisher2011}, consistent with
a signature of the spin nematic phase that breaks the in-plane
fourfold rotational symmetry ($C_4$) of the underlying
tetragonal lattice \cite{RMFernandes2014,cfang,JDai2009}.
NMR experiments on 1111 family of materials also indicate the presence of a nematic phase below $T_s$ \cite{MFu2012}.
However, recent scanning tunneling microscopy \cite{MPAllan2013}
and transport measurements \cite{SIshida2013} suggest that the resistivity anisotropy in Co-doped BaFe$_2$As$_2$
arises from Co-impurity scattering and is not an intrinsic property of these materials. On the other hand, ARPES
measurements on Co-doped BaFe$_2$As$_2$ \cite{MYi2011} and NaFeAs \cite{YZhang2012} reveal a splitting in energy between two orthogonal bands
with dominant $d_{xz}$ and $d_{yz}$ character at the temperature of resistivity anisotropy in uniaxial strain detwinned samples, thereby suggesting that
orbital ordering is also important for the electronic properties of iron pnictides
\cite{FKruger2009,WKu2009,WCLv2009}. Finally, since transport measurements were carried out on uniaxial strain detwinned samples \cite{IRFisher2011}, it is unclear if the uniaxial strain can modify the structural
and magnetic phase transitions in these materials.

\begin{figure}[t]
\includegraphics[scale=.25]{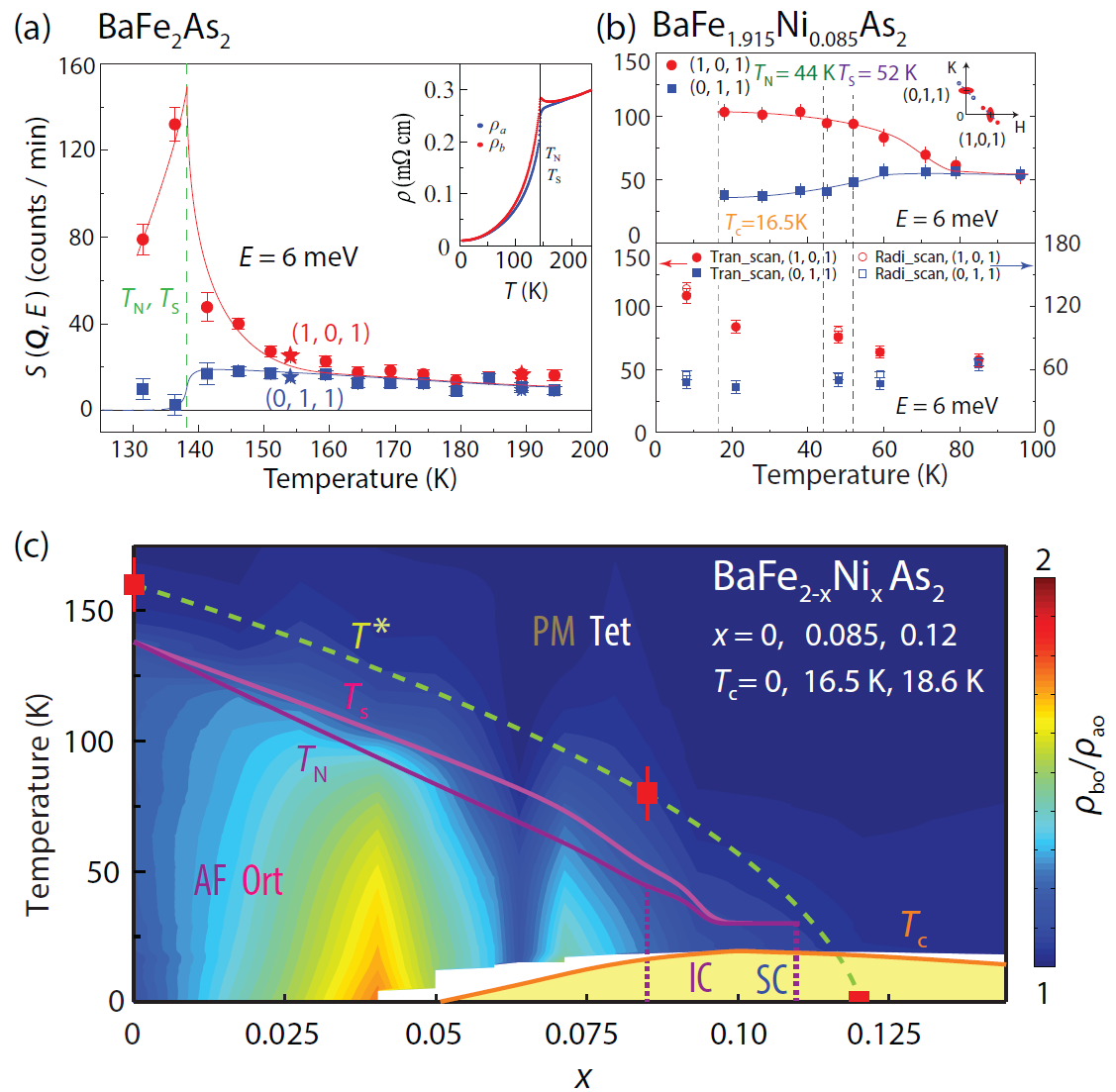}
\caption{
(Color online)
Temperature dependence of the spin excitation anisotropy at wave vectors
$(1,0,1)$ and $(0,1,1)$ and its comparison with transport measurements for BaFe$_{2-x}$Ni$_x$As$_2$.
(a) Temperature
dependence of spin excitations at $E=6$ meV for $(1, 0, 1)$ and $(0, 1, 1)$ under $P\sim 15$ MPa uniaxial pressure.
The anisotropy in spin excitations vanishes around $T = 160\pm 10$ K. The inset shows
the transport measurement of the in-plane resistivity anisotropy for the annealed BaFe$_2$As$_2$.
(b) Temperature dependence of $E=6$ meV spin excitations at $(1, 0, 1)$ and $(0, 1, 1)$ for BaFe$_{1.915}$Ni$_{0.085}$As$_2$.
The data in the top panel were obtained by subtracting the background
intensity from the peak intensity at every temperature; the data in the bottom panel were
obtained by fitting the wave vector scans.
(c) The electronic phase diagram of BaFe$_{2-x}$Ni$_x$As$_2$ from resistivity
anisotropy ratio $\rho_{bo}/\rho_{ao}$ obtained under uniaxial pressure.
The spin excitation anisotropy temperatures
are marked as $T^\ast$. The AF orthorhombic (Ort), incommensurate AF (IC), paramagnetic
tetragonal (PM Tet), and superconductivity (SC) phases are marked \cite{XYLu2014S}.
}
\end{figure}

The first neutron scattering experiment carried out under uniaxial strain was
on as-grown BaFe$_2$As$_2$ \cite{CDhital2012}.  The data show that
modest strain fields along the in-plane orthorhombic $b_o$ axis as shown in Fig. 32(a) can
induce significant changes in the structural and magnetic phase
behavior simultaneous with the removal of structural twinning effects.   Both
the structural lattice distortion and long-range spin ordering occur at temperatures far exceeding the strain free phase transition temperatures [Fig. 32(b)], thus suggesting that the resistivity anisotropy in transport measurements
is a consequence of the shift in $T_N$ and $T_s$ under uniaxial strain \cite{CDhital2012}.
In subsequent neutron scattering study of the effect of uniaxial pressure on $T_N$ and $T_s$ in NaFeAs,
as-grown and annealed BaFe$_2$As$_2$ \cite{YSong13b},
it was found that while the uniaxial strain necessary to detwin the sample indeed
induces a significant increase in $T_N$ and $T_s$ for as-grown BaFe$_2$As$_2$,
similar uniaxial pressure used to detwin NaFeAs and annealed BaFe$_2$As$_2$ has a very small
effect on their $T_N$ and $T_s$. These results would suggest that resistivity anisotropy observed in
transport measurements \cite{IRFisher2011} is an intrinsic property of these materials \cite{YSong13b}.

In a recent systematic study of magnetic and
structural transitions of the as-grown parent and lightly Co-doped Ba(Fe$_{1-x}$Co$_x$)$_2$As$_2$ under uniaxial pressure \cite{CDhital2014},
it was found that the uniaxial strain induces
a thermal shift in the onset
of AF order that grows as a percentage of $T_N$ as Co-doping is increased and the
superconducting phase is approached.  In addition, the authors find a
decoupling between the onsets of the $T_s$ and $T_N$ under uniaxial strain for parent
and lightly-doped Ba(Fe$_{1-x}$Co$_x$)$_2$As$_2$ on the first order side of the tri-critical point \cite{CDhital2014}.

\begin{figure}[t]
\includegraphics[scale=.4]{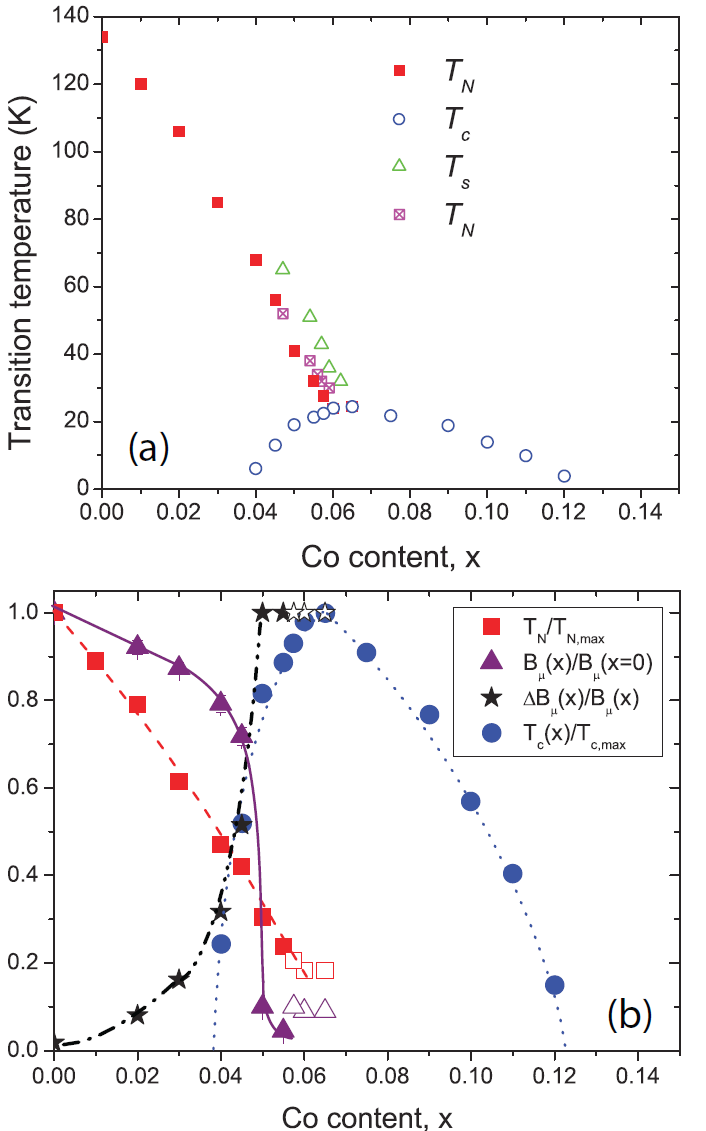}
\caption{
(Color online) The electronic phase diagram of Ba(Fe$_{1-x}$Co$_x$)$_2$As$_2$ as determined from combined
$\mu$SR and neutron diffraction experiments.
(a) Phase diagram of $T_N$, as determined with $\mu$SR and $T_c$, obtained from resistivity and magnetic
susceptibility measurements as well as from the specific heat data. (b) The Co dependence
of the normalized $T_N$ and $T_c$, and the normalized values of the average
magnetic field at the muon site, $B_\mu$ and of its relative spread, $\Delta B_\mu$.
The open symbols show the magnetic properties in the spatially
inhomogeneous magnetic state near optimum doping \cite{Bernhard2012}.
}
\end{figure}

At around the same time, elastic and inelastic neutron scattering experiments were carried out on the annealed BaFe$_2$As$_2$ ($T_N=138$ K),
BaFe$_{1.915}$Ni$_{0.085}$As$_2$ ($T_c=16.5$ K, $T_N=44$ K), and
BaFe$_{1.88}$Ni$_{0.12}$As$_2$ ($T_c=18.6$ K, tetragonal
structure without static AF order) to study the temperature dependence of the spin excitation anisotropy
at wave vectors $(1,0)$ and $(0,1)$ \cite{XYLu2014S}. By comparing the temperature dependence of the magnetic order parameters
at wave vectors $(1,0,1)$ and $(0,1,1)$ in zero and $\sim$15 MPa uniaxial pressure on annealed BaFe$_2$As$_2$ [Fig. 32(c)],
it was concluded
that the applied uniaxial strain is sufficient to completely detwin the sample, and does not affect
$T_N$.  Furthermore, the temperature dependence of the intensity at the $(2,-2,0)$ nuclear
Bragg reflection for the twinned and detwinned samples both show a dramatic jump at
$T_s = 138$ K arising from the neutron extinction release that occurs due to strain and domain
formation related to the orthorhombic lattice distortion, indicating that the uniaxial pressure
does not change the tetragonal-to-orthorhombic structural transition temperature [Fig. 32(d)].
Since the measurable extinction release at temperatures well above $T_s$ was
suggested to arise from significant structural fluctuations related to the
orthorhombic distortion \cite{Kreyssig10}, data for the
detwinned sample indicates that the applied uniaxial pressure pushes structural fluctuations to a
temperature similar to that at which resistivity anisotropy emerges.  These results are
different from those of Ref. \cite{CDhital2014} carried out on as-grown BaFe$_2$As$_2$.  It remains to
be seen how the different results in these experiments
can be reconciled \cite{CDhital2012,YSong13b,CDhital2014,XYLu2014S}.

In addition to determining the effect of uniaxial strain on structural and magnetic phase transitions
in annealed BaFe$_2$As$_2$, inelastic neutron scattering experiments on BaFe$_{2-x}$Ni$_x$As$_2$ also reveal that low-energy
spin excitations in these materials
change from fourfold symmetric to twofold symmetric in the uniaxial-strained tetragonal phase
at temperatures corresponding to the onset of
in-plane resistivity anisotropy \cite{XYLu2014S}.  The inset in Figure 33(a) shows the in-plane resistivity
anisotropy on annealed BaFe$_2$As$_2$ under uniaxial strain.  The temperature dependence
of the $E=6$ meV spin excitations (signal above background
scattering) at $(1,0,1)$ and $(0,1,1)$ is shown in Fig. 33(a).  In the AF ordered
state, there are only spin waves at the AF wave vector
${\bf Q}_{\rm AF}=(1,0,1)$. On warming to the paramagnetic
tetragonal state above $T_N$ and $T_s$, we see clear
differences between $(1,0,1)$ and $(0,1,1)$ that vanish above $\sim$160 K, the same temperature below
which anisotropy is observed in the in-plane
resistivity [inset in Fig. 33(a)].  Similar measurements on
underdoped BaFe$_{1.915}$Ni$_{0.085}$As$_2$ reveal that the $E = 6$ meV spin
excitations at the $(1,0,1)$ and $(0,1,1)$ wave vectors becomes anisotropic below $\sim$80 K [Fig. 33(b)], again
consistent with the in-plane resistivity anisotropy from uniaxial strain
detwinned BaFe$_{2-x}$Ni$_x$As$_2$ [Fig. 33(c)] \cite{IRFisher2011}. Finally, uniaxial strain on electron overdoped
BaFe$_{1.88}$Ni$_{0.12}$As$_2$ induces neither spin excitations
nor in-plane resistivity anisotropy at all temperatures \cite{IRFisher2011,XYLu2014S}.  Therefore,
resistivity and spin excitation anisotropies both vanish near optimal
superconductivity and are likely intimately connected, consistent with spin
nematic phase induced electronic anisotropy \cite{RMFernandes2014}.

\begin{figure}[t]
\includegraphics[scale=.2]{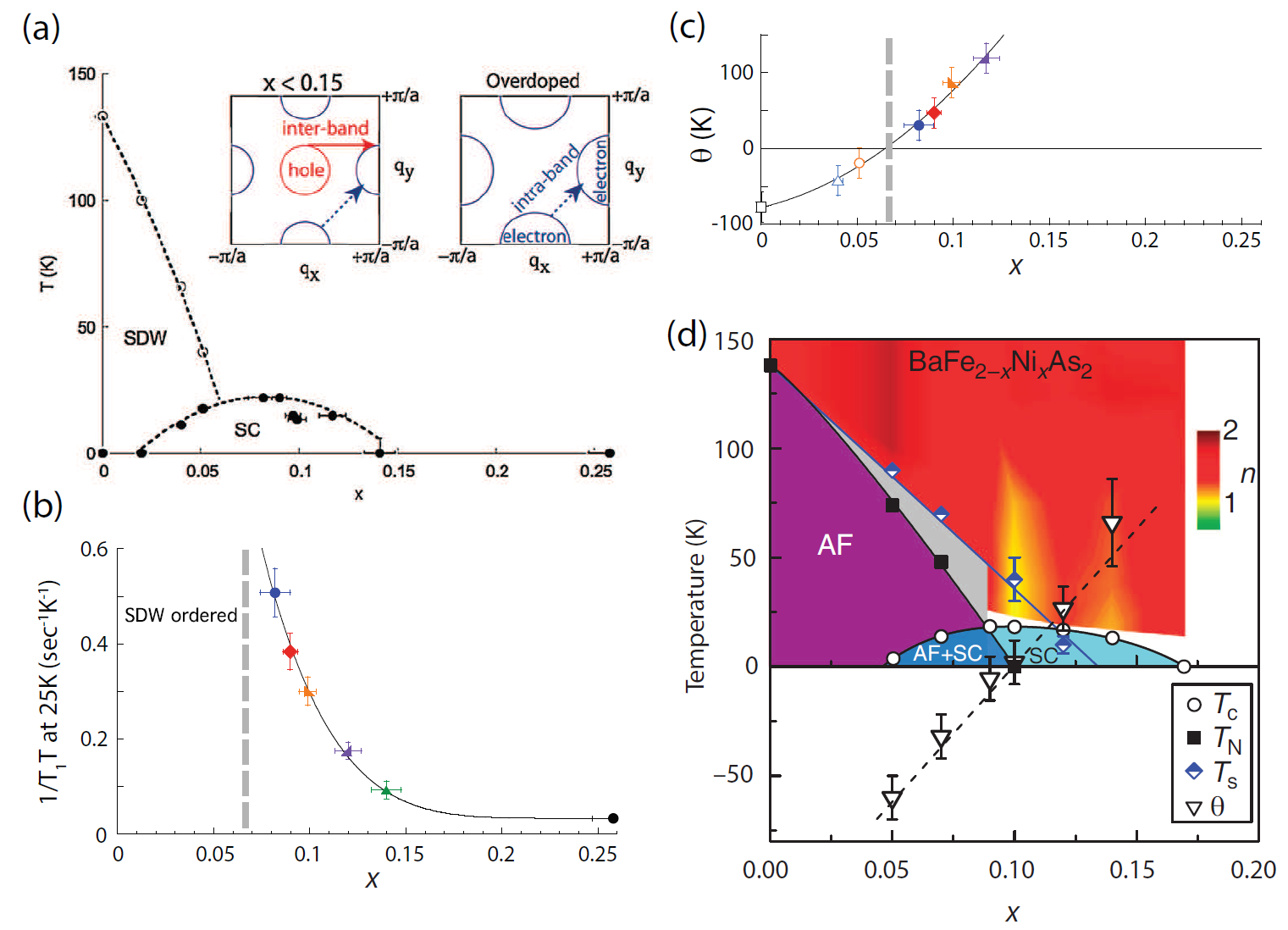}
\caption{
(Color online)
NMR determination of the phase diagram for electron-doped iron pnictides.
(a) The AF order $T_N$ and $T_c$ in Ba(Fe$_{1-x}$Co$_x$)$_2$As$_2$ as
determined from the NMR measurements.  The left and right
insets are schematic representations of the Fermi surfaces
in unfolded first Brillouin zone for the $x<0.15$ and overdoped
nonsuperconducting samples, respectively.  The absence of electron-hole
Fermi surface nesting is believed to be responsible for the suppression
of superconductivity in the overdoped regime.
(b) The Co-doping dependence of the strength of the paramagnetic spin excitations
as measured by $1/T_1T$ at 25 K ($\ge T_c$). (c) Weiss temperature $\theta$ obtained from
fitting the interband (electron-hole pocket excitations)
AF spin excitations $(1/T_1T)_{\rm inter}$ with a Curie-Weiss term
$(1/T_1T)_{\rm inter}=C/(T+\theta)$.  Here  $1/T_1T=(1/T_1T)_{\rm inter}+(1/T_1T)_{\rm intra}$, where
the intraband scattering can be fitted with a phenomenological form
$(1/T_1T)_{\rm intra}=\alpha+\beta \exp(-\Delta/k_BT)$ \cite{FLNing10}. (d) The electronic phase diagram
of BaFe$_{2-x}$Ni$_x$As$_2$ as determined from NMR and transport measurements \cite{GQZheng1}.
}
\end{figure}

\subsection{Comparison of magnetic order and spin excitations determined from neutron scattering
with those from $\mu$SR, NMR, and RIXS measurements}

In the discussions above, we have summarized recent progress in
the AF order and spin dynamics in iron-based high temperature superconductors
determined from elastic and inelastic neutron scattering. However, neutron scattering is not the only technique to study magnetism
in these materials. Other probes such as $\mu$SR, NMR, and RIXS experiments have also
been used to investigate their magnetic properties.
Compared with neutron scattering, which is a global probe determining the average magnetic properties of the
bulk solids, the $\mu$SR technique is a local probe
measuring the magnitude of the static random field
experienced by muons that reside on interstitial lattice
sites of the studied material. In addition to being able to detect static AF order and superfluid density,
it can determine the volume fractions of the magnetic and superconducting
phases and their temperature dependence \cite{uemura,Carretta2013}.  Similarly, NMR is also a local probe that
can detect magnetic and superconducting properties of the studied materials \cite{alloul}.
 In this section, we briefly summarize recent results from these techniques and
compare them with results obtained from neutron scattering.

\begin{figure}[t] \includegraphics[scale=.43]{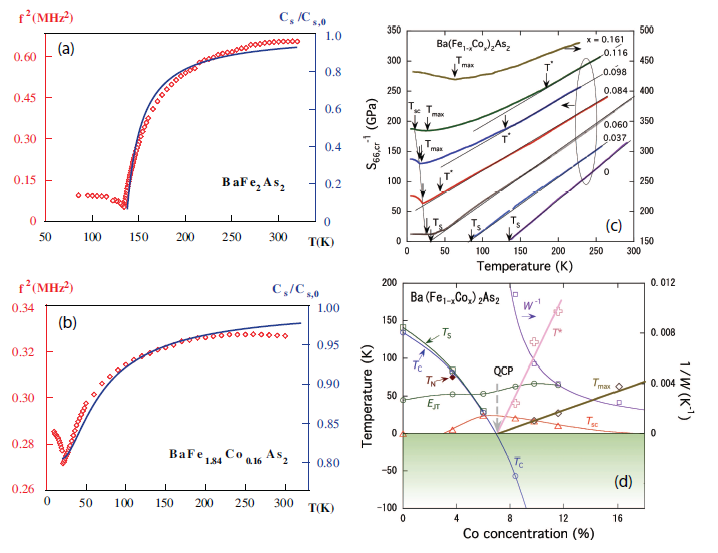}
\caption{
(Color online)
The elastic properties of the iron-based superconductor Ba(Fe$_{1-x}$Co$_x$)$_2$As$_2$ as
determined from ultrasonic spectroscopy measurements.
The temperature dependence of the anisotropic elastic stiffness $C_{66}$ for (a) BaFe$_2$As$_2$
and (b) BaFe$_{1.84}$Co$_{0.16}$As$_2$ as determined from resonant ultrasonic
spectroscopy. The solid lines are fits from a model considering spin nematic phase \cite{RMFernandes10}.
(c) Temperature dependence of the inverse of the
elastic stiffness  $S_{66}=1/C_{66}=S_{66}^0+S_{66,{\rm Cr}}$, where
$S_{66}^0$ is the normal (background) contribution, for Ba(Fe$_{1-x}$Co$_x$)$_2$As$_2$
with different $x$. (d) Structural and magnetic phase diagram of Ba(Fe$_{1-x}$Co$_x$)$_2$As$_2$
as determined from ultrasonic spectroscopy
measurements.  A structural quantum critical point is identified near optimal superconductivity \cite{MYoshizawa12}.
}
\end{figure}

We begin by comparing the electronic phase diagrams of electron doped Ba(Fe$_{1-x}$Co$_x$)$_2$As$_2$
determined from X-ray and neutron scattering \cite{snandi10,dkpratt11} with those determined from
 the $\mu$SR measurements [Fig. 34(a)] \cite{Bernhard2012}. For Co-doped samples in the
underdoped regime ($x\le 0.045$), the
$\mu$SR results find full volume AF ordered phase coexisting and competing
with superconductivity.  This is consistent with neutron diffraction
results indicating static commensurate
AF order coexists and competes with superconductivity in the underdoped regime (Fig. 13) \cite{dkbratt09,adchristianson09}.
For Ba(Fe$_{1-x}$Co$_x$)$_2$As$_2$ near optimal superconductivity with $x\ge 0.05$,
magnetic order only develops in parts of the sample volume and the normalized values of the average
magnetic field at the muon site, $B_\mu$, are reduced dramatically [Fig. 34(b)].
In addition, the
AF N\'eel temperature of the system saturates to a value near or slightly above $T_c$ before
vanishing in a first order fashion with increasing Co-doping [Fig. 34(b)] \cite{Bernhard2012}.
This region of Co-doping is consistent with the appearance of the transverse
incommensurate AF order seen in neutron diffraction experiments \cite{dkpratt11}.  However, instead of a
uniform
incommensurate spin-density-wave ordered phase,
the $\mu$SR data indicates a spatially inhomogeneous magnetic state
for which the volume fraction of the ordered phase decreases with
increasing Co-doping \cite{Bernhard2012}.  The NMR measurements on Ba(Fe$_{1-x}$Co$_x$)$_2$As$_2$ samples
near optimal superconductivity reveal that the system is in the spin-glass state
which competes with superconductivity \cite{apdioguardi}.
Instead of being a consequence of Fermi surface nesting \cite{dkpratt11},
the incommensurate magnetic order in Co-doped BaFe$_2$As$_2$
may arise from inhomogeneous short-range magnetic order similar to
 the moment modulating cluster spin glass in Ni-doped BaFe$_2$As$_2$ \cite{xylu14}.
Therefore, the electronic phase diagrams of the Co- and Ni-doped BaFe$_2$As$_2$ are similar,
showing a homogeneous commensurate long-range
static AF ordered phase competing with superconductivity in the underdoped region and a
short-range spin glass phase near optimal superconductivity [Fig. 5(b)] \cite{hqluo12,xylu13,apdioguardi,xylu14,Bernhard2012}.
There is no evidence for a conventional magnetic quantum critical point near optimal superconductivity, and the AF order disappears in the superconducting phase with increasing Co/Ni-doping in a first order fashion \cite{hqluo12,xylu13,apdioguardi,xylu14,Bernhard2012}.

The electron phase diagrams of the Co- and Ni-doped BaFe$_2$As$_2$ systems have also been mapped out
by NMR measurements [Figs. 35(a) and 35(d)] \cite{GQZheng1,FLNing09,FLNing10}. In the case of
Ba(Fe$_{1-x}$Co$_x$)$_2$As$_2$, the NMR experiments on electron overdoped samples
suggest that the absence of quasiparticle excitations
with momentum transfer ${\bf Q}_{\rm AF}$
 between the hole and electron Fermi surfaces results in complete
suppression of the low-energy spin fluctuations for $x\ge 0.15$.  The insets in Fig. 35(a) show
that the hole bands sink below the Fermi surface for $x\ge 0.15$, disallowing
inter-band quasiparticle transitions between the hole and electron Fermi surfaces \cite{FLNing10}.
These results are consistent with neutron scattering data on
electron-doped BaFe$_{2-x}$Ni$_x$As$_2$, where there is a
large ($\sim$50 meV) spin gap in the nonsuperconducting sample with $x=0.3$ \cite{MWang13}.

\begin{figure}[t]
\includegraphics[scale=.43]{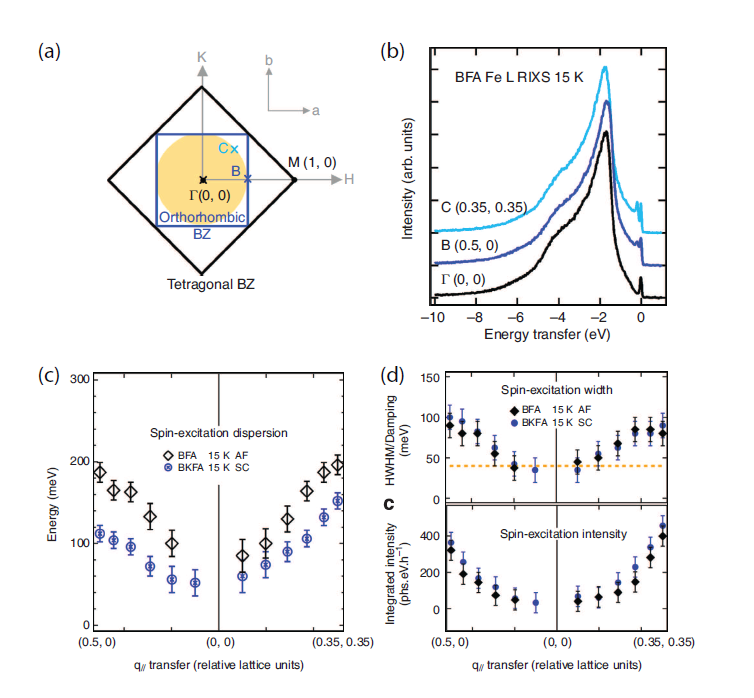}
\caption{
(Color online)
RIXS measurements of the high-energy spin excitations in optimally hole-doped BaFe$_2$As$_2$.
(a) Schematic view of the reciprocal space covered by Fe $L_3$ RIXS is
shaded by a yellow circle. $\Gamma$, B and C are the reciprocal space positions
at which RIXS spectra were collected. Black (blue) squares represent the tetragonal (orthorhombic) Brillouin
zone. All RIXS spectra use the orthorhombic Brillouin zone
convention for defining relative momentum transfer values. $\Gamma$ point is the
structural zone center, while $\Gamma$-M is the AF ordering wave vector.
(b) Three typical RIXS spectra of BFe$_2$As$_2$ collected at 15 K with $\pi$ polarized incoming light, at
$\Gamma$, B, and C positions in reciprocal space. (c) Dispersion of spin excitations
of BFe$_2$As$_2$ (BFA) in the AF phase, and Ba$_{0.6}$K$_{0.4}$Fe$_2$As$_2$ (BKFA)
in the superconducting phase. (d) Half Width Half Maximum (HWHM, damping) and integrated
intensity of spin excitations of BFA and BKFA. The horizontal
dotted line in upper panel marks the HWHM
of the total instrumental resolution of the RIXS experiment (40 meV) \cite{kjzhou}.
 }
\end{figure}

In addition to determining the electronic phase diagram of Ba(Fe$_{1-x}$Co$_x$)$_2$As$_2$ as shown in Fig. 35(a),
the $1/T_1T$ obtained by NMR measurements is also related to the wave vector
integral of the low-energy spin dynamic susceptibility $\chi^{\prime\prime}({\bf Q},f)$ via
$1/T_1T = A/(T-\theta) \sim \sum_{\bf Q} \left|A({\bf Q})\right|^2\chi^{\prime\prime}({\bf Q},f)/f$, where
$\theta$ is the Curie-Weiss temperature (the temperature at which a plot of the reciprocal
molar magnetic susceptibility against the absolute temperature $T$ intersects the $T$-axis),
$f$ is the NMR frequency, $a$ is the lattice constant, $\left|A({\bf Q})\right|^2=\left|A\cos(Q_xa/2)\cos(Q_ya/2)\right|^2$
is the form factor of transferred hyperfine coupling at the $^{75}$As sites, and the wave vector summation of ${\bf Q}$
is taken over the entire first Brillouin zone \cite{FLNing09}.  By measuring
the Co-doping dependence of $1/T_1T$, one can
fit the data with Curie-Weiss law and
obtain the electron doping dependence of $\theta$. Figure 35(b) and 35(c) shows the Co-doping dependence of the
$1/T_1T$ in the normal state ($T\ge T_c$) and $\theta(K)$, respectively \cite{FLNing10}.
The enhancement of $1/T_1T$ and the negative to positive crossing of $\theta(K)$ near vanishing AF order
suggest the presence of a magnetic quantum critical point at $x\approx 0.07$ \cite{FLNing10}.  Similar
NMR data and transport measurements
on BaFe$_{2-x}$Ni$_x$As$_2$ suggest the presence of two quantum critical points associated with AF order and the
tetragonal-to-orthorhombic lattice
distortions, respectively [Fig. 35(d)] \cite{GQZheng1}.
These results are in direct contrast to the conclusions of neutron scattering and $\mu$SR experiments as described earlier in the session, which suggests a weakly first order phase transition from static
AF order to superconductivity in electron-doped BaFe$_2$As$_2$.
We note that these NMR measurements have not been carried out
for the Co and Ni-doped samples with incommensurate magnetic order and near optimal superconductivity \cite{FLNing10,GQZheng1}.

Although determining whether a conventional magnetic quantum critical
point exists in the iron pnictides is important, it is equally important to understand what happens to
the tetragonal-to-orthorhombic lattice distortion in the Co and Ni-doped BaFe$_2$As$_2$ phase diagram near optimal superconductivity.  From initial high-resolution X-ray diffraction experiments, it was suggested that Ba(Fe$_{1-x}$Co$_x$)$_2$As$_2$ has reentrant behavior near optimal superconductivity, exhibiting a tetragonal-orthorhombic transition
above $T_c$ and orthorhombic-tetragonal structural transition below $T_c$ \cite{snandi10}.  While the overall trends of these results were confirmed by later high-resolution X-ray diffraction experiments on BaFe$_{2-x}$Ni$_x$As$_2$ near optimal superconductivity, the presence of low-temperature incommensurate AF order suggests that the system is still in the orthorhombic phase \cite{xylu13}.  Using resonant ultrasonic spectroscopy, one can measure the temperature dependence of the
anisotropic elastic stiffness $C_{ij}$ associated with the tetragonal-orthorhombic lattice distortion in
Ba(Fe$_{1-x}$Co$_x$)$_2$As$_2$ \cite{RMFernandes10,MYoshizawa12}. Figure 36(a) and 36(b) shows the temperature
dependence of the resonant ultrasonic spectroscopy measured squared resonant frequency $f^2$ (red points),
and of the calculated elastic stiffness (shear modulus) $C_s\equiv C_{66}$ of the tetragonal phase (solid lines)
for BaFe$_2$As$_2$ and BaFe$_{1.84}$Co$_{0.16}$As$_2$, respectively \cite{RMFernandes10}.  The dramatic softening of the $C_{66}$
shear modulus at temperatures above the tetragonal-orthorhombic lattice distortion temperature $T_s$ has been interpreted as due to the spin
nematic phase \cite{RMFernandes10}.  Figure 36(c) shows the temperature dependence of the inverse of $C_{66}$ for Co-doping levels
of $x=0, 0.037, 0.060, 0.084, 0.098, 0.116, 0.161,$ and 0.245 \cite{MYoshizawa12}.
The resulting phase diagram shown in Fig. 36(d) suggests
the presence of a structural quantum critical point near $x\approx 0.07$.  While this is consistent with the NMR results
of a magnetic quantum critical point \cite{FLNing09,FLNing10}, it differs from the first order nature
of the AF order to superconducting phase transition in electron-doped pnictides determined from
neutron and $\mu$SR experiments \cite{xylu13,apdioguardi,xylu14,Bernhard2012}.

In the past, the only way to explore the energy and wave vector dependence of the spin excitations
in materials
was via inelastic neutron scattering, which in principle can map out the spin excitations
in absolute units from low energy to high energy throughout the Brillouin zone as described in section III of this review.  However, such technique suffers from the need for large amounts of single crystals, which may not be available.  Recent advances in RIXS provide an alternative method to look for high-energy spin excitations in copper \cite{Tacon11} and iron \cite{kjzhou} based high-$T_c$ superconductors, although the precise RIXS cross section is difficult to calculate and includes couplings to orbital and electronic excitations in addition to magnetic excitations \cite{ament}.
For the hole-doped YBa$_2$Cu$_3$O$_{6+x}$ family of materials, RIXS measurements showed the existence of damped but well-defined dispersive magnetic excitations, deep in the Stoner continuum of hole-doped cuprates with doping beyond the optimal
 level \cite{Tacon11}.  The high-energy spin excitation spectral weights are found to be similar to those of spin waves in the undoped, antiferromagnetically ordered parent material \cite{Tacon11}.
So far, these measurements on hole-doped copper oxide superconductors have not been independently confirmed by inelastic neutron scattering experiments, which mostly probe spin excitations near the AF ordering wave vector instead of near the origin as in RIXS experiments due to the small neutron scattering cross section near $\Gamma$ (Fig. 4) \cite{Yamada,Tranquada2}.  Given that the RIXS cross section is not well-known \cite{ament}, it is extremely
important to compare inelastic neutron scattering and RIXS experiments on similar samples. Recently, such a
comparison has been made for high-energy spin excitations of Sr$_2$CuO$_2$Cl$_2$ \cite{KWPlumb2014}.

\begin{figure}[t]
\includegraphics[scale=.35]{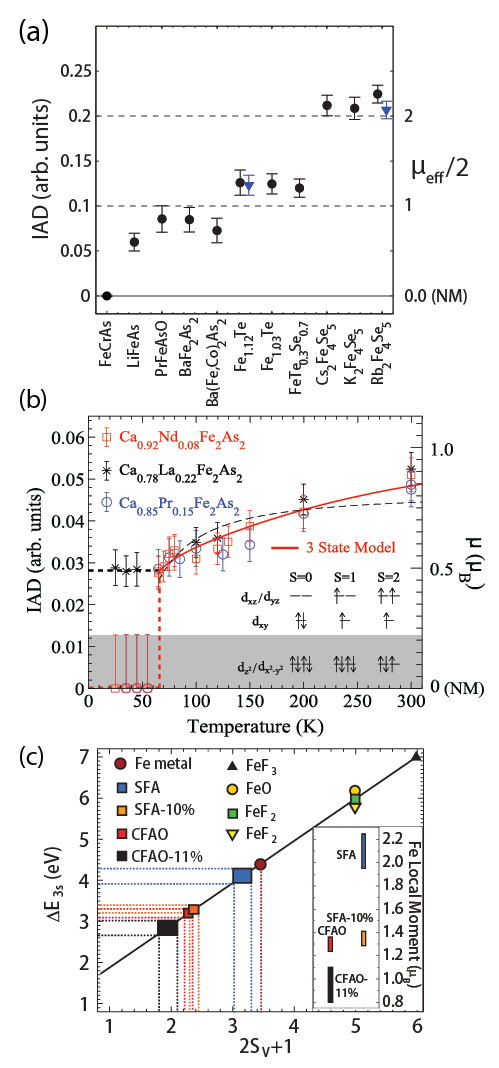}
\caption{
(Color online)
Local moments of various iron-based superconductors in their paramagnetic states determined by X-ray emission
spectroscopy and core level photoelectron spectroscopy.
(a) The IAD values derived from the XES spectra for various samples.
The room-temperature data are shown in
circles, and the low-temperature IAD values at $T = 15$ K are shown
in triangles for Fe$_{1.12}$Te and Rb$_2$Fe$_4$Se$_5$ \cite{HGretarsson2011}. The right hand axis is scaled to
the total (static+fluctuating) moments of Rb$_{0.89}$Fe$_{1.58}$Se$_2$ from \cite{MYWang11}
(Y.-J. Kim, private communication).
(b) Temperature dependence of the IAD values for
Ca$_{0.92}$Nd$_{0.08}$Fe$_2$As$_2$, Ca$_{0.78}$La$_{0.22}$Fe$_2$As$_2$,
and Ca$_{0.85}$Nd$_{0.15}$Fe$_2$As$_2$ \cite{HGretarsson2013}.
(c) Estimation of the spin moment on the Fe sites from the multiplet energy separation
$\Delta E_{3S}$ obtained from PES measurement.
The continuous line is the extrapolation
of the linear fit of the $\Delta E_{3S}$ values plotted against $(2S_V + 1)$ for
the Fe ionic compounds FeF$_3$, FeF$_2$, FeO, for which $S_V$ is known to
be 5/2 (FeF$_3$) and 2 (FeF$_2$, FeO).
Here SFA, SFA-10\%, CFAO, and CFAO-11\% are
SrFe$_2$As$_2$, SrFe$_{1.76}$Co$_{0.24}$As$_2$, CeFeAsO, and CeFeAsO$_{0.89}$Fe$_{0.11}$,
respectively.
The linear fit results in
the relation  $\Delta E_{3S} = 0.94 + 1.01 (2S_V + 1)$ \cite{Vilmercati2012}.
}
\end{figure}

Figure 37(a) shows schematic view of the reciprocal space that can be covered by Fe $L_3$ RIXS shaded by a yellow circle.
Black (blue) squares represent the tetragonal (orthorhombic) Brillouin
zone \cite{kjzhou}. Similar to the case of copper oxide superconductors, RIXS by Fe $L_3$ edge will not be able to cover the same region of the reciprocal space as that of inelastic neutron scattering [Fig. 3(e)].  Figure 37(b) plots Fe $L_3$ edge RIXS spectrum on BaFe$_2$As$_2$
collected at 15 K with $\pi$ polarized incoming light, at
$(Q_x, Q_y)=(0, 0)$ ($\Gamma$), $(0.5, 0)$ ($B$) and (0.35, 0.35) ($C$) using orthorhombic reciprocal lattice units [see Fig. 37(a) for
the $\Gamma$, $B$, and $C$ positions in reciprocal space].  In addition to the intense Fe $3d$ fluorescence at around $-2$ eV energy transfer, there are momentum dispersive excitations centered around 200 meV near the quasi-elastic peak at zero energy [Fig. 37(b)].
Since the dispersions of these excitations are identical to those of spin waves in BaFe$_2$As$_2$ determined from inelastic
neutron scattering [Fig. 37(c)] \cite{LWHarriger11}, they are believed to arise from the same spin waves in
BaFe$_2$As$_2$ measured by RIXS \cite{kjzhou}.  Similar RIXS measurements on optimally hole-doped superconducting
Ba$_{0.6}$K$_{0.4}$Fe$_2$As$_2$ (BKFA) indicate that while the dispersions of spin excitations are softened [Fig. 37(c)], the excitation intensity and widths are unchanged from its undoped parent BaFe$_2$As$_2$ [Fig. 37(d)] \cite{kjzhou}.
This is a surprising result, different from the hole-doping evolution of the high-energy spin excitations in copper oxides \cite{Tacon11}.
While the dispersion of the spin excitations in BKFA determined by RIXS
has been confirmed by later inelastic neutron scattering experiments, the absolute spin excitation intensity in BKFA measured by neutrons is much smaller than that of the spin waves in BaFe$_2$As$_2$ \cite{MWang13}. At present, it is unclear how to reconcile the
RIXS and neutron scattering results, although we note that region of the reciprocal space probed by RIXS is different from
that probed by inelastic neutron scattering.

As discussed in the introduction, neutron scattering can measure the overall strength of the
magnetic excitations through the local fluctuating moment $\left\langle{\bf m}^2\right\rangle$. For a local moment Heisenberg
model with spin $S$, the integrated spectral weight when integrated over all wave vector and energy space should yield
$\left\langle{\bf m}^2\right\rangle=(g\mu_B)^2 S(S+1)$.  Therefore, one can in principle determine the magnitude of $S$ by measuring
elastic and inelastic magnetic scattering of the system in absolute units throughout the Brillouin zone, and the outcome should be the same as the local moment $S$
determined from the X-ray emission
spectroscopy (XES) \cite{Rueff2010} and core level photo-emission spectra (PES) \cite{Vilmercati2012}.  However, the current unpolarized time-of-flight neutron scattering technology can only measure correlated spin excitations, and thus will underestimate the size of the effective spin $S$
when spin excitations become diffusive and broad as in the case of most iron pnictides [except for the true local moment system
Rb$_{0.89}$Fe$_{1.58}$Se$_2$ \cite{MYWang11}].  By measuring the local dynamic susceptibility
for electron-doped BaFe$_{2-x}$Ni$_x$As$_2$, we see that the magnitude of $\left\langle{\bf m}^2\right\rangle$
decreases from $\left\langle{\bf m}^2\right\rangle\approx 3.5$ for $x=0$ to $2.7\pm 0.1$ $\mu_B^2/$Fe, corresponding
to $S\approx 1/2$ \cite{HQLuo13,MWang13}.

Figure 38(a) shows the local moment $S$ for various iron-based superconductors in the paramagnetic
phase as determined from XES \cite{HGretarsson2011}. Here the material variation of the local moment of Fe can be extracted
using the overall shape of the Fe $K\beta$ emission spectra by
applying the integrated absolute difference (IAD) analysis.
The absolute values of $S$ were obtained by scaling the results to neutron scattering data.
Figure 38(b) plots the temperature dependence of the IAD for Ca$_{0.92}$Nd$_{0.08}$Fe$_2$As$_2$,
Ca$_{0.78}$La$_{0.22}$Fe$_2$As$_2$,
and Ca$_{0.85}$Nd$_{0.15}$Fe$_2$As$_2$ \cite{HGretarsson2013}.  The local moments
in the Nd- and Pr-doped samples disappear in the
cT phase \cite{AIGoldman3}, indicating that the Fe$^{2+}$ ions
go through a spin-state transition by taking on the low-spin state in the cT phase.
Inelastic neutron scattering experiments on CaFe$_2$As$_2$ also reveal vanishing spin excitations in
the cT phase \cite{JHSoh2013}, similar to the XES results \cite{HGretarsson2013}. Figure 38(c) shows the values
of local Fe moment for various iron pnictides estimated from PES measurements \cite{Vilmercati2012}.  We see that the values of
$\approx 2$ $\mu_B$ for BaFe$_2$As$_2$ and SrFe$_2$As$_2$ family of iron pnictides are much larger
than those obtained by inelastic neutron scattering, and decrease by about 40\% in the optimally doped
compound, also different from doping dependence results from neutron scattering \cite{Mannella2014}.
As emphasized in a recent
Review \cite{Mannella2014}, the PES experiments sample spin excitations on time scales shorter
than $10^{-15}$-$10^{-16}$ s, while inelastic neutron scattering measurements
probing spin excitations up to $\sim$300 meV correspond to time scales on the order
of 10-15 fs, at least one or two orders of magnitude slower than those in PES experiments.
Furthermore, PES measurements determine the total uncorrelated spin while present inelastic neutron scattering
measures correlated spin excitations near the AF ordering wave vector within the first Brillouin zone.

\begin{figure}[t]
\includegraphics[scale=.18]{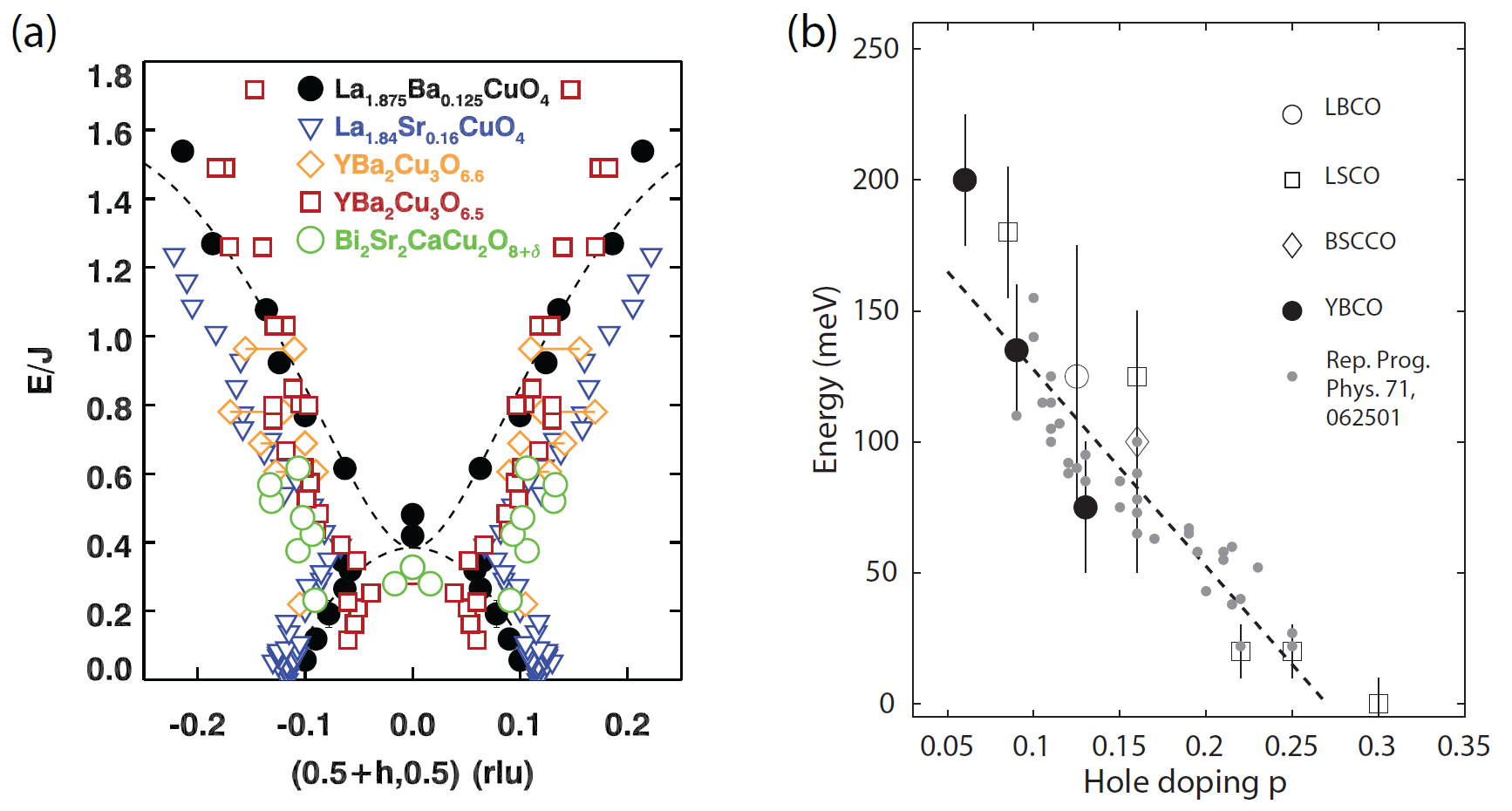}
\caption{
(Color online)
The dispersions of spin excitations in various copper oxide superconductors and the energy scale of
the spin excitations.
(a) Magnetic dispersion relation along the $(0.5+h,0.5,0)_{\rm tetra}$
in various cuprates, corresponding to wave vectors parallel to the Cu-O bonds. The energy is scaled by
$J\sim 110$ meV for the AF parent compound La$_2$CuO$_4$. (b) Large symbols: estimates of doping-dependent
energy scale at which magnetic spectral weight falls below
half that of the AF state based on inelastic
neutron scattering studies of various cuprates \cite{CStock2010B}.  Small gray
symbols: pseudogap energy from various electronic spectroscopies \cite{Tranquada2}.
}
\end{figure}

\subsection{Comparison of spin excitations in iron-based superconductors with those in
copper oxide and heavy fermion superconductors}

With the discovery of La$_{2-x}$Ba$_x$CuO$_4$ family of copper oxide superconductors in 1986 \cite{bednorz}, the field of high-$T_c$ superconductivity was born.  Although research in
high-$T_c$ copper oxide superconductors is still active with many exciting new results
almost 30 years later \cite{GGhiringhelli12,EFradkin2012}, the discovery of iron-based superconductors in 2008 provided
an entire new class of materials where high-$T_c$ superconductivity occurs \cite{kamihara}.
Since high-$T_c$ copper oxide and iron-based superconductors, as well as heavy fermion superconductors are close to
AF instability, magnetism may be
a common thread for unconventional superconductivity \cite{DJScalapino}. If this is indeed the case, it will be interesting to determine
the similarities and differences in spin excitations of these superconductors.

\begin{figure}[t] \includegraphics[scale=.24]{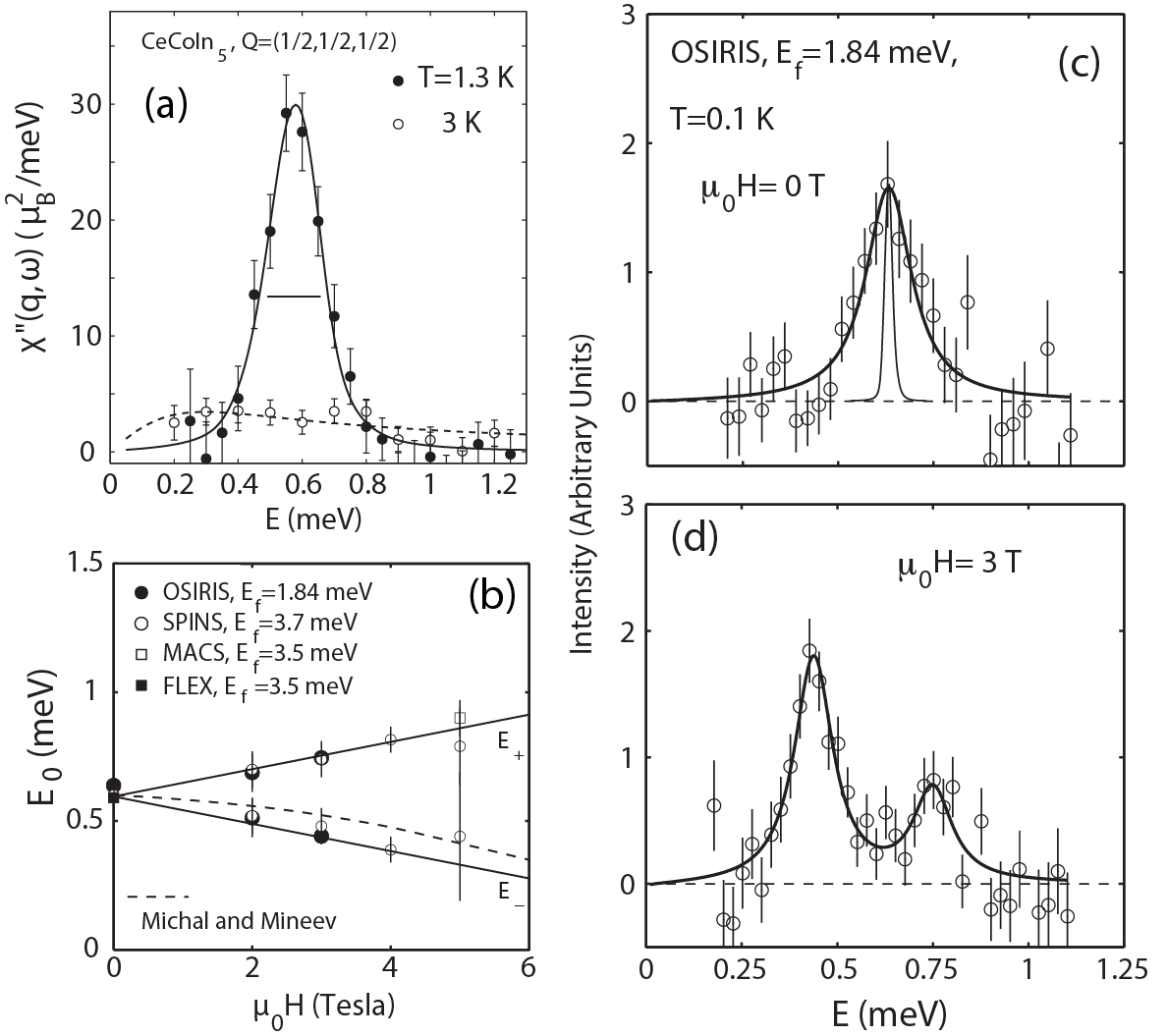}
\caption{
(Color online)
Neutron spin resonance in heavy fermion superconductor CeCoIn$_5$ and its magnetic field dependence.
(a) The imaginary part of the dynamic susceptibility at
${\bf Q}_{\rm AF}=(1/2,1/2,1/2)$ in the normal (3 K) and superconducting (1.3 K) states.  A
background scattering taken at ${\bf Q} = (0.3,0.3,0.5)$ and $(0.7,0.7,0.5)$ was subtracted.
The horizontal bar is the instrumental resolution width \cite{CStock2008}.
(b) The peak positions of the field spit resonance as a
function of applied field within the $a$-$b$ plane. The solid lines are
fits to $E_\pm =\hbar\omega_0 \pm g \mu_B \mu_0 H/2$ with $g=1.92 \pm 0.10$.
(c) Resonance at zero applied magnetic field. (d) Identical scan under an
applied field of 3-T.  A 10 K background was subtracted from the scans \cite{CStock2012}.
}
\end{figure}

As discussed in recent review articles \cite{Yamada,Tranquada2}, an important feature of the
spin excitations in copper oxide superconductors is the neutron spin resonance
and hourglass like dispersion of the spin excitation spectra.
An hourglass magnetic spectrum is also seen
in an insulating, hole-doped antiferromagnet La$_{5/3}$Sr$_{1/3}$CoO$_4$ \cite{ATBoothroyd2011}.
Figure 39(a) shows
the spin excitation dispersions of different copper oxide superconductors away from the AF ordering wave vector
${\bf Q}_{\rm AF}=(0.5,0.5)$, where $J$ is the nearest neighbor
magnetic exchange coupling ($J\approx 120$ meV) \cite{Yamada,Tranquada2}.
While the hourglass dispersion of spin excitations appears to be a ubiquitous feature of different families of
hole-doped copper oxide superconductors, they are clearly absent in spin excitation spectra of
electron-doped iron pnictides.
For optimally hole-doped iron pnictides, spin excitations change from longitudinally to transversely elongated
ellipses centered at ${\bf Q}_{\rm AF}=(1,0)$ on moving from below the resonance energy to above it \cite{MWang13}.  This is different from the hourglass
dispersion seen in hole-doped copper oxides.  Another important feature of the spin excitation spectra is the electron and hole-doping evolution of the local
dynamic susceptibility. For hole-doped copper oxide superconductors, the strength of the high energy magnetic
response near ${\bf Q}_{\rm AF}=(0.5,0.5)$ decreases with increasing doping level.
To quantify this behavior, Stock {\it et al.} evaluated the hole doping dependence
of the energy at which the local dynamic susceptibility $\chi^{\prime\prime}(\omega)$ falls
below half of that for an undoped AF parent compound \cite{CStock2010}.  The outcome suggests that the hole-dependence of
the magnetic energy scale corresponds very
well with the pseudogap energy determined from electronic spectroscopies [Fig. 39(b)]
\cite{SHufner2008}.
These results indicate that AF spin excitations in copper oxides decreases dramatically with increasing hole
doping.  Although there are only limited data available for hole-doped iron pnictides \cite{MWang13}, they
show a similar trend as that of hole-doped copper oxides. More inelastic neutron scattering experiments on hole-doped iron pnictides are necessary in order
to make a detailed comparison between spin excitations in iron and copper based high-$T_c$ superconductors.

\begin{figure}[t]
\includegraphics[scale=.37]{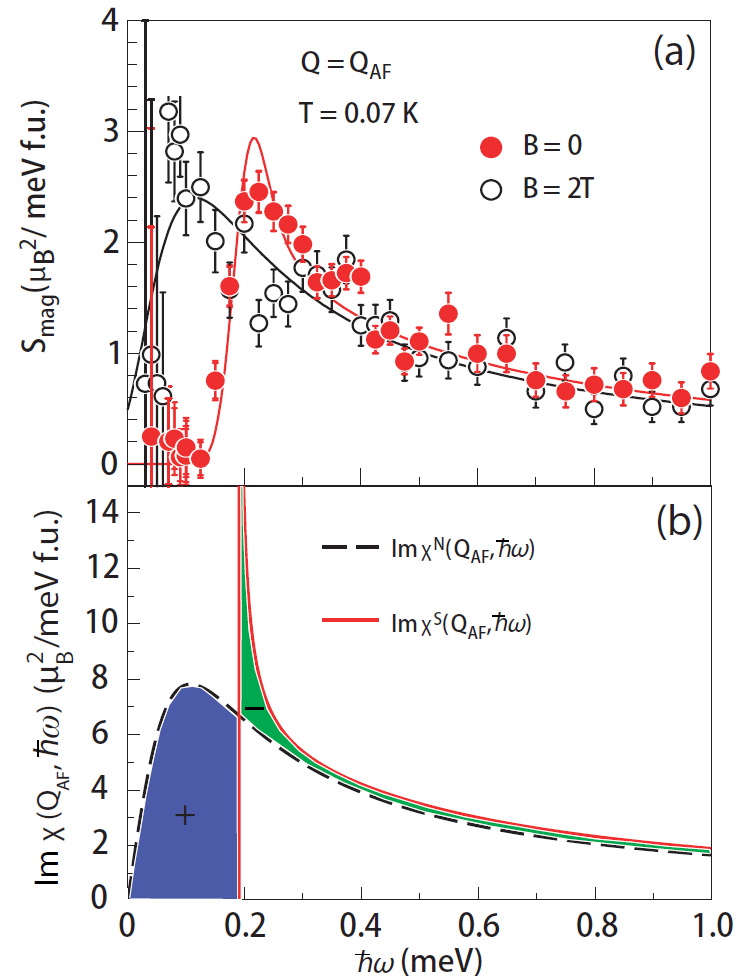}
\caption{
(Color online)
Superconductivity-induced changes in spin dynamic susceptibility and magnetic contribution to the superconducting
condensation energy in CeCu$_2$Si$_2$.
(a) Energy scans in $S$-type CeCu$_2$Si$_2$ at ${\bf Q}={\bf Q}_{\rm AF} = (0.215,0.215,1.458)$ in the superconducting
state at zero field and normal state at $B=2$ T ($T=0.07$ K). (b)
Imaginary part of the dynamic susceptibility at ${\bf Q}_{\rm AF}$
in the normal ($\chi^{\prime\prime}_N({\bf Q}_{\rm AF},\hbar\omega)$) and superconducting ($\chi^{\prime\prime}_S({\bf Q}_{\rm AF},\hbar\omega)$) states.
The blue area (marked with a $+$) leads to an increase in magnetic exchange energy $\Delta E_x$,
whereas the green area (marked with a $-$) leads to a decrease in $\Delta E_x$ \cite{OStockert2011}.
}
\end{figure}

As discussed in Section H, recent advances in RIXS have allowed a direct study of spin excitations in copper and iron-based superconductors. However,
the results obtained by RIXS for hole-doped iron pnictides disagree with
those obtained by inelastic neutron scattering \cite{kjzhou,MWang13}.
RIXS measurements on copper oxide superconductors also find excitations that soften no
more than 10\% with doping, with negligible change in integrated intensity.
While these results are very similar to those obtained
on iron pnictides \cite{kjzhou}, they are again in conflict with those of inelastic neutron scattering probing spin excitations
near ${\bf Q}_{\rm AF}=(0.5,0.5)$ \cite{Tranquada2}.  As the RIXS cross section includes
both charge and spin excitations \cite{ament}, it is not known how to compare directly the RIXS intensity with that of
the well-established magnetic cross section obtained from inelastic neutron scattering.

In the case of electron-doped copper oxide superconductors, recent RIXS experiments have confirmed
that magnetic excitations harden across the antiferromagnetism and superconductivity
boundary seen originally with neutron scattering experiments \cite{wilson2006,MFujita2006,WSLee2014}.  However, RIXS
finds a negligible change in the integrated magnetic intensity as a function of electron doping at high energies, again in conflict
with neutron scattering results on similar materials \cite{wilson2006,MFujita2006,WSLee2014}.
While one can directly compare spin waves measured in neutron scattering experiments with RIXS in the AF ordered phase of
copper oxide and iron-based materials due to symmetry of the equivalent Brillouin zones (see dashed box near $\Gamma$
and magenta shaded box near ${\bf Q}_{\rm AF}$ in Fig. 3), there is no physical justification
for assuming that the excitations measured by RIXS near the $\Gamma$ (${\bf Q}=0$) point are the
same as those near ${\bf Q}_{\rm AF}$ obtained by neutron scattering for a doped metallic sample \cite{Tranquada2}.
Whereas the neutron scattering cross section is well understood, the RIXS cross section is complicated and the
significance of the RIXS measurements is unclear at present.  Only future RIXS
and neutron scattering measurements performed on the same sample at the same
region of the reciprocal space will shed new light
to our understanding of the spin and electronic excitations as revealed by RIXS and determine
their significance to high-$T_c$
superconductivity.

In addition to copper and iron based superconductors, unconventional superconductivity
also includes heavy fermion superconductors \cite{steglich,Stewart,qmsi,HVLohneysen2007}.
Compared with iron and copper-based superconductors, the parent compounds of heavy fermion superconductors are also long-range ordered antiferromagnets but with a magnetic exchange coupling energy scale much lower than
that of the AF parents of copper oxide and iron-based superconductors.
For example, in a recent study of spin waves in AF ordered CeRhIn$_5$, the parent compound of the
CeCoIn$_5$ family of heavy
Fermion superconductors \cite{JDThompson2012}, the dominant in-plane nearest neighbor magnetic
exchange coupling is $SJ_0=0.74$ meV \cite{PDas2014}, much smaller than that
of the AF parents of iron based superconductors (see Table II).  In spite of the dramatically
reduced energy scale, heavy Fermion superconductors still have some interesting
features also seen in copper and iron-based superconductors.  A case in point is the neutron spin
resonance in heavy Fermion superconductors such
as UPd$_2$Al$_3$ \cite{NKSato2001} and CeCoIn$_5$ \cite{CStock2008}.

Figure 40(a) shows energy dependence of the spin excitations
of CeCoIn$_5$ below and above $T_c$ at ${\bf Q}_{\rm AF}=(0.5,0.5,0.5)$ \cite{CStock2008}.  While the normal state spin excitations are featureless
within the probed energy range, superconductivity induces a sharp peak reminiscent of the
resonance in iron and copper based superconductors.
From neutron polarization analysis of the resonance
discussed in Section F, we know that the mode in iron-based superconductors is
isotropic, consistent with the singlet-to-triplet excitation of electron Cooper pairs \cite{eschrig}.  If this is indeed the case,
application of a magnetic field should Zeeman split the resonance into three peaks arising from a single ground state
to a triplet excited state \cite{PCDai2}, where the central field independent peak is longitudinally
polarized while the field dependent peaks are transversely polarized \cite{JPIsmer2007}.  Unfortunately, magnetic field experiments
on copper oxide \cite{PCDai2} and iron-based superconductors \cite{JZhao50,JSWen50,SLLi11} have been unable
to determine the ground state of the resonance.  Surprisingly, the application of a magnetic field on heavy Fermion
superconductor CeCoIn$_5$ splits the resonance into two peaks as shown in
Figs. 40(b)-40(d) \cite{CStock2012}.  This would suggest that the
resonance in CeCoIn$_5$ is a doublet instead of a singlet-to-triplet excitation.
In separate polarized inelastic neutron scattering experiments under a magnetic field, the double peak nature of the resonance was confirmed under a
2 T applied field \cite{SRaymond2012}.  In addition, the resonance line shape is found to
depend on the neutron polarization, suggesting that the resonance is
a degenerate mode with three fluctuation channels: a Zeeman split contribution and an
additional longitudinal mode \cite{SRaymond2012}.  While these results on CeCoIn$_5$ are interesting,
they still have not established conclusively the doublet or singlet-to-triplet nature of the resonance.

Using established models to calculate the magnetic contributions to the superconducting condensation energy in
copper oxide superconductors \cite{scalapino98,demler98}, one can in principle estimate the lowering of the magnetic exchange energy
in YBa$_2$Cu$_3$O$_{6+x}$ family of materials using spin excitation spectra
above and below $T_c$ \cite{pcdai1999,HWoo2006,TDahm2009}.
However, the large energy scale of the spin excitations in high-$T_c$ copper oxide superconductors means that it is difficult to obtain the overall
 spin excitation spectra in the low-temperature normal state using a magnetic field
to eliminate superconductivity \cite{pcdai1999,HWoo2006,TDahm2009}.  This makes the estimation of the
lowering of the magnetic exchange energy in the low-temperature state ambiguous \cite{scalapino98,demler98}.
Since heavy Fermion superconductors generally have a much smaller energy scale, one can easily obtain the
low-temperature normal state spin excitation spectra using a magnetic field to suppress superconductivity.
Figure 41(a) shows the energy dependence of the spin excitation spectra in
the low-temperature normal and superconducting states for CeCu$_2$Si$_2$ \cite{OStockert2011}, which exhibits
superconductivity below $T_c\approx 0.6$ K \cite{steglich}.  One can see the opening of
a spin gap and clear enhancement in spin excitations above the gap.
Using the energy dependence of the dynamic susceptibility in absolute units in the low-temperature normal and
superconducting states [Fig. 41(b)],
Stockert and co-workers estimated the lowering of the magnetic exchange energy in CeCu$_2$Si$_2$ and found it to be
considerably larger than the superconducting condensation energy \cite{OStockert2011}.
Although the lowering of the magnetic exchange energy is also found to be much larger than the superconducting
condensation energy in the superconducting iron pnictides \cite{MWang13} and chalcogenides \cite{JLeiner2014},
these experiments also suffer
the problem of not being able to determine
the overall spin excitation spectrum in the low-temperature normal state.

\section{Theoretical descriptions of static AF order and
spin excitations in iron-based superconductors}

Shortly after the discovery of iron-based superconductors, band structure calculations predicted
that the Fermi surfaces of parent compounds consist of two quasi-two-dimensional near-circular hole pockets centered around
 the zone center $\Gamma$, and two quasi-two-dimensional elliptic electron pockets center around the (1,0) and (0,1) points in the orthorhombic
unfolded Brillouin zone (Fig. 14) \cite{mazin2011n,hirschfeld}.  The sign reversed quasiparticle excitations between the hole and electron pockets (nesting) can give rise to the spin-density-wave order
observed experimentally \cite{jdong,cruz}.  In addition, the same quasiparticle excitations in doped superconductors
are expected to induce a neutron spin resonance in the superconducting state, which is
confirmed by experiments (see Section III. B) \cite{MMKorshunov2008,TAMaier2008,TAMaier2009}.  Within this weak-coupling analysis,
iron-based superconductors and their parents are assumed to be good metals made of itinerant electrons with spin-density-wave type
AF order.  Spin waves and spin excitations can then be calculated using RPA in a multiband Hubbard model
with appropriate Fermi surfaces for hole and electron pockets \cite{JKnolle2013}. In this approach, the large in-plane
effective magnetic exchange coupling anisotropy (Table II) in the spin waves of iron pnictides \cite{jzhao09,LWHarriger11} can be understood as due to the ellipticity of the electron pockets (Fig. 14), which induces
frustration between the $(1,0)$ and $(0,1)$ wave vectors connecting the hole and electron pockets \cite{JKnolle2000,Kaneshita10}.
In the underdoped regime where the static AF order coexists and competes with superconductivity \cite{dkbratt09,adchristianson09,MYWang2010}, spin excitations at $(1,0)$ are determined by
the presence of the AF order and associated spin waves, while the excitations
at wave vector $(0,1)$ are dominated by the superconductivity and formation of the
resonance in the single domain sample \cite{JKnolle2011}.  However, since most neutron
scattering experiments in underdoped
iron pnictides were carried out on twinned samples that cannot distinguish
the wave vector $(1,0)$ from $(0,1)$ \cite{dkbratt09,adchristianson09,MYWang2010}, it is unclear
how the resonance associated with superconductivity
interacts with itinerant electrons contributing to the spin waves.
In a systematic study of spin excitations in BaFe$_{2-x}$Ni$_x$As$_2$, the electron-doping evolution of the low-energy spin excitations was found to qualitatively agree with RPA calculations of
the nested Fermi surfaces \cite{hqluo12}.  However, the high-energy spin excitations are weakly electron-doping independent, and
have values much different from that found by RPA calculations \cite{MSLiu12,HQLuo13,MWang13}.
These results suggest that the weak-coupling analysis based on purely itinerant electrons is insufficient to explain the entire spin excitation spectrum and its electron or hole doping evolution.

Although the weak-coupling approach using a
Fermi surface nesting picture provides a nice framework to understand static AF order, spin excitations, and
their connection with superconductivity in iron-based materials \cite{mazin2011n,hirschfeld},
calculations show that the ordered moment in the AF iron pnictides is around 2 $\mu_B$/Fe \cite{FMa2008}, much larger than the
largest experimental value ($\sim$0.9 $\mu_B$/Fe, see Table I).
Furthermore,  the ordering wave vectors of the bi-collinear AF structure in
iron chalcogenides shown in Fig. 2(c) do not match the nesting wave vector of the Fermi surfaces \cite{ASubedi}.  In the strong coupling limit,
all unpaired electrons, not just itinerant electrons near the Fermi surface, participate in forming magnetic order, much like the magnetic moment
of Cu$^{2+}$ in the insulating copper oxides \cite{qsi,cfang,ckxu}. Here, the AF ordered state
of iron-based superconductors can be described by a
local moment Heisenberg Hamiltonian
with the nearest neighbors ($J_{1a}$-$J_{1b}$) and next nearest neighbor ($J_2$) exchange interactions (Fig. 4)
\cite{cfang,ckxu,JDai2009,TYildirim2008,AMoreo2009,MJHan2009,RYu2013}.  In this picture, the large in-plane magnetic
exchange coupling anisotropy in the parent compounds of iron pnictides
is understood in terms the presence of the biquadratic
exchange coupling $K$ between the nearest spins in the AF ordered states, which can be
mapped onto the $J_{1a}$-$J_{1b}$ model with a specific relationship between $J_{1a}$-$J_{1b}$ and $J_1$-$K$
 \cite{ALWysocki11,RYu2012b}.
This means that the usual local moment
Heisenberg Hamiltonian will be modified to $H=J_1 \sum_{i,\delta} {\bf S}_i\cdot {\bf S}_{i,\delta}
+J_2\sum_{i,\delta} {\bf S}_i\cdot {\bf S}_{i+\delta} -K\sum_{i,\delta} ({\bf S}_i\cdot {\bf S}_{i,\delta})^2$,
where $J_1$ and $J_2$ are the nearest and next nearest neighbor exchange couplings, respectively,
and ${\bf S}_i$ is the spin at site $i$ \cite{ALWysocki11,RYu2012b}.
The calculated dynamical structure factor $S({\bf Q},\omega)$
forms ellipses in momentum space around ${\bf Q}_{\rm AF}=(\pm 1,0)$
at low excitation energies.
With increasing energy, the elliptic features expand towards the zone boundary and gradually split into two
parts, forming a pattern around the wave vector $(\pm 1,\pm 1)$ consistent with neutron scattering experiments \cite{jzhao09,LWHarriger11,ALWysocki11,RYu2012b}.

By comparing the AF exchange interactions determined from neutron scattering
in the parent compounds of iron-based superconductors with those of the
superconducting gap functions in the corresponding superconducting materials observed by ARPES,
it was argued that the key ingredients in determining the high-$T_c$ superconductivity of iron-based materials
are the next nearest neighbor local AF exchange interactions in real space and
a Fermi surface topology in reciprocal space that matches to the pairing
form factor provided by AF interactions \cite{JPHu2012}.  From analysis of spin excitation spectra in hole and
electron-doped iron pnictides
and magnetic contributions to the superconducting condensation energy \cite{DJScalapino},
we find that
high-$T_c$ superconductivity is associated with materials having
large magnetic exchange coupling
and strong itinerant electron-spin excitation interactions \cite{MWang13},
similar to the large Debye temperatures and strong electron-phonon interactions in high-$T_c$ BCS superconductors \cite{bcs}.
Finally, the strong-coupling approach predicts the presence of a quantum critical point separating a paramagnetic
metallic phase from an AF phase phase in phosphorus doped iron pnictides \cite{JDai2009}.  Indeed, neutron scattering experiments on powder samples of CeFeAs$_{1-x}$P$_x$O suggest the presence of a quantum critical point near $x=0.4$ controlled by the pnictogen height away from the Fe-plane [Fig. 7(a)] \cite{CCruz10}. Similarly,
thermodynamic and transport measurements
on BaFe$_2$(As$_{1-x}$P$_x$)$_2$ have identified
a quantum critical point near optimal superconductivity at $x=0.3$ [Fig. 6(d)] \cite{tshibauchi14}.
However, recent neutron powder diffraction measurements on BaFe$_2$(As$_{1-x}$P$_x$)$_2$ suggest that structural quantum criticality cannot exist at compositions higher than $x = 0.28$ due to the actual phase stability range \cite{JMAllred2014}.  It is therefore desirable to carry out neutron scattering and $\mu$SR experiments on single crystals of
BaFe$_2$(As$_{1-x}$P$_x$)$_2$ near $x=0.3$ to establish the nature of the AF phase transition with increasing P-doping.

While there are many phenomenological reasons for
using the strong coupling approach to understand the electrical
transport, spin and charge dynamical properties of iron
pnictides and chalcogenides \cite{cfang,ckxu,JDai2009,TYildirim2008,RYu2013}, such an approach is incompatible with some
spin dynamical properties.  For example, in a strict local moment Heisenberg Hamiltonian, single particle excitations or
spin waves should only have transverse components and
would not support longitudinal spin excitations in the AF ordered phase of iron pnictides as seen in polarized neutron
scattering experiments [Fig. 29(f)] \cite{CWang2013}.
Furthermore, the electron and hole-doping evolution of the low-energy spin excitations are consistent with
the Fermi surface nesting predictions, but it is unclear whether the data is also compatible
with a pure local moment Heisenberg Hamiltonian (see Section III. C).
Finally, spin waves of iron chalcogenides are heavily damped at high-energies near the zone boundary and exhibit a number of anomalous properties difficult to understand within a local moment model \cite{OJLipscombe,IAZaliznyak}.

Instead of a strong or weak coupling approach, the iron pnictides may be
Hund's metals where the interaction between the electrons is not strong enough
to fully localize them to form a Mott insulator, but is sufficient so that the low energy
quasiparticles have much enhanced mass \cite{KHaule2009}. Here the electron
correlation strength would be primarily controlled by the Hund's coupling $J_H$, which
depends on the pnictogen heights and
tends to align spins of all the electrons on a given Fe-atom, and hence enhances
spin excitations without appreciably affecting the charge excitations \cite{ZPYin2011}.  This is different
from the effect of large Coulomb repulsion $U$ in a Mott insulator, which hampers charge
excitations in order to enhance spin fluctuations \cite{PALee}.  The electronic excitations in
iron-based superconductors are neither fully itinerant nor fully localized, but have a dual nature that
can be realistically described by a combination of DFT and DMFT \cite{Kotliar2006}.
This idea is similar to the picture where
single electron spectral function is composed of
coherent and incoherent parts representing electrons near (itinerant electrons)
and far away (local moments) from the Fermi surface \cite{EAbrahams11}.

Using the combined DFT and DMFT method,
one can estimate the size of the ordered moment for different iron pnictides and find them to be
close to the observed value \cite{ZPYin2011}.  The same method has also been used to calculate the spin wave spectra
in BaFe$_2$As$_2$ and good agreement was found with neutron scattering experiments \cite{HPark2011,ZPYin2014}.
Finally, the combined DFT and DMFT method has been used to calculate the electron and hole-doping
dependence of the spin
excitation spectrum in absolute units for Ba$_{1-x}$K$_x$Fe$_2$As$_2$ and BaFe$_{2-x}$Ni$_x$As$_2$ \cite{MWang13}.
Compared with RPA calculations based on the weak-coupling approach \cite{hirschfeld}, combining DFT and DMFT
gives a more realistic estimation of the absolute intensity of the local dynamic susceptibility, and can
quantitatively model
the electron and hole-doping evolution of the spin excitations in absolute units \cite{MWang13}.  Furthermore,
it can account for the pnictogen height dependence of the spin-wave spectra \cite{clzhang14}.

The static AF order and spin excitations in iron-based materials can also be understood by hybrid models consisting of
local moments on each Fe site
 and itinerant electrons from the degenerate $d_{xz}$ and $d_{yz}$ orbitals
\cite{SPKou2009,WGYin2010}.  In this picture, the local moments interact with each other via $J_1$ and $J_2$ Heisenberg
exchanges, and they are coupled to the itinerant electrons
via Hund’s rule coupling.  Since itinerant electrons are only associated with
$d_{xz}$ and $d_{yz}$ orbitals that break the $C_4$ rotational symmetry of the
underlying $x$-$y$ lattice plane due to their different occupancies, these orbitals can form a Hamiltonian that drive the
in-plane magnetic anisotropy,  producing unfrustrated
collinear AF order and lifting the degeneracy of the $(1,0)$  and $(0,1)$ magnetic states \cite{WKu2009,FKruger2009,SPKou2009,WGYin2010,WLv2010,CCChen2010}. Here the magnetic anisotropy is due to purely
electronic ferro-orbital order that spontaneously
breaks the rotational symmetry of the underlying lattice and drives the observed
magnetic and structural transitions without Fermi surface nesting or magnetic
frustration \cite{WKu2009,FKruger2009}.  Using a fermionic representation of the local
moments and a generalized RPA framework, one can calculate the spin wave spectra of BaFe$_2$As$_2$
and find that the outcome is consistent
with spin excitations in the AF ordered and paramagnetic states
obtained from inelastic neutron scattering \cite{FYang2010,ZDLeong2014}.
In addition, the global phase diagram for the AF and superconducting states calculated from the hybrid model
on the mean-field level is qualitatively consistent with experiments
\cite{YZYou2011}. At high characteristic temperatures, electrons in more localized
orbitals of the multiband system may first form short-ranged AF order.  Upon cooling
to lower temperatures, the electrons in more itinerant orbitals can be driven
into a true static AF ordered or superconducting state via Hund's coupling
to the preformed localized AF state.  This is analogous to the orbital-selective Mott
transition, where itinerant and localized electrons in different orbitals may separate
as independent degrees of freedom \cite{SPKou2009,WLv2010,FYang2010,ZDLeong2014,YZYou2011}.

\section{Summary and outlook}

In this paper, we have reviewed recent progress of neutron scattering studies of the static AF order
and spin dynamics in iron-based high temperature superconductors.  Soon after the discovery of these materials in 2008 \cite{kamihara},
neutron diffraction measurements at the NIST center for neutron research
and high flux isotope reactor at Oak Ridge National Laboratory
 have determined the AF order and crystalline structures of the parent and superconducting compounds \cite{cruz,qhuang}.
These measurements have established the basis that superconductivity in iron-based materials arises from the  suppression of static long-range ordered antiferromagnets, much like copper oxide superconductors \cite{Tranquada2}.
When single crystals of the 122 family of iron pnictides and iron chalcogenides have become available, the advanced time-of-flight
neutron spectrometers at spallation neutron sources at Oak Ridge National Laboratory
and ISIS at Rutherford-Appleton Laboratory have allowed detailed mapping of the spin wave spectra throughout the Brillouin zone \cite{sodiallo09,jzhao09}.  This first occurred only slightly more than one year after the discovery of the iron pnictide superconductors.  In copper oxide superconductors \cite{bednorz}, the first complete spin wave spectrum was measured
15 years after its discovery \cite{RColdea}.  Using the
overall spin wave spectra in the AF ordered iron pnictides, one can fit the dispersion curves with a Heisenberg Hamiltonian, revealing the anisotropic in-plane effective magnetic exchange couplings.  This has inspired much discussion on the microscopic origin of the
effective magnetic anisotropy as described in Section IV.

Since high quality single crystals of electron and hole-doped BaFe$_2$As$_2$ are available, most of the
elastic and inelastic neutron scattering experiments have been carried out on these materials.  With elastic neutron scattering experiments, one can map out the electron and hole doping evolution of the structural and magnetic phase diagrams.  For electron-doped materials
obtained via Co and Ni substitution
of Fe, the tetragonal-to-orthorhombic lattice distortion precedes the AF phase transition, and the static long-range order coexists/competes with superconductivity in the underdoped regime.  However, the AF order becomes incommensurate with a
short-range correlation length
near optimal superconductivity, indicating that it is a spin-glass phase in the matrix of the
superconducting phase, coexisting and competing with superconductivity \cite{apdioguardi,xylu14,Bernhard2012}.
NMR measurements on 1111 family of materials also suggest nanascale electronic inhomogeneity \cite{GLang2010}.
For hole-doped
Ba$_{1-x}$K$_x$Fe$_2$As$_2$ and Ba$_{1-x}$Na$_x$Fe$_2$As$_2$, the structural and magnetic phase transitions are coupled at
at all doping levels \cite{savci12,savci13,savci14}.  Near optimal superconductivity, there is a new magnetic phase within the tetragonal
structure, possibly associated with a spin nematic phase, and the transition from AF order to superconductivity may also occur in the
first order fashion \cite{savci13,savci14}.  Finally, although transport and NMR measurements suggest the presence of a quantum critical point
near $x=0.3$ for isoelectronically doped BaFe$_2$(As$_{1-x}$P$_x$)$_2$ \cite{tshibauchi14},
neutron diffraction measurements have only mapped out the magnetic and structural phase diagram
in the underdoped regime \cite{JMAllred2014}, and much is not known about the evolution of AF order
near optimal superconductivity.

Similar to neutron diffraction work, most of the inelastic neutron
scattering studies of spin excitations in iron pnictides have been focused on
electron-doped BaFe$_2$As$_2$.  Compared with the undoped parent compounds, electron-doping appears to
modify spin excitations below $\sim$80 meV while leaving high energy spin excitations mostly unchanged \cite{HQLuo13}.  However, hole-doping suppresses high-energy spin excitations and transfers the spectral weight to low-energies \cite{MWang13}.  In addition, the wave vector dependence of the low-energy spin excitations in iron pnictides appears to be controlled by the quasiparticle
nesting between the hole and electron Fermi surfaces.  These results are consistent with the notion that spin excitations in iron pnictides have both local and itinerant character with the electron correlations controlled by the
pnictogen height and strength of the
Hund's coupling \cite{KHaule2009,SPKou2009,WGYin2010}.  The availability of large single crystals of
iron chalcogenides Fe$_{1+y}$Te$_{1-x}$Se$_x$
 means that spin excitations in these materials have been carefully mapped out \cite{OJLipscombe,IAZaliznyak,MDLumsden10}.  In particular, application of the sum rules of neutron scattering indicate
that the integrated spin excitation intensity of Fe$_{1+x}$Te is inconsistent with an
$S = 1$ Fe$^{2+}$ ground state expected in the presence of a strong crystalline electric field \cite{CStock2014}, suggesting the
importance of itinerant electrons even for the iron chalcogenides, which exhibit strong electron
correlations and localized moments \cite{ZPYin2011}.

Compared with electron-doped BaFe$_2$As$_2$, spin excitations in
hole-doped Ba$_{1-x}$K$_x$Fe$_2$As$_2$ and isoelectronically doped BaFe$_2$As$_{2-x}$P$_x$ iron pnictides have been much less studied \cite{CHLee11,CHLee13,CLZhang11,MWang13}.  Given the recent discovery of the possible spin nematic phase
in the tetragonal phase \cite{savci14}, it will be interesting to study the evolution of the overall
spin excitations in hole and isoelectronically doped iron pnictides.  In particular, since the electron pairing symmetry of the heavily hole-doped superconducting Ba$_{1-x}$K$_x$Fe$_2$As$_2$ is still unclear \cite{FFTafti2013,YOta2014}, it is important to carry out temperature dependent measurements to study the effect of superconductivity on low-energy spin excitations.  A determination
of the wave vector and energy of the superconductivity-induced neutron spin resonance will put considerable constraint on the nature
of the superconducting pairing state.

Although most neutron scattering work has focused on the 122 family of iron pnictides and iron chalcogenides, the 111 family, including
Co-doped NaFeAs and LiFeAs, is equally interesting since these materials may be more correlated than the 122 family \cite{ZPYin2011}.
At present, spin waves throughout the Brillouin zone have been mapped out for NaFeAs \cite{clzhang14}.  It is important to determine how electron-doping affects the spin excitations and to compare the outcome with the pure LiFeAs and Co-doped LiFeAs.  Similarly, it is important to study temperature and doping dependent spin excitations in Se-overdoped Fe$_{1+y}$Te$_{1-x}$Se$_x$ and pure FeSe.  The case of pure FeSe is particularly interesting as this is the system where the structural phase transition happens without static AF order \cite{johnston,GRStewart}. A complete understanding of this material may reveal spin or orbital driven electronic nematic phase.
Detailed experiments on other iron-based superconductors and associated materials are necessary
to establish the common features of the magnetism in various materials and their connection
to high-$T_c$ superconductivity.  Neutron scattering, together with RIXS, $\mu$SR, and NMR, can play a
unique role in our quest to find
the microscopic origin of high-$T_c$ superconductivity.

\begin{acknowledgments}

The author expresses sincere appreciation to Xingye Lu and Yu Song for preparing the figures of the present manuscript.
He is also in debt to David Tam, Yu Song, and Qimiao Si for a critical reading of the manuscript.
The former and current graduate students/postdocs in the author's group,
including Shiliang Li, Clarina de la Cruz, Jun Zhao, Songxue Chi, Oliver Lipscombe, Leland Harriger, Miaoyin Wang, Mengshu Liu, Meng Wang,
Xingye Lu, Huiqiang Luo, Chenglin Zhang, Scott Carr, and Yu Song, have played a key role in obtaining the neutron scattering results
reported here.  They continue to educate the author on the interesting physics in iron-based superconductors.
The author is in debt to his colleagues G. F. Chen, Elbio Dagotto, Takeshi Egami, DongLai Feng,
Kristjan Haule, Stephen Hayden, Jiangping Hu, Gabriel Kotliar, Jeff Lynn, Thomas Maier, Herb Mook, Andriy H. Nevidomskyy, Toby Perring, L.-P. Regnault, Qimiao Si, Yixi Su, Tomo Uemura, Nanlin Wang, Zhiping Yin, and Rong Yu for fruitful collaborations.
The author also thanks Xianhui Chen, Hong Ding,
 M. H. Fang, , D.-H. Lee, Tao Xiang, Haihu Wen, Z. Y. Weng, X. J. Zhou,
and friends in Beijing Forum for high-$T_c$ superconductivity
for many helpful discussions over the years.
The materials synthesis and neutron
scattering efforts on the hole-doped 122 and 111 family of materials are supported by the
U.S. DOE, Office of Basic Energy Sciences, under contract No. no. DE-SC0012311.
The neutron scattering work on the electron-doped 122 family of materials is supported by U.S. NSF, DMR-1362219.
The combination of RPA calculation and neutron scattering is supported by U.S. NSF, DMR-143606 and DMR-1308603.
Part of the materials work at Rice University is supported by the Robert A. Welch Foundation Grants No. C-1839.
\end{acknowledgments}

\bibliographystyle{apsrmp4-1}
\bibliography{rmp}
\end{document}